\documentstyle[preprint,aps,eqsecnum,axodraw]{revtex}
\tighten
\begin{document}
\baselineskip=14 true pt
\centerline{\it Title:}
\centerline{\large\bf {Selected Topics in Light Front Field Theory and
Applications to}}
\centerline{\large\bf {the High Energy Phenomena}}
\vskip 2cm
\centerline{\bf Abstract}
\vskip .2cm
In this thesis, we have tried to put 
forth some of the aspects of light-front (LF)
field theory through their successful application in the 
Deep Inelastic Scattering (DIS). 
We have developed 
a LFQCD Hamiltonian
description of the DIS structure functions starting from 
Bjorken-Johnson-Low limit of virtual forward 
Compton scattering amplitude 
and using LF current commutators.
We worked in the 
LF gauge $A^+=0$  and used the old-fashioned LFQCD 
perturbation theory in our calculations.
The importance of our work can be
summarized from two different viewpoints. 

Firstly, from  DIS point of view, our approach is convenient for it 
closely follows the physical intuitions of the parton model. 
Our approach addresses directly the structure functions, which are
experimental objects, instead of its moments which come naturally in the
usual way (i.e., using QCD improved parton model or OPE method). 
More importantly, our approach 
has the potential of incorporating
the non-perturbative contents of the structure functions. 
We have shown introducing a new factorization scheme that the
non-perturbative contents of the DIS structure functions can be obtained by 
solving the
LF bound state equations (which seems viable due to the ongoing
research activity towards this direction using the
similarity renormalization scheme), while  
the perturbative contents can be extracted by calculating 
the dressed parton structure functions as we worked them out explicitly  
to the leading order. 
In contrast, in the usual approaches nonperturbative information  
is only parametrized, putting the   
emphasis only on the $Q^2$-evolution of the moments of the structure
functions using perturbation theory. 

Simplicity of our approach becomes evident when we try to
describe structure functions in the context of the nucleonic helicity by
defining new structure functions. The ambiguity of gauge
invariance and interaction dependence in defining various parts of the 
helicity operator for
quarks and gluons in the usual way are absent in our gauge fixed theory
and the well known LF helicity operator  seems to provide the
consistent physical information as we have shown explicitly. Then, we
proceed to calculate anomalous dimensions relevant for the $Q^2$-evolution
of such structure functions which agree with recent calculation using other
methods. 

Secondly, our  study is important in view of establishing LF field theory as
a valid framework. Even though  LF field theory is a self-consistent theory, 
there is still some doubt whether it is {\it equivalent} 
to the more familiar equal-time one, 
since the formulation and the
methods used in two cases are so different, which we discussed in detail. 
Perturbative calculations of the dressed parton structure functions presented 
in this work show that,
to the leading order in coupling, our approach yields equivalent results as in
the usual way and hence the equivalence. 
Also, our investigation in the context of
coupling constant renormalization in LFQCD hamiltonian methods
shows the importance of Gallilean boost symmetry in understanding the
correctness of any higher order calculation using ($x^+$)-ordered 
LFQCD perturbation theory. 

\newpage
{\flushleft {\huge\bf Contents}}
\vskip 1cm 
\noindent {\large\bf Chapter 1} \hspace{13cm} {\bf 1}\\

\indent 1.1 Prologue\\ 
\indent 1.2 Motivation and the Organization of the Thesis.\\

\noindent {\large\bf Chapter 2} \hspace {13cm} {\bf 9}\\

\indent Light-Front Field Theory: A Review on Selected Topics\\

\noindent {\large\bf Chapter 3} \hskip 13cm {\bf 22}\\

\indent 3.1 A Brief Overview on Deep Inelastic Structure Functions\\
\indent 3.2 Light-Front Current Commutator and BJL Theorem\\
\indent 3.3 The Generalized Expressions for Deep Inelastic Structure 
Functions\\

\noindent {\large\bf Chapter 4} \hskip 13cm {\bf 44}\\

\indent 4.1 Unravelling the Complexities of Structure Functions\\
\indent 4.2 Unpolarized Dressed Parton Structure Functions \\
\indent 4.3 Structure Function of Hadron: Parton Picture, Scale Evolution
and Factorization\\
\indent 4.4 Physical Interpretation of the Structure Functions from Sum
Rule\\

\noindent {\large\bf Chapter 5} \hskip 13cm {\bf 77}\\

\indent 5.1 Transverse Boost Symmetry: Canonical Considerations\\
\indent 5.2 One Loop Calculations\\
\indent 5.3 Coupling Constant Renormalization\\
\indent 4.4 Discussion\\

\noindent {\large\bf Chapter 6} \hskip 13cm {\bf 94}\\

\indent 6.1 Light-Front Helicity Operator $J^3$ from the Manifestly Gauge
Invariant\\
\indent\indent Energy Momentum Tensor\\
\indent 6.2 Orbital Helicity Distribution Functions\\
\indent 6.3 Perturbative Calculation of Anomalous Dimensions\\
\indent 6.4 Verification of Helicity Sum Rule \\

\noindent {\large\bf Summary and Conclusions}  \hskip 8.9cm {\bf 114}\\

\noindent {\large\bf Appendix A}  \hskip 12.4cm {\bf 116}\\
\noindent {\large\bf Appendix B}  \hskip 12.4cm  {\bf 118}\\
\noindent {\large\bf Appendix C}  \hskip 12,4cm  {\bf 119}\\
\noindent {\large\bf Appendix D}  \hskip 12.4cm  {\bf 122}\\

\baselineskip 20 true pt
{\flushleft\huge\bf {Chapter 1}}
\section{Prologue}
The interest in the high energy community has become very much diverse and,
at the same time, very intense. It is so much so, that it becomes 
very difficult
sometimes for one  working in a particular field to find a connection
between his or her work with that of another one working in a 
different field, let alone understanding that. This is particularly true for
one relatively new in the field, to whom it may appear that a lot of
research activities are going on outside his or her future light cone (in a
figurative sense, of course) and hence, is unable to find the causal 
connection. 
In spite of all these ever-growing diversity and 
intricacy in modern 
day research, some of the very basic questions that intrigued our mind have
not changed over the years. Namely, what are the basic building blocks
of matter and how they conspire to build up the nature as we see it? As a
matter of fact, the answer to these questions did change from time to time. 

As far as the basic building blocks are concerned, all forms of matter were
thought to be 
built out of what is known as {\it chemical elements} in the early
days of Mendeleev, who successfully organized them by their various unique
properties. But the profusion of new elements coming into being led to the
suspicion that there might be something more fundamental out of which all
these chemical elements were composed of. Around 1932, Chadwick's discovery
of neutron put the Particle Physics on the solid premise for the first
time, with {\it electrons}, {\it protons} and {\it neutrons} 
being established as the
fundamental and elementary building blocks of matter. In fact, that was also
the last time when Particle Physics was so simple and complete (apparently).
As history repeated itself, all kinds of new particles like mesons, hadrons
and other leptons were discovered, which appeared no less elementary than
proton or neutron. That again hinted towards further elementary
constituents of matter. Now we know that the {\it quarks} and {\it 
leptons} and the {\it vector
bosons} that mediate interactions are the most elementary fundamental
particles. As it stands now,
in accordance to the Standard Model, 
already the number of elementary particles (considering different colours as
different species) 
amounts to 
sixty
(Higgs!)\cite{1grif}{\footnote {For each Chapter references are provided at
the end of that Chapter.} and the number can easily be increased 
if the nature prefers to have the symmetry like Supersymmetry.  
The story, of course, does not end here and
extends to the String Theory (supposed to be the theory of everything)
where the most elementary building block is a {\it string} and all the
present day particles are various excitation modes of the string. 
Only time can
tell whether naming quarks and gluons and so on as the elementary
constituents is premature. On the experimental side, one of the most
important high energy experiment devised to partly answer this question is
the {\it Deep Inelastic Scattering} (DIS) experiment  
where a high energy electron
probes the interior of a hadron. Thus, proper understanding of DIS 
phenomena is very important in order to find evidence regarding the basic
building blocks in nature. We will devote a major portion of our work
towards understanding the DIS phenomena. 

Now we go back and concentrate a little bit on the other aspect of the
question that we posed, 
namely, what actually controls the underlying behavior when  these
elementary 
constituents interact among themselves. From our experience, we know
that all the basic phenomena are governed by four kinds of underlying
interactions,  viz., strong, electromagnetic,  weak and gravitational
interactions. Classical principles of physics, which are so successful in
explaining, for example, planetary motion or the motion of a billiard ball,
fails to describe phenomena when it comes to electrons and protons. The fact
that simultaneous measurement of two variables, if they are conjugate to
each other, can not yield results with arbitrary level of accuracy (as first
pointed out by Heisenberg), revolutionized our understanding at the
microscopic length scale determined by the fundamental constant $h$, 
known as the 
Planck's constant. Thus, starting from atomic to further smaller length scale,
the description of physical phenomena is given in terms of {\it quantum
mechanical system} where the 
{\it Quantum Principles} take over from {\it Classical Principles}. Also,
for particles travelling with speed comparable to that of light, {\it
Special theory of relativity} (as discovered by Einstein) takes over from
{\it Gallilean relativity} in classical physics. Now, as it turns out, if one
tries to keep track of particles at very small length scale and at the same
time allows them to have very high velocities (as is the case in high energy
physics), then for the sake of consistency what emerges is known as {\it
Quantum field theory}, which actually describe the physical world.

In our work, we shall use quantum field theory, to be more precise, 
light-front Hamiltonian formulation of 
Quantum Chromodynamics and apply it to DIS, in
order to understand  the phenomena as well as the theory itself.
\section{Motivation and The Organization of the Thesis}
It is easier in explaining our motivation to introduce first the very
basic concepts involved in the light-front QCD. 
Let us briefly recall some of the features of
quantum field theory, in order to introduce the light-front QCD itself. 
After the success of Quantum Electrodynamics, it is now strongly believed
that a quantum mechanical system involving particles moving with
relativistic speed and their interactions can be best described by a local
quantum field theory\cite{1txtb}. 
The description of such a system involves infinitely
many degrees of freedom. Various particles are described by separate field
operators whose behavior under the Lorentz transformations determines what
kind of particle these are describing, i.e, scalar, vector or spinor.
Interactions among the particles are assumed to be local and given in terms
of a local Lagrangian or Hamiltonian.
In fact, we now know that strong, weak
and electro-magnetic interactions can be described by similar set of {\it gauge
theories}. In particular, the 
{\it Quantum Chromodynamics} (QCD) which is based on a 
non-abelian $SU(3)$-colour gauge theory describes the strong interactions
among the quarks and gluons,  
which are carrying special {\it colour charges} and are the basic fields
that enter into the  QCD Lagrangian.  In our work, 
we shall be mostly concerned with QCD.

Now, the basic algorithm in studying such a 
system is to start from a suitable local Lagrangian and get the equation of
motions satisfied by the fields using variational principle and then
quantize the system by assuming the values of the commutators of the fields
and their conjugate momenta on a space-like hypersurface (the so called
canonical quantization). This algorithm goes by the name of covariant
Lagrangian formulation, since it starts from a Lagrangian and does not need
to refer, in principle, to any particular Lorentz frame.
Free field equation of motion can be solved
exactly but these solutions are not very interesting for they describe
particles which are doing nothing but simply existing by themselves. On the
other hand, equation of motions involving non-trivial interacting fields are
in general highly nonlinear and rarely solved exactly. Only approximate
solutions can be obtained through iterative methods or by using perturbative
techniques. While the scattering problem can be treated using the whole
machinery of perturbation theory and well-studied, the bound state problems
are rarely so, let alone the analytic nonperturbative solution (save a few
exactly solvable models). In the case of QCD low 
energy bound state problems, 
we are unable to carry out perturbative calculations due to the
large value of the coupling in contrast to the situation in QED, where the
coupling  is small and still allows using perturbative techniques. 
The only hope for non-perturbative treatment of QCD bound state problem
seems to be the Lattice calculations which again has its own problem. Thus,
there is a need for an alternative to the covariant Lagrangian 
description where the QCD bound
state problem can be reasonably addressed.

On the other hand, there exists a parallel Hamiltonian formulation of field 
theory where the
starting point is an Hamiltonian obtained either from the Lagrangian or
constructed otherwise. Equations of motions are the usual Hamiltonian 
equations of motion as in the classical theory, with the Poisson brackets 
replaced by the 
commutators. Quantization conditions are set as in the earlier case on a
space-like hypersurface which serve as the initial conditions and the
Hamiltonian is responsible for the time evolution, i.e., gives the field
configurations at some later time, similar to the situation 
in nonrelativistic quantum
mechanics. Although the Hamiltonian formulation is somewhat natural and easy
to visualize especially due to the background experience in the study of 
quantum mechanics, one has to pay some price for that. A hint of
noncovariance creeps in while defining the Hamiltonian, which must be
specified on a particular time slice. In principle, both the formulations
should yield identical results {\it if} one can obtain the nonperturbative
solutions and which, at present, is not possible. 
In absence of the nonperturbative solutions, one has to rely on
perturbation techniques which not only are different for different
formulations but also depend on the particular frame in the second case.
Thus, one has to pay special attention in pursuing the Hamiltonian
formulation so that one does not confuse the real physics with what can just
be an artifact of a particular Lorentz frame.

In our work, we shall be concerned with the light front Hamiltonian
formulation where quantization surface is chosen to be the light-like
hypersurface and the light front Hamiltonian does the job of time evolution.
To be more precise, we define first the light-front co-ordinates
$\{x^+,x^-,x^\perp\}$ in terms of the 
Minkowskian co-ordinates $x^\mu$:
$$x^{\pm}\equiv x^0\pm x^3,\quad\quad x^\perp\equiv \{x^1,x^2\},$$ 
where
$x^+$ plays the role of {\it light-front time}. Inner product is defined in a
different way (see later), 
so that Lorentz invariant scalar remains the same: 
$$x^2\equiv x^\mu x_\mu \equiv x^+x^- - (x^\perp)^2=(x^0)^2 
-(x^1)^2-(x^2)^2-(x^3)^2\, .$$
Quantization surface is now chosen to be $x^+=0$, called the light-front, 
instead of $x^0=0$ in usual equal-time case and light-front hamiltonian
$P^-$ (which now 
is conjugate to $x^+$) does the job of new {\it time}-evolution.
It is hoped that in this description QCD bound states can be addressed in a
more tractable way than in other existing  description. 
We shall explain why it is so expected while introducing the 
light-front field theory 
in detail and reviewing its special features 
in Chapter 2, in the context of light-front QCD. 
There we also give an overview of the selected problems some of which we
attempt to understand in our work.

Now, that we know roughly what light-front field theory is (and details will
come in due course), it is time to spell out how we want to make use of it
and what has been our  motivation in this work. 
In field theory calculations, as for example in the higher order perturbative
calculations, the resulting expressions often become infinite. These
infinities are removed from the theory by redefining the physical parameters
in the theory. This is known as the {\it renormalization} of the
initial {\it bare} theory and  is unavoidable in field theory. 
As we shall see, the light-front {\it power counting}, which 
is quite different from that 
of the usual covariant description,  makes the  renormalization
procedure very different in light
front field theory. Renormalization procedure itself becomes somewhat
complicated in light-front QCD hamiltonian framework due to lack of
covariance. Also, due to the fact that light-front QCD essentially is the
dynamics of a constrained system, as we shall see, it is particularly
important to
work in the light-front gauge $A^+=0$ (see later). All these
differences, namely, quantization procedure, renormalization and
gauge-fixed noncovariant formulation, lead to the suspicion whether the
light-front field theory description is equivalent  
to the usual covariant one. As
is obvious, it is not possible to check this by comparing intermediate steps
of some calculation. The best way seems to be 
to try and apply both in a physical
phenomena where experiment can settle the issue. Keeping this objective 
in our mind, in this work, {\it we try to 
formulate a light-front hamiltonian description of DIS}  
where we have ample experimental evidence and theoretical understanding
using the usual covariant formulation of QCD for comparison.

There are, of course, other motivation for building up a parallel
description of DIS. First of all, DIS is a light cone dominated process and
it is strongly believed that light-front description may be ideally suited
here. For example, as we know, the understanding 
of gross features in DIS are most suitably realized in
terms of {\it parton distributions} as provided by Feynman's parton
model\cite{1feyn}, which  is simple and 
closely follows our physical intuition. 
In order
to define field theoretical generalization of these parton distributions,
people have used light-front description where it can be done most
conveniently  (as we shall also see).
But barring from a few attempt\cite{1brodsky}, there is no real extensive 
study exploring these possibilities. 
Also the fact that {\it constituent quark model} (CQM)\cite{1CQM}, where hadrons are 
thought of as composites of minimal number of quarks and anti-quarks and 
is so
successful in describing non-relativistic properties of hadrons, is not
realizable in QCD due to the complicated structure of the vacuum state in
its covariant formulation. As we shall see in Chapter 2, 
the simplicity of light-front
QCD vacuum, a unique property in light-front theory, gives us the hope of
reconciling CQM with the QCD.

Secondly, all the state-of-the-art
calculations using OPE methods or QCD improved parton model to understand
DIS up till now, concentrate solely on the evolution of the DIS structure
functions using perturbative QCD. As a consequence of using usual equal-time
covariant field theoretic formulation in describing the structure functions,
what naturally comes up in this analysis is the {\it moments} of structure
functions (instead of the structure functions themselves). This point is
clearly explained in Chapter 3. On the other hand, in DIS experiment, it is
the structure functions that are directly measured. Thus, for comparing
theoretical analysis with the experimental data, one has to extrapolate the
measured structure functions (to a domain of small $x$-values, 
which is still
not accessible to DIS experiments) in order to compute the moments bringing
in little uncertainty. Very little emphasis is put actually on studying the
structure functions themselves. In contrast, our approach addresses the 
structure functions themselves. 

Thirdly, as we shall explain in detail in Chapter 3, 
we have used the
equal light-front current commutators and light-front version of
Bjorken-Johnson-Low (BJL) theorem to get explicit expression for various 
structure functions closely following the current algebra methods of pre-QCD
era. These expressions are obtained without recourse
to the perturbation theory, and hence contain perturbative as well as the 
nonperturbative information contained in the  structure functions. 
In Chapter 4, we will show introducing a new factorization
scheme, how nonperturbative and perturbative contents of the structure
functions can be addressed in the same light-front Hamiltonian framework. As
it turned out, nonperturbative dynamics involved in the structure functions
can be obtained from the light-front bound-state equation and the
perturbative pieces can be easily obtained by working out the dressed parton
structure functions. This is particularly important from the viewpoint of
ongoing development towards solving QCD nonperturbatively using newly
proposed similarity renormalization scheme in the light-front QCD 
Hamiltonian
formalism, since any breakthrough here can now easily be incorporated to the
DIS structure functions giving rise to a complete picture. 
This is in contrast to the usually employed methods, which are not properly 
equipped to address the nonperturbative 
information contained in the structure functions. Instead, the
nonperturbative informations  are 
parametrized and the main theme of these studies has been the
$Q^2$-evolution of the structure functions in terms of moments, assuming its
form at a lower energy scale $Q_0^2$. 

Lastly, since this approach is going to be 
quite different from the commonly used ones, we can  expect some new results
and clear understanding of certain phenomena which otherwise are hard to
explain. As far as the light-front QCD is concerned, such an example, is
shown in Chapter 5, which shows the implication of certain special symmetry
in light-front theory in {\it simplifying} 
the higher order old-fashioned
time-ordered perturbative
calculations usually employed in light-front QCD. As far as DIS is
concerned, the examples are scattered in Chapter 3, 4, and 6 as well as  
some of the references mentioned therein. Specifically, in Chapter 6, we
shall discuss in detail, how our approach can be most naturally  extended 
towards understanding the distribution of nucleonic total helicity among its
constituents and gives us new insights to the helicity of a composite
dynamical system such as the  hadrons. 

We reemphasize the fact that our original works are presented in Chapter 4,
5 and 6; and also partially in Chapter 3.
\nopagebreak


\newpage
\setcounter{equation}{0}
\renewcommand{\theequation}{2.\arabic{equation}}
{\flushleft\huge\bf {Chapter 2}}
\vskip 1 cm
In this Chapter, we are going to introduce the basic features of light-front
field theory, in the context of light-front QCD (LFQCD), 
its advantages and the
problems that one must understand for a successful 
practical application of the theory.
Here we shall only be concerned with the Hamiltonian formulation of
LFQCD. 
In the usual Hamiltonian formulation of field theory, quantization
conditions in the form of commutator of dynamical fields 
and their conjugate momenta
are specified on the space-like hypersurface $x^0=0$ and the equal-time
Hamiltonian is responsible for the 
time-evolution of the system, i.e., gives the
field configurations at a later time. 
Dirac\cite{2dirac} first showed that consistent field theory
can be formulated by specifying the quantization conditions on a light-like
hypersurface $x^+=x^0+x^3=0$ (called a light front) and using a different
(light front-) Hamiltonian for a new time ($x^+$) evolution. This
formulation (in Dirac's terminology, ``front form") is known as the 
light-front 
Hamiltonian field theory. On the other hand, 
in an attempt
to see what happens if one traveling at the speed of light tries to
formulate field theory, by boosting the covariant field theory results to
a so called infinite momentum frame (IMF), following observation was
made\cite{2susk}.
Although the Lorentz transformation required to arrive at IMF is evidently
singular ($\gamma=1/\sqrt{1-\frac{v^2}{c^2}}\rightarrow \infty \,{\rm as}
\,v\rightarrow c$), the
singularity cancels in the calculation of physical objects (like Poincare
generators) and results in
an effective coordinate change given by
\begin{equation}
x^{(\pm)}=x^0 \pm x^3,\quad\quad {\bf x}^\perp=(x^1,x^2),
\label{lfcord}
\end{equation}
same as the light-front co-ordinate we defined in Chapter 1. Thus, one can see the fact
that what one obtains after going through singular limiting procedure in IMF
is built in quite naturally in the light front field theory. Naively, 
using quantum mechanics analogy, one can
expect for particles moving with relativistic speed where the particle 
world lines
cluster near the light cone, the quantization on a light front and evolving the 
system along $x^+$-axis by light-front Hamiltonian may be helpful. For this
reason light-front field theory is sometimes also referred as the field
theory of Infinite
Momentum Frame. But, we should reemphasize that the formulation here is as
prescribed by Dirac and has no connection with any singular limiting
procedure\cite{2lfb}. 
\newpage
\noindent {\bf 1. LF dispersion relation}
\vskip .2cm
Light-front co-ordinates given in eq.(\ref{lfcord}) are defined 
in terms of $x^\mu$ and should not be thought of as a Lorentz transformation.
The  
inner product between two four-vectors is defined on the light front as 
\begin{equation}
a\cdot b = {a^+b^-\over 2} + {a^-b^+\over 2} - a^\perp\cdot b^\perp.
\end{equation}
We define the light-front four momenta as
\begin{equation}
k^{(\pm)}=k^0 \pm k^3,\quad\quad {\bf k}^\perp=(k^1,k^2),
\end{equation}
 with $k^-$ being
conjugate to $x^+$ is the light front energy and $k^+$ is the longitudinal
momentum. With the above definitions, the  dispersion relation, i.e., the
relation between light-front energy $k^-$ and the spatial components of 
momenta $(k^+,{\bf
k}^\perp)$, for an on mass-shell  particle of mass $m$, is given by, 
\begin{equation}
 k^- =
{{\bf k}^2_\perp + m^2\over k^+}\, .
\label{ekd}
\end{equation}
First thing that one notices is the fact that there is no square root involved 
in contrast to  
the usual case $E=\sqrt{\vec{k}^2+m^2}$. This may provide great
simplification if one tries to solve eigenvalue equation like
$\hat{H}\mid
\psi\rangle = E\mid \psi \rangle$. Secondly, the numerator in eq.(\ref{ekd})
being always positive implies that the particles with positive light-front 
energy ($k^-$) always carries positive longitudinal momentum ($k^+$). 
As usual, the 
particles with negative $k^-$ which must have negative $k^+$
are mapped to antiparticles with positive $k^-$ and $k^+$. As a consequence,
we always have $k^+\ge0$ for real particles. 
This has great significance for the light front
vacuum which we shall discuss shortly. Finally, $k^-$ becomes large for the
large value of $k^\perp$ as well as very small values (near zero) of $k^+$.
This makes light front renormalization aspects very different 
from the usual one. We shall see later in this Chapter how these aspects
crucially dictate the build up of light-front theory. 
These drastic differences in the basic dispersion relation on the 
light-front makes light-front theory so different from the usual
covariant theory. 
\vskip .3cm
\noindent{\bf 2. Simplicity of LF vacuum}
\vskip .3cm
We first emphasize the uniqueness of light front vacuum as dictated by the
above dispersion relation. Vacuum state is
always an eigenstate of the longitudinal momenta $\hat{P^+}\mid0\rangle =0$.
Now, the condition $k^+\ge0$ implies that the vacuum $\mid 0\rangle$ is
either devoid of particles or, at most can have particles with longitudinal
momenta exactly equal to zero, $k^+=0$ (so called zero modes). 
Now, zero modes will have infinite energy unless both ${\bf k}^\perp=0$ and
$m=0$. This makes it sensible to replace the zero modes by an effective
interaction, since this exactly is the strategy used when renormalizing the
divergences away. Thus, we always consider a cut-off theory where
longitudinal momentum is restricted to be $k^+>\epsilon$ for all the states. With
this prescription, $\mid 0\rangle$ becomes completely devoid of any particle
and therefore, an eigenstate of the full interacting hamiltonian with zero
eigenvalue, i.e., in cut-off theory
\begin{equation}
 \hat{P^+}\mid0\rangle =0 \quad\Rightarrow\quad
\hat{P^-}\mid0\rangle =0.
\end{equation}
Thus, the light-front vacuum has become  {\it trivially} simple. 
At the same time, it puts a restriction on a state with finite $P^+$, which
can now contain at most $P^+/\epsilon$ constituents. 

This is very different from the equal-time 
case where the vacuum has highly complicated structure, since it can
contain, in principle,  infinite number of particles moving 
with positive and negative momenta adding up to
zero. This
actually makes the the idea of minimal constituents of CQM impracticable in
covariant QCD. On the other hand, this complicated vacuum structure  is
supposed to be responsible for spontaneous chiral symmetry breaking or
confinement in QCD. It seems that with the trivial vacuum structure in
light-front theory after removing the zero modes, we may lose these
important aspects in our theory. 
It should be emphasized
that we have not simply removed the zero modes from our theory. The
longitudinal momentum cut-off ($\epsilon$) 
should be removed from the theory at the end of
any calculation by adding necessary counter terms in the effective
Hamiltonian to render the observables
independent of $\epsilon$. Thus, we expect to get back all the effects of
zero mode as an effective interaction in the  Hamiltonian through
renormalization. 

\vskip .3cm
\noindent{\bf 3. Poincare generators in LF}
\vskip .3cm
Before we proceed further, let us first highlight the dynamical structure of
any light-front theory, which again is unique and very different from
the equal-time version of the theory. 
Any dynamical system
can be described by {\it ten} dynamical variables (Hamiltonian ($P^0$), {\it 
three}  linear
momenta ($\vec P$), {\it three} angular momenta ($\vec J$) and {\it three} 
boosts 
($\vec K$)). As is well-known, 
in field theory they 
become operators that generate the corresponding changes of the state  
vectors and known as Poincare generators, which 
satisfy Poincare algebra. In equal time theory, we know that six of
them are kinematical operators $ \{{\vec P}, {\vec J}\}$ 
(i.e., do not depend on the dynamics) and the 
rest
are dynamical $\{ P^0,  {\vec K}\}$.  
In light-front theory, boost operators become kinematical. Longitudinal
boost is like a scale transformation and the transverse boosts behave like
Gallilean boosts in the nonrelativistic theory. On the other hand, two
rotations (about transverse axes) which are kinematical in equal-time case
become dynamical in light-front theory, other than the Hamiltonian itself. 

Let us consider an example which elucidates further the difference between  
light-front case and the usual equal-time one.  
Specifically, we consider the effect of boost on the space-time co-ordinates 
along 3-axis ($K^3$) with a speed $v$:
\begin{eqnarray}
{\tilde x}^0= \gamma(x^0-\beta x^3),~~~
{\tilde x}^3= \gamma(x^3-\beta x^0),~~~
{\tilde x}^{1,2}= x^{1,2},
\end{eqnarray}
where $\beta={v\over c}$ and $\gamma={1\over \sqrt {1-\beta^2}}$.
It is clear from the above equation that under $K^3$, the quantization
surface $x^0=0$ (in equal-time case) evolves to something else, which
requires dynamical information and hence, $K^3$ is a dynamical generator.  
Introducing the parameter $\phi$ such that $\gamma=\cosh\phi$ and 
$\beta\gamma=\sinh\phi$, we see that,
\begin{equation}
{\tilde x}^{+}= {\tilde x}^0+{\tilde x}^3=
e^{-\phi}~x^+,~~~{\tilde x}^{-}={\tilde x}^0-{\tilde x}^3 = e^{\phi}~x^-,~~~
{\tilde x}^{1,2}= x^{1,2}\, .
\end{equation}
It clearly shows that $K^3$, which is  known as generator of 
longitudinal boost in
light-front, behaves like 
a scale transformation. In particular, it keeps the
quantization surface $x^+=0$ invariant. Therefore, it is a kinematical
generator in light-front theory. 

Light-front Poincare generators are obtained in the same way as in
equal-time case. Namely, starting from Lagrangian density we obtain the
energy momentum stress tensor $T^{\mu\nu}$ and then integrate over a
space-time hypersurface. Only difference being the following: the role of
time $x^0$ in equal-time is replaced by the new time $x^+$ in the light-front 
and consequently, the integration surface is changed to a light-like surface
which is normal to the new time $x^+$-direction. Thus 
\begin{equation}
P^\mu = {1\over 2}\int dx^-d^2x^\perp T^{+\mu}\, ,
\end{equation}
\begin{equation}
M^{\mu\nu} = {1\over 2}\int dx^-d^2x^\perp \big[ x^\nu T^{+\mu}-x^\mu
T^{+\nu}\big]\, .
\end{equation}
Note that $M^{\mu\nu}$ is antisymmetric and hence has six independent
components. In light-front dynamics $P^-$ is the Hamiltonian and $P^+$ and
$P^i$ with $(i=1,2)$ are the longitudinal and transverse momenta. 
$M^{+-}=2K^3$ and $M^{+i}=E^i$ are the
boosts. $M^{12}=J^3$ and $M^{-i}=F^i$ are rotations. For details of 
the Poincare algebra
in light-front see Ref.\cite{2lfb}. For our purpose, we 
notice that the boost
generators form a closed algebra among themselves:
\begin{equation}
\big[ E^1, E^2\big]=0,~~\big[ K^3, E^i\big]= iE^i,~~
\end{equation}
and 
\begin{equation}
\big[J^3, E^i\big]=i\epsilon^{ij}E^j,
\end{equation}
which is similar  to the generators of non-relativistic dynamics in
a plane where $K^3$ has, of course, no role. 
This suggests the fact that there are built-in 
non-relativistic structures
in the relativistic light-front dynamics, as was also evident from the
dispersion relation in eq.(\ref{ekd}). In Chapter 5, we will see the
consequences of such underlying structure in simplifying  the relativistic
field theory calculations using old-fashioned perturbation theory, which is 
appropriate for the light-front QCD Hamiltonian formulation. 

To sum up, we see
that kinematical subgroup of the Poincare group enlarges and contains seven
generators in light-front theory. This may prove easier in 
understanding a dynamical system, 
since now we can fix more variables of the system, irrespective of any
knowledge regarding the real dynamics. Moreover, since different set of
generators are kinematical in light-front compared to the equal-time theory,
it is worth pursuing this theory, for certain things difficult to study in
equal-time may just become simpler here. One such example is the feasibility
of representing the QCD-bound states in terms of just a few 
boost invariant multi-particle wave-function in the Fock-space expansion, 
which we discuss next. 
\vskip .3cm
\noindent{\bf 4. Fock expansion for the bound state} 
\vskip .3cm
Since the Fock-states form a complete basis, any state vector, in
principle, can be
expanded in terms of that basis introducing corresponding amplitude for each
Fock-basis. 
For example, the bound state of a hadron on light-front can be simply 
expanded in terms of the Fock states as
\begin{equation}
        |PS \rangle = \sum_{n,\lambda_i} \int' dx_i d^2\kappa_{\bot i} 
		 | n, x_iP^+,x_iP_{\bot}+ 
		\kappa_{\bot i}, \lambda_i \rangle \Phi^S_n 
		(x_i,\kappa_{\bot i}, \lambda_i) \, , \label{2lfwf} 
\end{equation}
where $n$ represents $n$ constituents contained in the Fock state 
$|n, x_i P^+, x_i P_{\bot} + \kappa_{\bot i}, \lambda_i \rangle$, 
$\lambda_i$ is the helicity of the i-th constituent. $\int'$ denotes 
the integral over the space:
\begin{equation}  \label{2lfspc}
        \sum_i x_i = 1, ~~ {\rm and} ~~~ \sum_i \kappa_{\bot i} = 0,
\end{equation}
while $x_i$ is the fraction of the total longitudinal momentum 
carried by the $i$-th constituent, and $\kappa_{\bot i}$ is its relative 
transverse   momentum with respect to the center of mass frame:
\begin{equation}
        x_i = { p_i^+ \over P^+}~~, ~~~ \kappa_{i\bot} = p_{i\bot} - x_i 
P_{\bot}\, , 
\end{equation}
with $p_i^+, \,p_{i\bot}$ being the longitudinal and transverse momenta
of the $i$-th constituent. $\Phi^S_n (x_i,\kappa_{\bot i},\lambda_i)$ 
is the amplitude of the Fock state $| n, x_iP^+,x_iP_{\bot}+ \kappa_{\bot 
i},\lambda_i \rangle $, i.e., the {\it multi-parton wave function},
which is boost invariant and satisfies the normalization condition: 
\begin{equation}
        \sum_{n,\lambda_i} \int' dx_i d^2\kappa_{\bot i} 
		|\Phi^S_n (x_i,\kappa_{\bot i},\lambda_i)|^2 = 1,
\end{equation}
and is, in principle, determined from the light-front bound state 
equation,
\begin{equation}
        \Big(M^2 - \sum_{i=1}^n { \kappa_{i\bot}^2 + m_i^2 \over x_i} 
		\Big) \left[\begin{array}{c} \Phi^S_{qqq} \\
                \Phi^S_{qqqg} \\ \vdots \end{array} \right]
                  = \left[ \begin{array}{ccc} \langle qqq
                | H_{int} | qqq \rangle & \langle qqq | H_{int}
                | qqqg \rangle & \cdots \\ \langle qqq g
                | H_{int} | qqq \rangle & \cdots & ~~  \\ \vdots &
                \ddots & ~~ \end{array} \right] \left[\begin{array}{c}
                \Phi^S_{qqq} \\ \Phi^S_{qqqg} \\ \vdots \end{array}
                \right] . \label{bdseq}
\end{equation}
Here $H_{int}$ is the interaction part of the light-front QCD 
Hamiltonian given later.

In any practical application of this Fock-expansion we may face two
problems. Firstly, as we know, each Fock-state is obtained by operating
various creation operator(s) on the vacuum of the theory. Now, it can so
happen that the vacuum
already has a complicated structure (as is the case in equal-time theory),
which may contain arbitrary number of particles and thereby, needs a 
Fock-expansion in itself. This, in effect, render the Fock-expansion in
equal-time theory meaningless for any practical application. This is not the
case in light-front theory due to the simplicity of the vacuum.
Specifically, in the cut-off theory that we are going to use, it is trivial
as mentioned earlier and the Fock-expansion can be used meaningfully.
Also, the restriction $k^+>0$ makes the Fock-space smaller. 

Secondly, the expansion is still infinite and it is  
impossible to solve the bound state equation, eq.(\ref{bdseq}), which is an 
infinite dimensional coupled
equation. To make any practical
calculation viable using Fock-expansion, one needs to truncate the
expansion at a suitable maximum particle number (Tamm-Dancoff truncation,
TDF), with the hope that 
a first few terms in the
expansion may give useful information. 
This truncation in light-front theory violates rotational invariance about the
two transverse axes while in equal-time theory boost invariance is 
lost, since
by restricting the particle number the possibility of creating or
annihilating particles under these operations are restricted. It is
argued\cite{2cji} that the restoration of rotational invariance in the
light-front case
could be easier than the boost invariance in equal-time, 
since the rotation forms a
compact group compared to the boosts which are noncompact. 
Thus, the simplicity of vacuum and the kinematical nature of boost
transformation enable us to make use of  Fock-expansion in a practical
calculation in the  light-front QCD and we shall see such applications later
in our work.
\vskip .3cm
\noindent{\bf 5. Light-front QCD: Two component formalism}
\vskip .3cm
Here we briefly introduce the two component formalism of light-front QCD 
in order to introduce the basic features of the hamiltonian field theory. 
For details see Refs.\cite{2zhari1}\cite{2zhari2}. 
We start from the QCD Lagrangian
\begin{equation}
{\cal L}= -{1\over 2}{\rm Tr}
( F^{\mu\nu}F_{\mu\nu}) + {\bar \psi}(i\gamma_\mu
D^\mu -m )\psi \, ,
\end{equation}
where $F^{\mu\nu}=\partial^\mu A^\nu -\partial^\nu A^\mu -ig[A^\mu ,A^\nu ]$,
$A^\mu=\sum_aA^{\mu a}T^a$ is a $(3\times 3)$ gluon field colour matrix, and
$T^a$ are the generators of 
the SU(3) colour group: $[T^a, T^b]= if^{abc} T^c$
and ${\rm Tr}(T^aT^b)={1\over 2}\delta^{ab}$. The field variable $\psi$ describes
quarks with three colours and $N_f$ flavours, $D^\mu=
(\partial^\mu-igA^\mu)$ is the covariant derivative, and $m$ is an
$(N_f\times N_f)$ diagonal quark mass matrix. The Lagrange equations of
motion are:
\begin{eqnarray}
&& \partial_\mu F^{\mu\nu a}+gf^{abc}A_\mu^bF^{\mu\nu c} +g{\bar
\psi}\gamma^\nu T^a\psi=0\, ,\\
&&(i\gamma_\mu\partial^\mu -m +g\gamma_\mu A^\mu)\psi =0.
\end{eqnarray}
We always work in the light-front gauge $A^+=0$. We also define
$\psi=\psi^+ +\psi^-$, where $\psi^\pm=\Lambda^\pm \psi$ with
$\Lambda^\pm={1\over 2}\gamma^0\gamma^\pm$. (For details of notation and
convention, see Appendix A.) Now, in terms of these variables and in this
gauge, we get the following from the above equations of motion,
\begin{eqnarray}
{1\over 2}(\partial^+)^2A^{-a} &&~= \partial^+\partial^iA^{ia} + gf^{abc}
A^{ib}\partial^+A^{ic} +2g\psi^{+\dagger}T^a\psi^+\, ,\label{consrg}\\
i\partial^+\psi^- &&~= \big[ \alpha^\perp\cdot(i\partial^\perp +gA^\perp) +
\gamma^0 m \big]\psi^+ \, ,
\label{consr}
\end{eqnarray}
which are constrained equations, since these do not involve time derivative
$\partial^-$. Thus, in light-front variables and in light-front gauge,
$A^{-a}$ and $\psi^-$ are constrained fields. Their dynamics is constrained 
by the dynamics of the rest of the fields $A^{ia}$ and $\psi^+$, which are 
dynamical fields in light-front. Therefore, the dynamics involved in LFQCD
is  that of a constrained system, which is a general feature in light-front
field theory. 

To quantize such a system one may proceed according to the Dirac's method.
Alternatively, if one can solve the constrained fields in terms
of the dynamical fields and write down the Hamiltonian of the system
completely in terms of these dynamical fields, then the canonical 
quantization procedure goes through considering only the dynamical fields.
We shall follow the second path here. Conjugate momenta of the dynamical
fields are given by
\begin{eqnarray}
&&E^{ia}(x)={\partial {\cal L}\over \partial (\partial^-A^a_i)} =-{1\over
2}F^{+i\, a}(x)\, ,\\
&&\pi_{\psi^+}={\partial {\cal L}\over \partial (\partial^-\psi^+)}
={i\over 2} \psi^{+\dagger}\, ,\\
&&\pi_{\psi^{+\dagger}}={\partial {\cal L}\over \partial
(\partial^-\psi^{+\dagger})}
=-{i\over 2} \psi^{+}\, .
\end{eqnarray}
Now,  we separate the time derivative
terms in the Lagrangian, which helps us identifying the Hamiltonian of the
system. We rewrite the QCD Lagrangian as
\begin{eqnarray}
&&{\cal L}=\Big\{ {1\over 2}F^{+i\, a}(\partial^-A^{ia}) +{i\over
2}\psi^{+\dagger}(\partial^- \psi^+)-(\partial^-\psi^{+\dagger})\psi^+\Big\}
\nonumber\\
&&~~~~~~~~~~~~~~~
-{\cal H} -\Big\{ A^{-a}{\cal C}_a +{1\over 2}(\psi^{-\dagger}{\cal C}
+{\cal C}^\dagger\psi^-\Big\}\, ,
\end{eqnarray}
where 
\begin{eqnarray}
&&{\cal H} = {1\over 2} [(E^{-a})^2 +(B^{-a})^2] +{1\over 2}\big(
\psi^{+\dagger}\{\alpha^\perp\cdot(i\partial^\perp +gA^\perp) + \gamma^0 m\}
+ H.c.\big) \nonumber\\
&&~~~~~~~~~~~~~~~~+ [{1\over 2}
\partial^+(E^{-a}A^{-a}) -\partial^i(E^{ia}A^{-a})]
\label{hpre}
\end{eqnarray}
and ${\cal C}={\cal C}_a=0$ are exactly identical to the constrained
equations, eq.(\ref{consrg}) and eq.(\ref{consr}).
In eq.(\ref{hpre}) we have defined $E^{-a}=-{1\over 2}F^{+-\, a}$ and
$B^{-a}= F^{12\, a}$ as the longitudinal component of electric and magnetic
colour fields respectively. The reason for writing the Lagrangian in the
above form is to make the Hamiltonian density and the constraints manifest,
where constrained fields $A^{-a}$ and $\psi^-$ serves as the Lagrange's
multiplier.
Note that the Hamiltonian density ${\cal H}$ depends on the constrained
fields $A^{-a}$ and $\psi^-$. To obtain ${\cal H}$ in terms of the dynamical
fields alone, one needs to solve the constrained equations to eliminate 
$A^{-a}$ and $\psi^-$ in favour of $A^{ia}$ and $\psi^+$. In order to solve 
the constrained equations, 
we require a suitable definition of inverse longitudinal
derivatives $\big({1\over \partial^+}\big)$ and $\big({1\over
\partial^+}\big)^2$. We use the following:
\begin{eqnarray}
&&\big({1\over\partial^+}\big) f(x^-)= {1\over 4}\int^{+\infty}_{-\infty}
dy^- \epsilon(x^- -y^-)f(y^-) \, ,\nonumber\\
&&\big({1\over\partial^+}\big)^2 f(x^-)= {1\over 8}\int^{+\infty}_{-\infty}
dy^- \mid x^- -y^-\mid^2 f(y^-) \, ,
\end{eqnarray}
which amounts to assuming antisymmetric boundary conditions for the
dynamical fields at the longitudinal infinities, $x^-\rightarrow \pm\infty$.
See Ref.\cite{2zhari1} for detail discussion on boundary conditions.
We define two component quark fields $\xi$ as follows:
\begin{equation}
\psi^+= \left[ \begin{array}{c} \xi \\ 0\end{array}\right]~~\Rightarrow~~ 
\psi^-= \left[ \begin{array}{c} 0 \\ \big({1\over
i\partial^+}\big)[\sigma^i(i\partial^i +gA^i)+im]\xi\end{array}\right]
\end{equation}
Thus, the LFQCD Hamiltonian can now be expressed in terms of two component
dynamical quark and gluon fields:
\begin{equation}
H=\int dx^-d^2x^\perp ({\cal H}_0+{\cal H}_{int})\, ,
\end{equation}
where 
\begin{eqnarray}
&&{\cal H}_0= {1\over 2}(\partial^iA^{ja})(\partial^iA^{ja})+
\xi^\dagger\left(
{-(\partial^\perp)^2+ m^2\over i\partial^+}\right)\xi,\\
&&{\cal H}_{int}= {\cal H}_{qqg}+{\cal H}_{ggg}+ {\cal H}_{qqgg}+ 
{\cal H}_{qqqq}+ {\cal H}_{gggg}\, .
\end{eqnarray}
Here interaction Hamiltonian density is split into various pieces, which
give rise to various interaction vertices in light-front perturbation
theory. Notice that ${\cal H}_{qqqq}$ gives rise to four quark interaction
which is not present in the covariant version. For complete expressions 
of various interactions see Ref.\cite{2zhari2}. Here we give only those
expressions which we shall use later on.
\begin{eqnarray}
&&{\cal H}_{qqg}= g\xi^\dagger\Big\{-2\left({1\over
\partial^+}\right)(\partial^\perp\cdot A^\perp) + \sigma^\perp\cdot A^\perp
\left({1\over\partial^+}
\right)(\sigma^\perp\cdot \partial^\perp +m)\nonumber\\ 
&&~~~~~~~~~~~~
+ \left({1\over\partial^+}\right)(\sigma^\perp\cdot \partial^\perp -m) 
\sigma^\perp\cdot A^\perp \Big\}\xi \, ,\\
&&{\cal H}_{ggg}= gf^{abc}\Big\{\partial^i A^{ja}A^{ib}A^{jc} +
(\partial^\perp\cdot A^{\perp a})\left({1\over\partial^+}\right)
A^{jb}A^{jc}\Big\}\, .
\end{eqnarray}
The quantization conditions are specified on the light-front as follows:
\begin{eqnarray}
&&[A^{ia}(x), A^{jb}(y)]_{x^+=y^+} = -i\delta_{ab} \delta_{ij}{1\over
4}\epsilon(x^- -y^-)\delta^2(x^\perp -y^\perp)\, ,\label{acom}\\
&&\{\xi(x),\xi^\dagger(y)\}_{x^+=y^+} = \delta^3(x-y)\, ,
\end{eqnarray}
where $\delta^3(x-y)=\delta(x^--y^-)\delta^2(x^\perp-y^\perp)$.
Eq.(\ref{acom}) shows that the commutator between gluon fields themselves
are nonvanishing, which is very different from the usual equal-time case.
Also the presence of $\epsilon(x^--y^-)$ makes it nonlocal without violating
causality as will be discussed later on. Note the nonlocality is only in the
longitudinal direction $x^-$ and is one of the very important consequences
of quantizing the theory on the light-front.  

In the interaction picture, the equations of motion of the dynamical fields
are those of free fields and given by
\begin{eqnarray}
\partial^-A^{ia}(x)&=& {1\over i}[ A^{ia}(x), H_0]\nonumber\\
~~~&=& {1\over 4}\int^\infty_{-\infty} dy^- \epsilon(x^--y^-)
(\partial^\perp)^2 A^{ia}(x^+,y^-,x^\perp)\, ,\\
\partial^-\xi(x)&=& {1\over i}[ \xi(x), H_0]\nonumber\\
~~~&=& {1\over 4}\int^\infty_{-\infty} dy^- \epsilon(x^--y^-)
[(\partial^\perp)^2-m^2] \xi(x^+,y^-,x^\perp)\, ,
\end{eqnarray}
and their solutions are
\begin{eqnarray}
A^i(x)&=&\sum_\lambda \int {dq^+d^2q^\perp\over
2(2\pi)^3q^+}[\varepsilon^i_\lambda a(q,\lambda)e^{-iqx} + H.c.],\label{axi}\\
\xi(x)& =& \sum_\lambda \chi_\lambda\int {dp^+d^2p^\perp\over
2(2\pi)^3\sqrt p^+}\big [ b(p,\lambda)e^{-ipx} + d^\dagger
(p,-\lambda)e^{ipx}\big]\,
,\label{xix}
\end{eqnarray}
with $q^-={(q^\perp)^2\over q^+}$ and $p^-={(p^\perp)^2+m^2\over p^+}$. In
eq.(\ref{axi}) and eq.(\ref{xix}), $\lambda$ is defined to be
\begin{equation}
\lambda=\left\{\begin{array}{l} 1\\-1 \end{array}\right.
~~{\rm for~ gluons,}~~~~~~~
 \lambda = \left\{\begin{array}{l} 1/2\\-1/2 \end{array}
\right.~~{\rm for~ quarks.}
\end{equation}
The gluon polarization vectors are $\varepsilon^i_1={1\over \sqrt 2}(1,\,i)$
and $\varepsilon^i_{-1}={1\over \sqrt 2}(1,\,-i)$. The quark spinors are
simply the eigenstates of a spin-1/2 non-relativistic particle,
$\chi_{1\over 2}=\left(\begin{array}{c}1\\0\end{array}\right)$ and 
$\chi_{-{1\over 2}}=\left(\begin{array}{c}0\\1\end{array}\right)$.
The creation and annihilation operators in eq.(\ref{axi}) and eq.(\ref{xix})
satisfy the basic commutation relations
\begin{eqnarray}
[a(q,\lambda), a^\dagger(q^\prime,\lambda^\prime)]&=&
2(2\pi)^3q^+\delta^3(q-q^\prime)\delta_{\lambda\lambda^\prime}\nonumber\\
\{b(p,\lambda), b^\dagger(p^\prime,\lambda^\prime)\}&=&\{d(p,\lambda), 
d^\dagger(p^\prime,\lambda^\prime)\}\nonumber\\
&=&2(2\pi)^3p^+\delta^3(p-p^\prime)\delta_{\lambda\lambda^\prime}
\end{eqnarray}

The above few paragraphs introduced the basic features of light-front field
theory in the context of LFQCD and the two component formalism we use for
LFQCD. The perturbative calculations in LFQCD are that of old-fashioned time
ordered Hamiltonian perturbation theory, for a overview of which we refer
Ref.\cite{2zhari2}. We will mention them as and when used.

\vskip .3 cm
\noindent{\bf 7. Renormalization Aspects}
\vskip .3 cm

In light-front field theory in the Hamiltonian framework, the renormalization 
is a more 
complicated issue mainly due to the noncovariant structure of the theory
and quite different compared to the usual covariant one. 
This is due to the fact that the {\it power 
counting in light front
is very different}. For a detail discussion on light-front power counting, 
see the Ref.\cite{2Wilson}. Here we notice the fact that only transverse
directions $x^\perp$ carry the mass dimension, while the longitudinal
direction $x^-$ has no mass dimension. Thus, one has to treat transverse and
longitudinal directions  separately in determining the superficial degree of divergence of a
divergent integral by power counting, in contrast to the covariant case
where all the space-time directions are treated democratically. This is
also evident in the single particle dispersion relation
$k^-={(k^\perp)^2+m^2\over k^+}$, which shows that there are two sources 
of divergences: $k^+\rightarrow 0+$ and $k^\perp\rightarrow\infty$. The
divergence coming from $k^+\rightarrow 0+$ is referred as infrared (IR) 
divergence, whereas $k^\perp\rightarrow\infty$ is known as the 
ultraviolate divergence (UV) 
in light-front theory. 
 
For the above reason, dimensional regularization, which is so elegant and
commonly used in covariant theory, is of very little importance in
light-front theory. Only in the transverse direction, one may use
dimensional regularization. But this is now no better than putting a 
simple cut-off to the transverse momenta, as shown, at least,  in the
context of wave-function renormalization in the Ref.\cite{2zhari2}.   
In fact, it turns out that to regularize the UV divergences in light-front
theory cut-off regularization is the most convenient method. IR divergences
are also regularized by putting a small longitudinal momentum cut-off, 
which is equivalent to
using principal value prescription for the integration over longitudinal
momenta. Also the fact that the light-front theory being gauge fixed and
noncovariant, leads to new type of divergences like quadratic divergences 
(if we are using cut-off instead of transverse dimeninsional regularization) 
in 
mass renormalization or mixed divergences involving both IR and UV ones. To
remove these divergences one has to add counter terms to the canonical
Hamiltonian, which are often nonlocal and help restoring the invariance of
the theory that might be broken in the process of manipulation. For detail
discussion on this subject, which is still an unsettled issue, see the
Refs.\cite{2zhari2}, \cite{2zhari3},\cite{2Wilson}, \cite{2Perry}. Another
method specially designed to address the bound state problem in light-front, 
is that of
{\it similarity renormalization} introduced by Glazek and Wilson, where
first an effective Hamiltonian is obtained perturbatively, by giving a
similarity transformation to the original Hamiltonian. A discussion on that
is beyond the scope of this work (see for an overview, Refs.
\cite{2Wilson}, \cite{2glaz}). 

For our purpose, we shall use small longitudinal momentum cut-off and an UV
cut-off  for transverse momenta in light-front perturbative calculations
in the context of DIS, as discussed later on.

\section* {COMMENTS}
In this Chapter, we provided an overview of 
some of the special features of light-front field  theory and what
can possibly be the advantages in describing a dynamical system in this
language\cite{2none1}. 
It all emerged from the fact that we defined our coordinates and
momenta as well as the inner product in a different way and subsequently,
quantized the theory on a light front while using a different ``time" 
evolution for the system. This makes light-front  power counting very 
different and hence, the renormalization. 
In effect, the formulation here becomes so different from the usual
covariant field theory that it is not obvious {\it apriori} 
whether the light-front  theory is equivalent to
the usual one. 
Lot of investigations are currently on to establish
the equivalence\cite{2bakk}. 
The best way seems to  be the {\it application of the theory in
a physical problem} where experiment can settle the issue.

At the same time such an application, for example in DIS, gives us
the  opportunity to study  QCD in a new but consistent way, which is
worth pursuing in the quest for a better understanding, especially for the
case of bound states. 
Feasibility of describing the hadrons in terms of Fock states
gives the hope of reconciling constituent quark model (which has been so
successful in explaining the hadronic spectra) with QCD. Perturbative
analysis in this context is also interesting  not only to see 
the predictive power of such a
theory but also to understand the Hamiltonian renormalization  itself, which is
so important for building up a consistent light-front Hamiltonian field
theory description.

 Towards this goal,
we have taken up the project of investigating DIS, in an attempt to obtain
 the nonperturbative and perturbative picture involved therein, 
in one consistent language, as
discussed  in the rest of this thesis.



\newpage

\setcounter{section}{0}
\renewcommand{\thesection}{3.\arabic{section}}
\setcounter{equation}{0}
\renewcommand{\theequation}{3.\arabic{equation}}

{\flushleft\huge\bf {Chapter 3}}
\vskip 1cm
The basic goal in the deep inelastic scattering experiments is to probe the
interior of a hadron target, supposedly a bound state in QCD that is
readily available in the nature. Although the hadrons are what we have at
our disposal, only quarks and gluons are the true dynamical degrees of
freedom in QCD. Details of how these quarks and gluons conspire to form a
low energy bound state hadron are not known and generically attributed to
some non-perturbative confinement mechanism. On the other hand, due to
asymptotic freedom, quarks and gluons are accessible in the very high energy 
in the sense that their dynamics can be understood in the perturbation
theory. Experiments like DIS give us the opertunity to
parameterize the non-perturbative effects in terms of some structure functions
and study their behavior due to the quark gluon dynamics in the high energy
regime as can be calculated using perturbative QCD.

Historically, the observation of Bjorken scaling in the early SLAC
experiments on DIS prompted the prediction of point like constituents in the
hadron and gave birth to the quark-parton model as a valid framework to
interpret the data in terms of parton distributions inside the hadron. Thus,
it turned out that deep inelastic structure functions can be used to measure
hadron's parton distributions. Later observation of logarithmic 
violation of scaling indicated that the non-abelian gauge theory of QCD
might be the correct theory of the strong interactions. More and more 
accurate measurements now left very little doubt regarding the unique 
description of the structure functions in terms of perturbative QCD. Parton 
distributions may still be measured, but one must account for their 
evolution with $Q^2$.  The very fact that the interpretation 
in terms of parton distributions was successful in explaining the early data
can now be attributed to the asymptotic freedom, one of  the unique 
feature of the non-abelian gauge theories like QCD. 
  
Thus, at present, the goal of studying DIS is twofold -- studying strong
interactions in terms of perturbative QCD and measuring various parton
distributions inside the hadrons. Keeping these two things in mind, we
attempt to formulate a description of DIS in the light front hamiltonian
framework which at the one end follows closely the intuitive partonic
interpretation and at the same time, takes into account QCD with its full
glory.
In this Chapter, we first review the basic 
ingredients of DIS and thereby introducing the notations. And then we
mention very briefly how one usually goes about to deal with them using
perturbative QCD, emphasizing the need for building up an alternative 
description of DIS what we are up to. Then we will discuss the light front
current algebra and the BJL theorem extended to the light front framework,
which happen to be our starting point towards building up the alternative
description. Lastly, we will discuss how we obtain various structure
functions as the Fourier transform of hadronic matrix elements of the bilocal
currents.
\section{A brief overview on deep inelastic structure functions}
We begin with a brief review of the basic ingredients of  
lepton-nucleon deep inelastic scattering (DIS):
\begin{equation}
	e(k) + h(P) \longrightarrow e(k^\prime)  + X(P+q) \, ,
\end{equation}
where we have specified the four momenta of the particles explicitly 
and $q=k-k^\prime$
is the momentum transfer in the process through the virtual photon. 
The inclusive cross section for the above scattering process is given by 
\begin{equation}
	{d \sigma \over d\Omega dE'} = {1\over 2M}{\alpha^2 \over q^4}
		{E'\over E} L_{\mu \nu} W^{\mu \nu} \, ,
\end{equation}
where $E$ ($E'$) is the energy of the incoming (outgoing) lepton,
$L_{\mu \nu}$ is the leptonic tensor,
\begin{eqnarray}
	L_{\mu \nu} &=& {1\over 2}\sum_{s'}[\overline{u}(k,s)\gamma_\mu 
		u(k',s') \overline{u}(k',s')\gamma_\nu u(k,s)] \nonumber \\
	&=& 2(k'_\mu k_\nu + k'_\nu k_\mu) - 2 g_{\mu \nu} k \cdot k'
		- 2i \epsilon_{\mu \nu \rho \sigma} q^\rho s^\sigma \, ,
\end{eqnarray}
and $W^{\mu \nu}$ is the hadronic tensor which contains all the hadronic 
dynamics involved in DIS process, 
\begin{equation}
	W^{\mu \nu} = {1\over 4\pi} \int d^4 \xi~ e^{iq \cdot \xi} 
		\langle PS |[J^\mu(\xi), J^\nu(0)]|PS \rangle  \, ,
\label{wmunu}
\end{equation}
where $P$ and $S$ are the target four-momentum and polarization 
vector respectively ($P^2=M^2, S^2=-M^2, S\cdot P=0$), $q$ is 
the virtual-photon four momentum, and $J^\mu(x)=\sum_\alpha 
e_\alpha \overline{\psi}_\alpha(x) \gamma^\mu \psi_\alpha(x)$ the 
electromagnetic current with quark field $\psi_\alpha (x)$ carrying
the flavor index $\alpha$ and the charge $e_\alpha$. 

The above hadronic tensor can be decomposed into independent
Lorentz invariant scalar functions:
\begin{eqnarray}
	W^{\mu \nu} &=&\Big(-g^{\mu \nu} + {q^\mu q^\nu 
		\over q^2} \Big) W_1(x,Q^2) + \Big(P^\mu - {\nu 
		\over q^2} q^\mu\Big)\Big(P^\nu -{\nu \over q^2} 
		q^\nu\Big)W_2(x,Q^2) \nonumber \\
	& & - i \epsilon^{\mu \nu \lambda \sigma}q_\lambda \Big[
		S_\sigma W_3(x,Q^2)+ P_\sigma S\cdot q W_4(x,Q^2) 
		\Big] \nonumber \\
	&=&  \Big( g^{\mu \nu} - {q^\mu q^\nu \over q^2} \Big)\Big({1\over 2}
		F_L (x,Q^2) -{M^2\over \nu}F_2(x,Q^2)\Big)\, 
		+ \Big [P^\mu P^\nu - {\nu \over q^2} \Big(
		P^\mu q^\nu + P^\nu q^\mu \Big)  \nonumber \\
	& & \quad\quad\quad\quad + g^{\mu \nu} {\nu^2 
		\over q^2} \Big] {F_2(x,Q^2) \over \nu}
		- i \epsilon^{\mu \nu \lambda \sigma}{q_\lambda \over \nu}
		\Big[ S_{\sigma L} g_1(x,Q^2) + S_{\sigma T} g_T
		(x,Q^2) \Big] \, . \label{wfun}
\end{eqnarray}
Here, in the first step, we have parametrized $W^{\mu\nu}$ in terms of four
scalar functions $W_i$'s with $\{i=1,2,3,4\}$ (as is usual in the parity
conserving cases), which are known as the structure functions. These structure
functions are again functions of two independent scalar variables present
in the problem, $Q^2=-q^2$ (the negative of momentum transfered
square) and the Bjorken scaling variable $x={Q^2\over 2\nu}$ with 
$\nu = P\cdot q$. In the next step, it is reparametrized in terms of
experimentally more accessible structure functions which can be written in
terms $W_i$'s as follows. 
\begin{eqnarray}
F_L(x,Q^2)&=& 2 \Big [-W_1 + \big [ M^2 -{(P.q)^2 \over q^2}\big ] W_2
\Big ]\\
F_2(x,Q^2) &=& \nu W_2(x,Q^2)\\
g_1(x,Q^2) &=&\nu \Big[ W_3(x,Q^2) + \nu W_4 (x,Q^2) \Big]~~~\\	
g_T(x,Q^2) &=& g_1(x,Q^2) + g_2(x,Q^2) = \nu W_3(x,Q^2)~~ 
\end{eqnarray}
$F_L(x,Q^2)$ and $F_2(x,Q^2)$ contained in the symmetric part of the 
$W_{\mu\nu}$ are known as the unpolarized structure functions, since only the
symmetric part of hadronic tensor contributes in the scattering from
unpolarized target. Whereas $g_1(x,Q^2)$ and $g_T(x,Q^2)$ are 
known as the longitudinal
and transverse polarized structure functions respectively. 
The longitudinal and transverse polarization vector 
components are given by
\begin{equation}
	S_{\mu L} = S_\mu - S_{\mu T}~~, ~~~~~
	S_{\mu T} = S_\mu - P_\mu {S \cdot q \over \nu} \, .
\end{equation}

As mentioned earlier, these structure 
functions provide a probe to explore various aspects of 
the intrinsic structure of the hadrons. It may be worth noting that in the 
literature, $g_1$ and $g_2$ are usually used to characterize the 
longitudinal and transverse polarized structure functions. However, 
$g_2$ is not really a transverse polarized structure function. 
It also has no clear physical interpretation. Only $g_T$ which can
be directly measured when the target is polarized along the transverse
direction characterizes the full information on
the transverse polarization structure.
\subsection{ Scaling and scaling violation}
As noted, DIS structure functions are scalar functions of $x$ and $Q^2$. But
in the DIS regime (i.e., $Q^2\rightarrow\infty$ and $\nu\rightarrow\infty$
with $x$ fixed), the $Q^2$ dependence of the structure functions fades away
and roughly speaking depends only on $x$. This phenomena as was first noted
by Bjorken from current algebra approach and also was in reasonable
agreement with the early SLAC data, is known as Bjorken scaling. In the
parton model\cite{3feyn}, where the hadron is supposed to be a cluster of
collinearly moving, non-interacting, massless, point-like particles known as
parton, electron-hadron scattering is viewed as the
incoherent sum of electron-parton scattering and the scaling of the
structure functions comes out automatically. Partons are actually quarks
and gluons in reality as described by QCD and are not free. Thus in an
asymptotically free theory like QCD scaling is expected to be violated making
structure functions $Q^2$-dependent as was also confirmed in the later
accurate measurement. In this subsection we briefly mention how
$Q^2$-dependence is usually addressed using QCD. Our aim here is to provide
the basic picture and highlight only those aspects which make these 
approaches 
very different from that we are going to adopt and the motivation behind
building up such an alternate approach in the first place. 

\noindent {\bf QCD improved parton model.} This approach of incorporating
QCD into the DIS picture is based on the factorization of the cross-section
into `soft' and `hard' part which is deeply 
rooted into the parton model. The basic
idea behind  the intuitive parton model in understanding DIS is the
following. If we consider the electron-hadron scattering mediated 
by a virtual photon 
with high energy and momentum transfer (which is the case in DIS), two major
things happen to the hadron when looked from the centre-of-mass frame. The
hadron gets Lorentz contracted in the direction of collision and its
internal dynamics gets {\it almost} frozen due 
to time dilation. Then the virtual
photon sees the hadron like a `pancake' composed of a bunch of nearly 
non-interacting partons. Each of these partons may be thought of as carrying
a definite fraction $x$ of the parent hadron's momentum satisfying $0<x<1$,
since otherwise one or more partons would have to move in the opposite
direction to that of the hadron, an unlikely configuration. Thus the hadron
could be modeled in terms of these parton distributions and the  
cross-section of the electron-hadron scattering can be computed
approximately by summing all possible electron-parton cross-section folded
with the probability of finding such a parton inside the hadron. So,
\begin{equation}
\sigma_{eH}(x,Q^2) = \sum_a \int_x^1 dy f_{a/H}(y) \sigma_{ea}(x/y,Q^2)
\label{facs}
\end{equation}
where $f_{a/H}(y)$ gives the probability to find the parton of type $a$ or
the parton distribution inside the hadron and the summing over all possible
$y$ starts from $x$ (the Bjorken variable) by simple kinematics.
Eq.(\ref{facs}) expresses the basic theme of factorization where all
the long-distance nonperturbative effects of the dynamics known as the 
`soft' part, are dumped into the unknown parton distributions and the `hard'
part $\sigma_{ea}(x/y,Q^2)$ can be computed in the perturbation theory. 

In the QCD improved parton model calculations this `hard' part gets QCD
correction from the diagrams, some of which are shown in the figure below.  
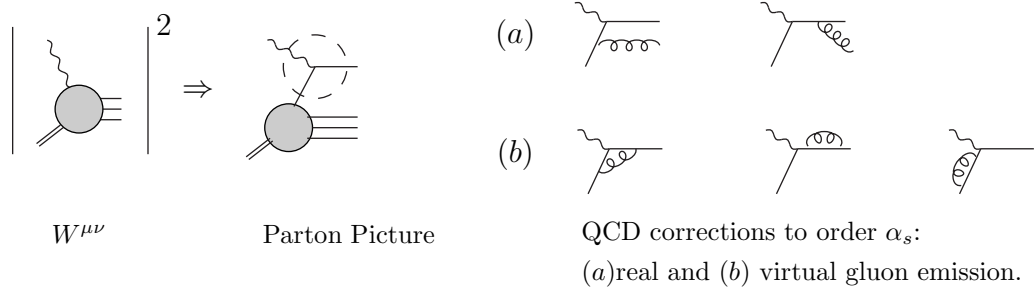
\begin{center}
\begin{picture}(300,110)(0,0)
\Text(165,80)[]{$(a)$}
\Text(165,35)[]{$(b)$}
\Text(0,5)[]{\footnotesize $W^{\mu\nu}$}
\Text(102,5)[]{\footnotesize Parton Picture}
\Text(190,5)[l]{\footnotesize QCD corrections to order $\alpha_s$:}
\Text(190,-10)[l]{\footnotesize ($a$)real and ($b$) virtual gluon emission.}
\Photon(-12,78)(-3,60){1}{3}
\Line(-15,38)(-6,45)
\Line(-16,39)(-7,46)
\Line(7,48)(16,48)
\Line(9,52)(16,52)
\Line(7,56)(16,56)
\GCirc(0,52){9}{.8}
\Line(-25,83)(-25,35)
\Line(26,83)(26,35)
\Text(32,84)[]{$2$}
\Text(45,59)[]{$\Rightarrow$}
\GCirc(79,45){9}{.8}
\Line(64,33)(73,40)
\Line(63,34)(72,41)
\Line(81,53)(89,68)
\Photon(89,68)(71,78){1}{3}
\Line(89,68)(105,68)
\Line(88,45)(105,45)
\Line(86,49)(105,49)
\Line(86,41)(105,41)
\DashCArc(89,68)(12,20,352){5}
\SetOffset(40,2)
\Line(151,66)(159,83)
\Photon(159,83)(147,90){1}{2}
\Line(159,83)(179,83)
\Gluon(156,74)(179,74){2}{3}
\SetOffset(70,2)
\Line(191,66)(199,83)
\Photon(199,83)(187,90){1}{2}
\Line(199,83)(219,83)
\Gluon(222,72)(209,83){2}{3}
\SetScale{.5}
\SetScaledOffset(362,-30)
\SetWidth{1}
\Line(162,66)(178,100)
\Photon(178,100)(154,114){2}{2}
\Line(178,100)(218,100)
\GlueArc(175,80)(14,93,213){4}{2}
\SetScaledOffset(224,-30)
\SetWidth{1}
\Line(162,66)(178,100)
\Photon(178,100)(154,114){2}{2}
\Line(178,100)(218,100)
\GlueArc(198,95)(14,30,150){4}{2}
\SetScaledOffset(82,-30)
\SetWidth{1}
\Line(162,66)(178,100)
\Photon(178,100)(154,114){2}{2}
\Line(178,100)(218,100)
\Gluon(170,83)(198,100){-4}{2}
\end{picture}
\end{center}
\noindent 
\vskip .5cm
\noindent and the statement of
factorization changes to the following.
\begin{equation}
\sigma_{eH}(x,Q^2) = \sum_a \int_x^1 dy f_{a/H}(y, \mu)
\sigma_{ea}(x/y,Q^2,\mu^2,\alpha_s(\mu^2)) + ...
\label{facsi}
\end{equation}
Notice that only the short
distance QCD effects can be calculated utilizing asymptotic freedom.
Details of how one proceeds and calculates are unimportant for our discussion
and can be found, for example, in Ref.\cite{3field}. It is important,
however, to
notice that the cross-sections in the figure above 
become singular due to gluon mass
going to zero (infrared singularity) 
or due to the possibility of collinear emission 
of gluon (mass singularity) if quarks and gluons are assumed to be massless  
(which one generally does in such calculations). So, one needs to
regularize them. Infrared singularities get canceled among the real and
virtual contributions, while the mass singularity is dumped into the
unknown parton distribution functions bringing in the renormalization 
scale $\mu$ in the picture as shown in eq.(\ref{facsi}). 
Obviously, one needs to choose this scale $\mu$ to be large enough to ensure
the validity of perturbative calculation by keeping $\alpha_s(\mu^2)$ small.
The $Q^2$-dependence of the hard scattering cross-section 
turns out to be like $\alpha_s lnQ^2$ and
since $\alpha_s\sim {1\over lnQ^2}$, one has to sum all the terms of the
form $(\alpha_slnQ^2)^n$. This is generally done 
by what is known as leading log
approximation (LLA) \cite{3lla}. One thus eventually arrives at the 
Altarelli-Parisi equation, which can also be obtained otherwise using more
intuitive but less-rigorous method as was first obtained by Altarelli and
Parisi \cite{3ap}. The $Q^2$-evolution of the structure functions are then
studied usually by taking moments or using convolution method (see
ref.\cite{3field}). One important point in the calculation of these hard
processes
is that the cross-section (and hence the structure functions) 
become renormalization 
scheme dependent due to the presence
of finite terms ($Q^2$-independent terms) which are different in different
renormalization scheme.
This ambiguity does not bother people simply because all the studies so far
are directed towards how the structure functions evolve with $Q^2$ (where this
finite $Q^2$-independent terms play no role) and not the
structure functions themselves. The ellipsis in eq.(\ref{facsi}) stands for
the higher twist terms which are suppressed by ${1\over Q^2}$ and needs
separate consideration (beyond the purview of this simple factorization
technique) for their studies.

\noindent {\bf OPE method.} The OPE method\cite{3rey} 
starts from considering the 
virtual Compton scattering
amplitude $T^{\mu\nu}$ defined by
\begin{equation}
T^{\mu\nu}= i \int d^4 z e^{iq\cdot z} \langle PS| T ( J^\mu (z) J^\nu
(0))|PS \rangle ,
\end{equation}
and uses the optical theorem 
\begin{equation}
W^{\mu\nu}= {1\over 2 \pi} {\rm Im} T^{\mu\nu}
\end{equation} 
to get predictions for the structure functions. As is well known, the
products of currents only in the region near the light-cone $z^2\sim0$
contribute to the DIS. So, one uses the OPE for the products of currents in
$T^{\mu\nu}$ at $q^2\rightarrow -\infty$ or $z^2\rightarrow 0$. Pictorially
it can be naively depicted as the following.
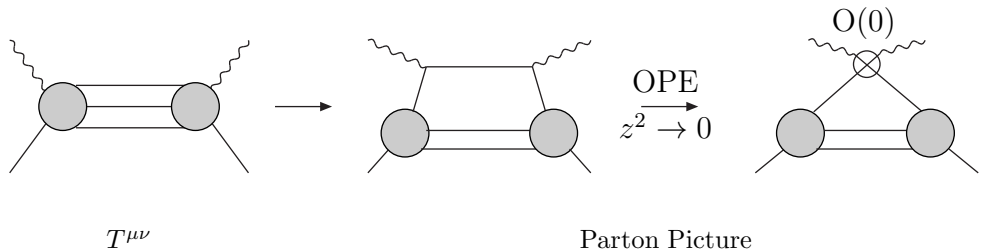
\begin{center}
\begin{picture}(300,100)(0,0)
\Text(20,10)[]{\footnotesize $T^{\mu\nu}$}
\Text(223,10)[]{\footnotesize Parton Picture}
\Text(223,53)[]{$z^2\rightarrow 0$}
\Text(223,70)[]{OPE}
\Photon(-25,85)(-11,66){1}{4}
\Line(-25,35)(-11,54) 
\Line(0,68)(44,68)
\Line(0,52)(44,52)
\Line(3,60)(42,60)
\GCirc(-5,60){9}{0.8}
\Line(51,54)(65,35) 
\Photon(51,66)(65,85){1}{4}
\GCirc(45,60){9}{0.8}
\LongArrow(75,60)(95,60)
\Photon(110,85)(132,75){1}{4}
\Line(110,35)(118,44) 
\GCirc(124,50){9}{0.8}
\Line(132,75)(172,75)
\Line(132,75)(127,58)
\Line(172,75)(177,58)
\Line(132,51)(172,51)
\Line(130,44)(174,44)
\GCirc(180,50){9}{0.8}
\Photon(172,75)(194,85){1}{4}
\Line(194,35)(186,44)
\LongArrow(213,60)(233,60)
\Photon(276,85)(293,80){1}{3}
\Line(256,35)(267,44) 
\GCirc(273,50){9}{0.8}
\BCirc(298,76){5}
\Text(298,92)[]{O(0)}
\Line(303,80)(278,58)
\Line(293,80)(319,58)
\Line(282,51)(315,51)
\Line(279,44)(317,44)
\GCirc(322,50){9}{0.8}
\Photon(303,80)(320,85){1}{3}
\Line(340,35)(329,44)
\end{picture}
\end{center}
\vskip .3cm
\noindent 
The handbag diagram here corresponds to the parton model where applying OPE
(naturally in the free theory) gives perfect scaling with the matrix
elements of the local operators $O(0)$ 
between the target states remain unknown and
are related to the parton distributions that has to be inferred from the
experiment. Now, adding appropriate radiative gluon corrections to the above
diagram and using the renormalization group equation, one can calculate all
the leading $ln Q^2$ corrections to all orders in $\alpha_s$. 

To elucidate further the basic theme, we consider the light-cone 
behavior of the product of two scalar
operators $A$ and $B$.
OPE near light-cone is given by the following expansion 
\begin{equation}
A(z)B(0)= \sum_{i,n} C_i^n(z^2)z_{\mu_1}....z_{\mu_n}
O_i^{\mu_1....\mu_n}(0),
\label{ope}
\end{equation}
where the sum is over $i$, the various types of local operators $O(0)$ that may
contribute and $n$ denotes the {\it spin} of the operator $O(0)$, 
determined by
its Lorentz transformation property. The string of local operators 
$O_i^{\mu_1....\mu_n}(0)$ are considered to be non-singular, local,
symmetric and traceless operators to ensure definite spin ($n$). The
expansion parameter $C_i^n (z^2)$, the so called Wilson coefficients, are
c-number singular functions at $z^2\sim 0$ and controls all the singular
behavior of the product of the currents. From naive dimensional argument
$C_i^n$s may be taken to behave as 
\begin{equation}
C_i^n(z^2)\,\,{\stackrel{z^2\rightarrow 0}{\sim}} \,\,\,
\Big ({1\over z^2}\Big
)^{[d_A+d_B-(d_{O_i}-n)]/2}\,\, ,
\end{equation}
where $d_{O_i}$ 
denotes the naive mass dimension of the appropriate operator in
eq.(\ref{ope}). Thus the strongest singularity in the expansion is obtained
for the operator with minimum {\it twist} $\tau$ defined as 
\begin{equation}
\tau \equiv d_{O_i} -n \, ,
\end{equation}
whereas less-singular terms do not contribute, as we shall see shortly, to
the leading power behavior in $Q^2$. It should be emphasized that the
Wilson coefficients are process independent and can be calculated using
perturbative QCD, while  all the specific information regarding the
particular process is buried in the matrix elements of the local operator,
implying 
\begin{equation}
\langle P |A(z)B(0)|P\rangle = \sum_{i,n} C_i^n(z^2)z_{\mu_1}....z_{\mu_n}
\langle P|O_i^{\mu_1....\mu_n}(0)|P\rangle.
\label{opem}
\end{equation}

Now we go back to the Compton amplitude and use the above information there.
We suppress all the obvious Lorentz indices as well as avoid the  
unnecessary complication coming out of the target spin in the  the
following, 
in order to  highlight the basic logic. 
\begin{eqnarray}
T(x,Q^2) &=& i \int d^4z e^{iq\cdot z} \langle P| T(J(z)J(0))|P\rangle
\nonumber\\
~~~  &{\stackrel{z^2\rightarrow 0}{=}}& \int d^4z e^{iq\cdot z}
 \sum_{i,n} C_i^n(z^2)z_{\mu_1}....z_{\mu_n}
\langle P|O_i^{\mu_1....\mu_n}(0)|P\rangle\nonumber\\
~~~  &=& \sum_{i,n}2 q_{\mu_1}....2q_{\mu_n}{\partial^n \over \partial
(iq^2)^n}\int d^4z e^{iq\cdot z} C_i^n(z^2)
\langle P|O_i^{\mu_1....\mu_n}(0)|P\rangle
\end{eqnarray}
The last step is obtained by replacing $z_\mu$s by ${\partial\over \partial
q_\mu}$ and using the relation
\begin{equation}
{\partial\over \partial
q_{\mu_1}}....{\partial\over \partial
q_{\mu_n}} = 2q_{\mu_1}....2q_{\mu_n}{\partial^n\over \partial
(iq^2)^n} +~{\rm trace}~{\rm terms}\, ,
\end{equation}
and ignoring the trace terms for their contributions will be suppressed for
large $Q^2$. 
The most general Lorentz structure for the matrix elements of $O_i$ is given
as 
\begin{equation}
\langle P|O_i^{\mu_1....\mu_n}(0)|P\rangle = A_i^n \big (
P_{\mu_1}....P_{\mu_n}- M^2 g^{\mu_1\mu_2}P_{\mu_3}....P_{\mu_n}+ {\rm
permutations} \big )
\label{omx}
\end{equation}
where $M$ is the target mass and the terms proportional to $g^{\mu_1\mu_2}$
are the so called trace terms. $A^n_i$s are the numbers containing the
nonperturbative information of the process.  Thus, using eq.(\ref{omx}) 
we get,
\begin{eqnarray}
T(x,Q^2) &=& \sum_{i,n} \big [ \Big({2P\cdot q\over Q^2}\Big)^n (Q^2)^n
{\partial^n \over \partial
(iq^2)^n}\int d^4z e^{iq\cdot z} C_i^n(z^2)A_i^n + O({1\over Q^2})\big]\, ,
\end{eqnarray}
where we have multiplied $Q^{2n}$ in the numerator and denominator. Defining
the Fourier transform of Wilson coefficient as 
\begin{equation}
C^n_i(Q^2)\equiv (Q^2)^n
{\partial^n \over \partial
(iq^2)^n}\int d^4z e^{q\cdot z} C_i^n(z^2)
\end{equation}
we get,
\begin{equation}
T(x,Q^2) = \sum_{i,n} C_i^n(Q^2) x^{-n} A^n_i + O(x^{-n+2} M^2/Q^2)\, .
\label{tx}
\end{equation}
The terms suppressed by ${1\over Q^2}$ are either coming from target mass
effect (the trace terms) or due to higher twist contributions not included
in the simple handbag kind of diagram shown here. Now, these Wilson
coefficients $C^n_i(Q^2)$ (which also depends on the renormalization scale
$\mu$ and the coupling $g(\mu)$) can 
be shown to obey the renormalization group
equation, 
\begin{equation}
\Big( \mu {\partial\over \partial \mu} +\beta {\partial\over \partial g}
-\gamma_{O^n_i}\Big) C^n_i(Q^2/\mu^2, g(\mu)) = 0 \,\,,
\label{rmg}
\end{equation}
which allows us to calculate the leading $Q^2$-dependence in
eq.(\ref{tx}) to all orders in $\alpha_s$ provided we know the first
non-trivial (1-loop) order anomalous dimension ($\gamma_{O^n_i}$) 
for  the operator
$O^{\mu_1...\mu_n}_i$  and QCD $\beta$-function \cite{3muta}. For instance,
the solution of eq.(\ref{rmg}), which is of the form 
\begin{equation}
C^n_i(Q^2/\mu^2, g(\mu)) = C^n_i(1, {\bar g(Q^2)}) exp \Big [
-\int_0^{{1\over 2} ln Q^2/\mu^2}
 \gamma_{O^n_i}dt^\prime \Big] \,\, ,
\label{solrmg}
\end{equation}
can be used in eq.(\ref{tx}). 

After everything said and done for
$T^{\mu\nu}$, we can use the optical theorem to pass over to the
structure functions. 
Now, this passing over to the structure functions ($W^{\mu\nu}$) from 
$T^{\mu\nu}$ is not that straightforward. Notice that the physical region
for DIS is $0\leq x\leq 1$ and clearly the expression that we have obtained
for $T(x,Q^2)$ in eq.(\ref{tx})  
diverges for this physical region. What is therefore needed 
is an analytic continuation of $T(x,Q^2)$ in complex $x$ and get the result
corresponding to the physical region as a limit of the analytic function in
the unphysical region. This  naturally leads to the 
moments of  structure functions, instead of the structure functions
themselves, as is shown below.
Now, $T(x,Q^2)$ (eq.(\ref{tx})) is good enough as a
function of complex $x$ for it is analytic as $|x|\rightarrow \infty$, 
and has a cut from $-$1 to +1, since this region is connected to particle
production in the elastic or inelastic scattering. Thus, we can {\it only}  
isolate 
the coefficient of $x^{-n}$ by taking the Mellin transform:
\begin{equation}
{1\over 2 \pi i}\int_{\cal C} dx x^{n-1} T(x,Q^2) = \sum_i C^n_i(Q^2) A_i^n
\, \,\, ,
\end{equation}
where the contour ${\cal C}$ is chosen as in the figure below. 
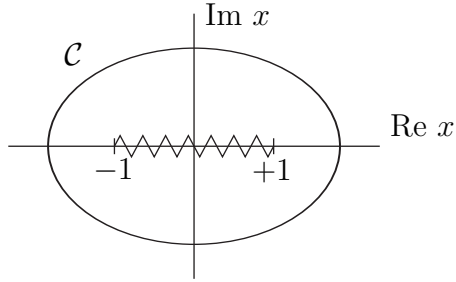
\begin{center}
\begin{picture}(300,100)(0,0)
\Text(167,100)[]{Im~$x$}
\Text(237,58)[]{Re~$x$}
\Text(105,85)[]{${\cal C}$}
\Text(180,41)[]{$+1$}
\Text(120,41)[]{$-1$}
\Oval(150,50)(37,55)(0)
\ZigZag(120,50)(180,50){4}{7}
\Line(120,47)(120,53)
\Line(180,47)(180,53)
\Line(80,50)(220,50)
\Line(150,100)(150,0)
\end{picture}
\end{center}
\noindent Using the analyticity of $T(x,Q^2)$, the contour ${\cal C}$ 
can be
shrunk to the cut, to obtain
\begin{eqnarray}
{1\over 2\pi i}&\big[& \int_{-1}^{+1} dx x^{n-1} T(x+i\epsilon, Q^2) -
\int_{-1}^{+1} dx x^{n-1} T(x-i\epsilon, Q^2)\big] \nonumber\\
~~~~ &=& {1\over 2\pi i}\int_{-1}^{+1} dx x^{n-1} \big[T(x+i\epsilon,
Q^2)-T^*(x+i\epsilon,Q^2)\big]\nonumber\\
~~~~ &=& {1\over 2\pi i}\int_{-1}^{+1} dx x^{n-1} 2i {\rm Im}
T(x+i\epsilon,Q^2)\nonumber\\
~~~~ &=& \int_{-1}^{+1} dx x^{n-1} 2W(x,Q^2)\nonumber\\
~~~~ &=& 4 \int_{0}^{+1} dx x^{n-1} W(x,Q^2)\quad {\rm or,} \quad 0.
\end{eqnarray}
The result in the last line depends on the crossing symmetry ($x\rightarrow
-x$) of $W$s and the value of $n$ (whether odd or even). So we get for the
nonvanishing moments,
\begin{equation}
4 \int_{0}^{+1} dx x^{n-1} W(x,Q^2)= \sum_i C^n_i(Q^2) A_i^n\,\,.
\end{equation}
To get back the structure functions one has to invert this equation which is
done very seldom. Instead, one uses the solution for $C^n_i$ (as given in 
eq.(\ref{solrmg})) in eq.(\ref{tx}) and takes ratios of moments 
at different $Q^2$, in order to get rid
off $A^n_i$s and compare moments at
various scale $Q^2$  (see, for example,
Ref.\cite{3muta} for details).

Thus, we see that the methods that are usually employed in studying DIS
structure functions mainly concentrate on the studies of moments and the
$Q^2$-evolution thereof, using perturbative QCD. 
Structure functions themselves are either ambiguous or not properly 
addressed at all, whereas these are the objects that really go into the
cross-section which is measured in the experiment. 
These methods are framed in such a way that  
the nonperturbative informations contained in the
parton distributions are separated from the very beginning and paid very
little attention to. In QCD improved parton model, which enjoys the
assumptions of collinearity and massless partons of the original model, this
separation is done through factorization, 
while OPE does the job in the other method.
Also, the intuitive meaning of the parton distributions gets buried in the
OPE method which involves cumbersome mathematics. On the other hand, a
complete understanding of hadrons crucially depends on the knowledge of
nonperturbative QCD informations contained in these parton distributions. So,
an approach which treats nonperturbative as well as perturbative
descriptions of DIS structure functions in the same framework (which is really
missing) will be of great importance in understanding the hadrons
better. 
This is exactly what we attempt to build up in our work.

Towards building up this alternate approach, in the next section, we first
review the basic ideas behind the current algebra (in particular, 
the characteristics of
the light cone version of it)  which plays a crucial role in
our approach. We then show how one can get relatively simple expressions for 
the structure functions using these ideas and light-cone version of the
Bjorken-Johnson-Low theorem. We should emphasize that our approach closely
follows the one proposed earlier before the advent of QCD \cite{3jackiw}, 
but now
built within the framework of QCD and hence, free of most of the assumptions
employed earlier as we shall see below.
\section{Light front current commutators and BJL theorem}
In the pre-QCD era, current algebra approach was proposed to study the strong
interactions based on various observed symmetries of these interactions
without touching upon the real dynamics that are involved. In particular, 
it was
introduced to circumvent two major difficulties that hindered the progress 
in particle physics. Firstly, the lack of proper knowledge in those 
precise laws
which govern the processes except electromagnetism and secondly, an
inability to solve any of the realistic models which had been proposed to
explain the dynamics. It was first proposed by Gellman \cite{3gel} and then
extended and used heavily by him and others to produce various sum rules and
low energy theorems which had experimental consequences. The basic idea
comes from the fact that electromagnetic and weak interactions of hadrons 
could be successfully described in terms of a current-current interaction
Lagrangian with experimental quantities like decay width or scattering
cross-sections being intimately related to the matrix elements of operators
involving these currents. Although the form of these currents are unknown
(or, at best, model dependent), a knowledge of current commutators can be
exploited to obtain sum rules which heavily constrain the experimental
quantities. As is obvious, one can have exact knowledge of these commutators
in a model (for example, the 
 quark model) but can only postulate in reality.

To illustrate further, let us consider 
the currents $J_a^\mu(x)$ corresponding to some
approximate internal symmetry observed in the strong interaction processes.
The charges $Q_a(t)$'s that generate the symmetry  are defined as 
$\int d^3x J_a^0(x)$. 
If the symmetry is exact, then the currents are conserved 
(i.e., $\partial_\mu J^\mu_a(x)=0$) and
the charges $Q_a$'s are time-independent
and if we assume that the
currents themselves transform in a known 
fashion under the symmetry transformations, 
we have the following relations,
\begin{eqnarray}
[Q_a,Q_b] &=& if_{abc}Q_c \label{c1} \\
~[Q_a,J^\mu_b(0)] &=& if_{abc}J^\mu_c(0)\label{c2}
\end{eqnarray}
where $f_{abc}$ are the structure constants defining the characteristic 
algebra of the symmetry group. It was 
postulated that even
if the symmetry is not exact in reality (thereby making the charges 
time-dependent), {\it the equal time versions of eqs.(\ref{c1},\ref{c2}) 
would still remain valid}. One
just needs to go one step further and assume the {\it local} 
version of eqs.(\ref{c2}) in the form of an equal time
closed algebra among the currents themselves. 
\begin{equation}
[J^0_a(x),J^\mu_b(0)]_{x^0=0}=~if_{abc} J^\mu_c(0)\delta^3(\vec
x )
\label{jcom}
\end{equation}
Such assumptions can be generalized for spatial components of the vector
currents and also be extended for the axial vector currents if assumed to
be present. Notice the presence of $\delta^3(\vec x )$ in the RHS of
eq.(\ref{jcom}) which reflects the fact
that if $\vec x \neq 0$ (i.e., two points are separated by a
space-like distance), the commutator vanishes due to causality. Details of
how one gets various sum rules and low energy theorems by assuming such
algebra of currents are, of course, beyond the scope of present discussion
and can be found in the Ref.\cite{3gel}. 

One of the most important point found in such studies in current algebra was 
the following. The assumption of the local version of 
the algebra in terms of currents as in eq.(\ref{jcom})
entails further complication due to the {\it Schwinger terms}
which may be present there.  
The exact form of the Schwinger terms are unknown except that they are 
like total spatial-derivatives and thus compatible with eq.(\ref{c2}). Thus, 
the algebra no longer remains closed. This 
ambiguity makes the current algebra approach less predictive than it was
thought to be and lot of work has been done towards how to live with them.
That discussion is again beyond the scope 
of the present discussion. For our purpose,
it is sufficient to notice that, (i)  by postulating equal time commutators of
hadronic currents in the form of a closed algebra one could predict certain
sum rules having experimental consequences, 
(ii) the algebra is compatible with causality, 
(iii) the presence of 
Schwinger terms
further complicates the issue rendering 
the algebra no more closed and (iv) the
exact form of the currents as well as the schwinger terms are not known. 

Now, one introduced another 
concept of going to infinite momentum frame by taking  
infinite momentum limit ($p\rightarrow \infty$) of the matrix elements of
equal time current commutators in deriving, for example, the fixed mass 
sum rules like Fubini-Dashen and Gellman sum rule. Although the limiting
procedure was not always straight-forward and free of ambiguity, it was
realized that the light cone behavior of the current commutators played the
important role there. It was also shown 
that the DIS structure functions directly measure
the matrix elements of the current commutators on the light cone. 
These observations gave a new direction to the current algebra studies. 
Thus, it
was suggested and shown that assuming the algebra of new charges defined on
the light cone ($Q_a(x^+)= \int dx^-d^2x^\perp J^+(x)$) and hence the light 
cone current algebra, one could directly obtain the fixed mass sum rules
without going through the cumbersome $p \rightarrow \infty$ limit. 
Details of light cone current algebra and their consequences can be found in
the ref.\cite{3jackiw,3jac}. Here we want to emphasize the most striking difference
between equal-time and light cone algebra of currents as dictated by 
causality.
In the light cone version of the current algebra that corresponds to 
eq.(\ref{jcom}), we have
\begin{equation}
[J^+_a(x),J^\mu_b(0)]_{x^+=0}=~if_{abc} J^\mu_c(x)\delta(x^-)\delta^2(
x^\perp ) + \partial^+S_{ab}^\mu +\partial^\perp S^{\perp,\mu}_{ab}\quad,
\label{lfjcom}
\end{equation}
where, the Schwinger terms $S^\mu_{ab}$'s are explicitly shown. As pointed out
earlier, current commutator vanishes if the points concerned are separated
by a space-like distance, i.e., $x^2 < 0$. Thus, restricting to $x^0=0$  in
the equal-time case, causality forces the commutator to be local, i.e.,
nonvanishing only when $\vec x =0$ as depicted
by the $\delta$-function in eq.(\ref{jcom}). On the other hand, restricting to
$x^+=0$ in the case of light cone algebra, $x^2=x^+x^- -x^2_\perp=0$ is
maintained if $x^\perp=0$, irrespective of $x^-$. Implying that the
causality only enforces locality in $x^\perp$ in the light cone commutators 
and non-locality in $x^-$ creeps in. We will see shortly how this
non-locality, which is one of the unique feature of LF field theory, 
is reflected in the equal-$x^+$ commutator of currents giving
rise to generalization from local to {\it bilocal currents} in the
context of DIS, and the important role played by it. 

After the advent of QCD as the underlying theory of strong interactions,
current algebra approach was mostly abandoned in favour of operator product
expansion (OPE) and perturbative calculation of Wilson co-efficients. It was
largely due to the fact that most of the current algebra sum rules turned out
to be invalid due to the perturbative QCD corrections, except those which
are protected by some conservation laws. In our way of addressing the
problem, we closely follow the current algebra methods employed earlier
taking QCD as the guideline. For example, now since 
we know that QCD is the underlying
theory, we know the exact form of the currents in terms of the field
variables and are able to calculate the 
equal-$x^+$ commutators of currents (instead of postulating),  
using that among the field variables themselves. Thus, we attempt to 
supplement the current algebra approach to DIS by incorporating QCD, 
as will be discussed here and in the later Chapters. In this section, we
first show how equal-$x^+$ current commutators come into the DIS picture and
then derive the relevant commutators.
\subsection{An expansion in inverse power of light-front energy of
	 the virtual photon}
The hadronic tensor $W^{\mu\nu}$ is given in terms of 
hadronic matrix elements
of the current commutator as in eq.(\ref{wmunu}). Notice that it is not
equal-time or equal-$x^+$ commutator. 
To see how DIS structure functions can be related to equal-$x^+$ commutators
of current, we start with $T^{\mu\nu}$, the well known 
forward virtual photon-hadron Compton scattering amplitude:  
\begin{equation}
T^{\mu \nu} = i \int d^4\xi e^{iq\cdot \xi} \langle PS | 
T(J^\mu(\xi) J^\nu(0)) |PS \rangle .
\end{equation}
As it 
is already noted, the hadronic tensor is related to the forward 
virtual-photon hadron Compton scattering amplitude as 
\begin{equation}
	W^{\mu \nu} = {1\over 2\pi}{\rm Im} T^{\mu \nu} .
\end{equation}
Similar to the case of hadronic tensor, 
$T^{\mu\nu}$ can be parametrized as 
\begin{eqnarray}	
\label{tfun}	
T^{\mu \nu}&=&\Big(-g^{\mu \nu} + {q^\mu q^\nu 
		\over q^2} \Big) T_1(x,Q^2) + \Big(p^\mu - {\nu 
		\over q^2} q^\mu\Big)\Big(p^\nu -{\nu \over q^2} 
		q^\nu\Big)T_2(x,Q^2) \nonumber \\
	& & ~~~~~~~~~~~~~ - i \epsilon^{\mu \nu \lambda \sigma}q_\lambda 
		\Big[ S_\sigma T_3(x,Q^2)+ P_\sigma S\cdot q T_4(x,Q^2) 
		\Big] \, .
\end{eqnarray}
Using the optical theorem, we have
\begin{equation}	\label{twr}
	T_i (x,Q^2) = 2 \int_{-\infty}^\infty d{q'}^+
		{W_i (x',Q^2) \over {q'}^+ - q^+} ~, ~~
		i = 1, 2, 3, 4	\, . 
\end{equation}
The above relations give us the opportunity to connect any information
regarding 
$T^{\mu\nu}$ to the structure functions themselves and 
as we shall see below, these structure functions can really be connected 
to the light-front bilocal currents through the $1/q^-$ expansion 
of $T^{\mu \nu}$ and using equal-$x^+$ current commutators.

An expansion of $T^{\mu \nu}$ in  ${1/q^-}$ was originally 
proposed by Jackiw {\it et al.} \cite{3jackiw} based on BJL theorem.  
The general expansion in $1/q^-$ is given by (see Appendix B for a 
derivation) 
\begin{equation} \label{lfcc}
	T^{\mu \nu} = - \sum_{n=0}^\infty \Big({1\over q^-}
		\Big)^{n+1} \int d\xi^- d^2 \xi_\bot e^{iq\cdot \xi}
	  \langle PS | [(i\partial_\xi^-)^nJ^\mu(\xi), J^\nu(0)]_{
		\xi^+=0}| PS\rangle \, ,
\end{equation}
where $q^-=q^0 - q^3$, the light-front energy of the virtual photon,
and $\partial^-=2{\partial \over \partial \xi^+}$ is a light-front
time derivative and $(\xi^+,\xi^-,\xi^i)$ are the light-front space-time 
coordinates. 
The above expansion shows that the time-ordered matrix element in
$T^{\mu\nu}$ can 
be expanded in terms of an infinite series of equal light-front time 
(i.e., equal-$x^+$) commutators.

For large $Q^2$ and large $\nu$ limits in DIS, theoretically without 
loss of generality we can always select a Lorentz frame 
such that the light-front energy $q^-$ of the virtual photon 
becomes very large. Explicitly, in terms of light front variables, we can
choose $q^+$ to be negative and finite for the virtual photon. 
Also, keeping $q^i$ to be finite, one can get large space-like $q^2$ 
($Q^2 \rightarrow \infty$) by taking $q^-\rightarrow \infty$. Thus, DIS
regime can simply be obtained for $q^-\rightarrow \infty$, such that,
\begin{equation}
Q^2 \sim -q^+q^- \rightarrow \infty,\quad \quad \nu \sim {P^+q^- \over
2}\rightarrow \infty , \quad\quad 
x \sim - {q^+ \over P^+}.
\end{equation}
Notice that $x$ is positive and finite, since $q^+$ is negative while both 
$q^+$ and $P^+$ are finite.
Then, in the DIS regime, 
only the leading term in the above expansion of $T^{\mu\nu}$ given in
eq.(\ref{lfcc}) is dominant, i.e.,
\begin{equation} \label{lfccl}
	T^{\mu \nu} \stackrel{{\rm large}~q^-}{=} - {1\over q^-}
		\int d\xi^- d^2 \xi_\bot e^{iq\cdot \xi}
	  \langle PS | [J^\mu(\xi), J^\nu(0)]_{\xi^+=0}| PS\rangle \, .
\end{equation}
Here we have assumed the 
fact that the convergence of the expansion for large
$q^-$ is not spoiled by
the integrals of the matrix elements that occur in the subsequent terms of
the expansion. 
As a result,
the leading contribution to the deep inelastic structure functions 
is determined by the light-front current algebra. Of course, as mentioned
earlier, we can compute the light-front 
current commutator directly and exactly from QCD 
(where QCD should be quantized on the light-front time surface $\xi^+
=\xi^0+\xi^3=0$ with the light-front gauge $A_a^+=0$)
\cite{3brodsky80,3zhang93}. Hence, all the subsequent derivations 
are exact within the light-front QCD and without further assumptions 
or approximations of the collinear and massless partons that were 
used in the derivations as discussed earlier.

At this point, we should emphasize the following facts which necessitate the
use of light front description in this context. Firstly, the above
exercise with the BJL expansion can be performed in terms of ${1 \over q^0}$
as well, thereby obtaining usual equal-time current commutator instead. But,
now taking $q^0\rightarrow \infty$ limit gives $q^2 > 0$, i.e., time-like
$q^2$ which is unphysical for DIS. One needs to circumvent this problem by
going to complex $q^0$-plane and taking $iq^0\rightarrow\infty$ or
otherwise, bringing in complexities. Secondly, as is well known, DIS is 
a light-cone dominated process. A knowledge of the current commutator on the 
$x^+=0$-surface, which shares a whole line with the light-cone, is rich in
information and suitable for DIS compared to that on $x^0=0$, which touches
only the tip of the light-cone. Thirdly, it
comes out, as a consequence of using light-front commutator and will be
shown shortly, that the structure functions in the inelastic scattering is
directly related to the matrix element of bilocal currents in a similar  
way as the
form factor in elastic scattering is to that of local currents. On the other
hand, since DIS needs dynamical information starting from current
commutators defined on $x^0=0$-surface,
one can only get an infinite set of relations between each moment of
structure function and the corresponding term in the BJL expansion
\cite{3jackiw}.
Lastly, before proceeding further, it should be reemphasized that we do not
need to bother about the intricacies associated with the current
algebra assumptions and their validity, since the current commutators that
we shall be using comes directly as a consequence of QCD, believed to be the
underlying theory. 
\subsection {Light front current commutator}
Our next task is to evaluate the light front current commutators. Here we
shall outline the derivation of only those current commutators which may be
necessary later on.  The 
hadronic current that takes part in the electromagnetic interaction with the
virtual photon is given by,
\begin{equation}
J^\mu(x)=\sum_\alpha 
e_\alpha \overline{\psi}_\alpha(x) \gamma^\mu \psi_\alpha(x) \, ,
\end{equation}
where $\psi_\alpha (x)$ is the quark field carrying
the flavor index $\alpha$ and the charge $e_\alpha$. We also define the
axial vector current for later purpose:
\begin{equation}
J_5^\mu(x)=\sum_\alpha 
e_\alpha \overline{\psi}_\alpha(x) \gamma^\mu 
\gamma_5\psi_\alpha(x) \, .
\end{equation}
Various components of the vector currents explicitly expressed in terms of
dynamical and constrained quark fields are as follows.
\begin{eqnarray}
J^+(x) &=& \sum_\alpha e_\alpha \,2 \psi_{+\alpha}^\dagger(x) 
\psi_{+\alpha}(x) \\
J^i(x) &=& \sum_\alpha e_\alpha \,\big[ \psi_{+\alpha}^\dagger(x)
\alpha^i\psi_{-\alpha}(x)+ 
\psi_{-\alpha}^\dagger(x)\alpha^i \psi_{+\alpha}(x)\big] \\
J^-(x) &=& \sum_\alpha 2e_\alpha  \psi_{-\alpha}^\dagger(x) 
\psi_{-\alpha}(x) \, .
\end{eqnarray}
In the literature, sometimes $J^+$ is denoted as the {\it good} component of
the current for it does not contain any interaction, $J^i$ as the {\it bad} 
component
for it contains interaction through constrained field
$\psi_{-\alpha}$ and so on. Degree of
{\it badness} is referred by the number of constrained fields present and is
supposed to be connected with the twist of these operators\cite{3jaff}.

For our purpose, we need to calculate the commutator between $J^+(x)$ and
$J^-(0)$ on the light front $x^+=0$.
As usual in light front field theory, the quark 
field $\psi(x)$ is decomposed into dynamical and
constrained components, 
\begin{equation}
	\psi(x) = \psi_+(x) + \psi_-(x)~~, ~~~~~ \psi_{\pm}(x) = 
		\Lambda^{\pm}\psi(x)~~, ~~~~ \Lambda^{\pm} = 
		{1\over 2}\gamma^0\gamma^\pm  \, .
\end{equation}
We use the basic commutation relation on light-front for the dynamical 
quark fields,
\begin{equation}  \label{qkcom}
	\{ \psi_+(x)~, ~\psi_+^\dagger (y) \}_{x^+=y^+}
	 = \Lambda^+ \delta(x^--y^-) \delta^2(x_\bot-y_\bot) \, ,
\end{equation}
where flavour indices are implicit. 
The minus component $\psi_-(x)$ is determined from $\psi_+(x)$
using the constraint equation:
\begin{equation}	\label{qkmc}
	\psi_-(x) = {1\over i\partial^+}\Big(i\alpha_\bot \cdot
		D_\bot + \beta m_q \Big) \psi_+(x) \, .
\end{equation}
Here we have already used the light-front gauge $A^+_a=0$, and $D_\bot
=\partial_\bot - ig A_\bot$ is the transverse component of the covariant 
derivative, $\alpha_\bot^i = \gamma^0 \gamma^i$, $\beta=\gamma^0$.

Because of the above special property of quark (or more generally 
fermion) fields on the light-front, the light-front current explicitly 
depends on interaction of the theory, which is very different from 
the usual equal-time formulation. In other words, the fundamental
interaction is manifested explicitly in the light-front
current commutators. 

From Eqs.(\ref{qkcom}) and (\ref{qkmc}), we have
\begin{equation}
	\{ \psi_+(x)~, ~\psi^*_-(y) \}_{x^+=y^+} = {\Lambda^+ \over 4i} 
		\epsilon (x^--y^-) \Big[i \alpha_\bot \cdot D^*_\bot -
		\beta m \Big] \delta^2(x_\bot - y_\bot) \, .
\label{qkpmcom}
\end{equation}
Thus, using eq.(\ref{qkcom}) and eq.(\ref{qkpmcom}), 
after a tedious but straightforward calculation, one can find
that
\begin{eqnarray}
	\Big[ J^+(x)~ , ~J^-(y) \Big]_{x^+=y^+} = \sum_\alpha e_\alpha^2
		\Bigg\{ && \partial^+_x \Big[ -{1\over 2}\epsilon(x^--y^-) 
		\delta^2(x_\bot -y_\bot) V_\alpha^-(x|y) \Big] \nonumber \\
		&&  + \partial^i_x \Big[{1\over 2}\epsilon(x^--y^-)
		\delta^2(x_\bot -y_\bot)\Big[V_\alpha^i(x|y) \nonumber \\
		&& ~~~~~~~~~~~~~~~~~ ~~~ + i 
		\epsilon^{ij}A_\alpha^j(x|y)\Big]\Big] - h.c. \Bigg\} \, ,
		\label{c+-c}
\end{eqnarray}
where $V_\alpha^\mu$ and $A_\alpha^\mu$ are defined as the bilocal 
vector and axial vector currents, which are straightforward generalization
of the corresponding local currents: 
\begin{eqnarray}
	V_\alpha^\mu(x|y) &=& \overline{\psi}_\alpha(x) \gamma^\mu 
		\psi_\alpha(y) \, , \\
	A_\alpha^\mu(x|y) &=& \overline{\psi}_\alpha(x) \gamma^\mu 
		\gamma_5 \psi_\alpha(y) \, . 
\end{eqnarray}

As we can see, the light-front current commutators are very different 
from the equal-time current commutators. Here the commutator is indeed 
given by terms involving spatial derivatives (Schwinger terms in the
current algebra language). These  space-derivatives
come from the non-locality of $\psi_-(x)$ on the light-front. In the 
equal-time formulation, there is no such nonlocality involved 
in connection  with the fermion  
field.  Therefore one cannot derive such a commutator from the naive 
canonical equal-time commutators. Note that the nonlocality is only in the
longitudinal direction $x^-$ as argued earlier in the current algebra
context, but now it comes out of direct calculation. As we will soon see 
in the next
section it is these nonlocalities that lead to the simple 
expressions of the structure functions in terms of bilocal current 
matrix elements.  This is  an 
essential feature in the present approach that make the light-front 
current algebra specially useful in the exploration of the deep 
inelastic structure functions. Also note the fact that these bilocal
currents explicitly depend on the interaction. But the way it is written,
the form of the commutator is the same as in the case of free
theory. Thus, the scaling of the structure functions which we know is 
exact in the free theory, will come out unaffected 
unless we really grind these bilocal
operators and try to calculate their matrix elements. 

The commutators for other current components, for example, $J^+$ and $J^i$ 
can also be found 
straightforwardly, 
\begin{eqnarray}  
	\Big[ J^+(x)~, &~& J^i(y) \Big]_{x^+=y^+} = \sum_\alpha e_\alpha^2
		\Bigg\{ \partial^+_x \left[ -{1\over 4}\,\epsilon(x^--y^-) 
		\delta^2(x_\bot -y_\bot)\Big[ V_\alpha^i(x|y)
		-i\epsilon^{ij}A^j_\alpha(x|y)\Big ]\right] \nonumber \\
		& & + \partial^j_x \left[-{1\over 4}\,\epsilon(x^--y^-)
		\delta^2(x_\bot -y_\bot)\Big[g^{ij}V_\alpha^+(x|y) + i 
		\epsilon^{ij}A_\alpha^+(x|y)\Big]\right] - h.c. \Bigg\} \, .
		\label{c+ic}
\end{eqnarray}
Thus, one can use eq.(\ref{c+-c}) to extract the structure
functions and then use eq.(\ref{c+ic}) to make a consistency check.
	
\section{The generalized expressions for Deep inelastic structure functions}
Now, the Compton scattering amplitude in the large $q^-$ limit can be
immediately expressed in terms of the hadronic matrix elements of the
bilocal vector and axial vector currents. We consider the $(+-)$
component the Compton amplitude and using eq.(\ref{c+-c}) we get,
\begin{eqnarray}
	T^{+-} & \stackrel{{\rm large}~q^-}{=} & -{1\over q^-} 
		\int d\xi^- e^{iq^+\xi^-/2} \epsilon
		(\xi^-) \langle PS| \sum_\alpha e_\alpha^2 \Big\{ {i\over 2} 
		q^+ V_\alpha^-(\xi^- |0)  \nonumber \\
	& & ~~~~~~~~~~~~~ - {i \over 2} q_\bot^i [ V_\alpha^i (\xi^-|0) 
		+ i \epsilon^{ij} A_\alpha^j(\xi^-|0)] \Big\}- h.c.
		~ | PS \rangle  \, . \label{t+-}
\end{eqnarray}
Notice that here we have first used partial integration 
which brings in $q^{+,i}$ in
the place of $\partial^{+,i}$ in the commutator (eq.(\ref{c+-c})) and
neglected the surface term and then used the $\delta^2(\xi^\perp)$ to
perform 
the $\xi^\perp$-integration. 
We introduce the form factors for the 
bilocal current matrix elements using Lorentz covariance,
\begin{eqnarray}  
	\langle PS| V_\alpha^\mu(\xi|0) - V^\mu_\alpha(0|\xi) |PS \rangle 
		&=& P^\mu \overline{V}_{1\alpha}(P^2,\xi \cdot P) +
		\xi^\mu \overline{V}_{2 \alpha}(P^2,\xi \cdot P) \, , 
		\label{bcff} \\
	\langle PS| A_\alpha^\mu(\xi|0) + A^\mu_\alpha(0|\xi) |PS \rangle 
		&=& S^\mu \overline{A}_{1\alpha}(P^2,\xi \cdot P) +
		P^\mu \xi \cdot S \overline{A}_{2\alpha}(P^2,\xi \cdot P) 
		\nonumber \\
	& & ~~~~~~~~~~~~~ + \xi^\mu S \cdot \xi \overline{A}_{3\alpha}(P^2,
		\xi \cdot P) \, . \label{bacff}
\end{eqnarray}
Using the definition
\begin{equation}
	\epsilon(\xi^-) = - {i \over \pi} \int_{-\infty}^\infty
		{d q^+ \over q^+} e^{i q^+\xi^-/2} \, ,
\end{equation} 
we find that
\begin{eqnarray}
	T^{+-} = -{1 \over \pi q^-} \int_{-\infty}^\infty {d{q'}^+ \over {q'}^+ 
		- q^+ } \int_{-\infty}^\infty & & d\xi^- e^{iq^+\xi^-/2}
		\sum_\alpha e_\alpha^2 \Bigg\{{1\over 2}(P^-q^+ - P_\bot
		\cdot q_\bot) \overline{V}_{1 \alpha} \nonumber \\
	& & + {1\over 2} q^+ \xi^- \overline{V}_{2 \alpha} + {i\over 2}
		\epsilon^{ij}q_i \Big[S_j \overline{A}_{1 \alpha} + P_j
		{S^+ \xi^-\over 2} \overline{A}_{2 \alpha} \Big] \Bigg\} 
		\, , \label{t+-lq}
\end{eqnarray}
where the bilocal current form factors are determined from eq.(\ref{bcff})
and eq.(\ref{bacff}):
\begin{eqnarray}
	\overline{V}_{1\alpha} &=& {1\over P^+} \langle PS |
		\overline{\psi}_\alpha (\xi^-) \gamma^+ \psi_\alpha (0) - 	
		\overline{\psi}_\alpha (0)\gamma^+ \psi_\alpha (\xi^-) |PS 
		\rangle \label{v1+} \\
	&=& {1\over P^i} \langle PS | \overline{\psi}_\alpha (\xi^-) 
		\gamma^i \psi_\alpha (0) - \overline{\psi}_\alpha (0)
		\gamma^i \psi_\alpha(\xi^-) |PS \rangle \, , \label{v1i} \\
	\overline{V}_{2\alpha} &=& {1\over \xi^-} \langle PS | 
	      \overline{\psi}_\alpha (\xi^-) \Big(\gamma^--{P^-\over P^+}
		\gamma^+ \Big)\psi_\alpha (0) - h.c. |PS \rangle \nonumber \\
	&=& {1\over \xi^-} \langle PS | \overline{\psi}_\alpha (\xi^-) 
		\Big(\gamma^--{P^-\over P^i}\gamma^i \Big)
		\psi_\alpha (0) - h.c. |PS \rangle \, , \\
	\overline{A}_{1\alpha} &=& {1\over S^i_T} \langle PS| 
		\overline{\psi}_\alpha (\xi^-) \Big(\gamma^i
		-{P^i\over P^+}\gamma^+ \Big)\gamma_5\psi_\alpha (0) 
		+ h.c. |PS \rangle \, , \\
	\overline{A}_{2\alpha} &=& {-2\over P^+\xi^- S^i_T} \langle PS| 
		\overline{\psi}_\alpha(\xi^-) 
		\Big(\gamma^i -{S^i\over S^+}\gamma^+ \Big)
		\gamma_5 \psi_\alpha(0) + h.c. |PS \rangle \, . \label{a2}
\end{eqnarray}
Notice that, since $\xi^{+, \perp} =0$ in the above expressions,  
it follows that the matrix elements of the plus
and transverse components of the bilocal current yield the same form factor
$\overline{V}_{1\alpha}$ as is evident from eq.(\ref{v1+}) and eq.(\ref{v1i}) 
(and similarly for $\overline{V}_{2\alpha}$). 

Now, let us pick up the same $(+-)$ component of the hadronic
tensor eq.(\ref{wfun}) in the large $q^-$ limit: 
\begin{eqnarray}
	W^{+-} &=& {1\over 2}F_L +  (P_\perp)^2 {F_2 \over \nu}
                  - 2 P_\bot \cdot q_\bot{F_2\over q^2} \nonumber \\
	       & & ~~~~~~~~~ + 2i\epsilon^{ij}q_i
		\Big[S_{jL} {g_1\over \nu} + S_{jT}{g_T\over \nu} \Big] \, , 
		\label{w+-lq}
\end{eqnarray}
where $S_{jT} = S_j - S^+ {P_j \over P^+}$ and $S_{jL} = S_j -S_{jT}
= S^+ {P_j \over P^+}$, and  $\nu = {1\over 2}P^+q^-$ in the large $q^-$ 
limit. To find the deep inelastic structure
functions, we compare eq.(\ref{w+-lq}) with eq.(\ref{t+-lq}) 
through eq.(\ref{twr}). 
Thus, we obtain the structure functions as
given below (here in the following we 
have used the notation $\eta \equiv {1\over 2}P^+ \xi^-$). 
\begin{eqnarray}
	{F_2(x,Q^2)\over x} &=& {1\over 4\pi} \int d\eta e^{-i\eta x} 
		\sum_\alpha e^2_\alpha \overline{V}_{1\alpha} 
		\label{f20} \\
	&=& {1\over 4\pi P^+} \int d\eta e^{-i\eta x} \sum_\alpha 
		e^2_\alpha \langle PS| \overline{\psi}_\alpha (\xi^-) 
		\gamma^+ \psi_\alpha (0) - \overline{\psi}_\alpha (0)
		\gamma^+ \psi_\alpha (\xi^-) |PS \rangle  \label{3f2+} \\
	&=& {1\over 4\pi P^i_\bot} \int d\eta e^{-i\eta x} \sum_\alpha
                e^2_\alpha \langle PS | \overline{\psi}_\alpha (\xi^-) 
		\gamma_\bot^i \psi_\alpha (0) - \overline{\psi}_\alpha (0)
		\gamma_\bot^i \psi_\alpha(\xi^-) |PS \rangle \, . \label{3f2i}
\end{eqnarray}
Note, the last equality is found for the first time in our work and it is 
discussed thoroughly in the work by A. Harindranath and W. Zhang 
\cite{3Hari97}. We shall come back to it later.
\begin{eqnarray}
	F_L(x,Q^2) &=& -{q^+\over \pi P^+q^-} \int d\eta e^{-i\eta x} 
		\sum_\alpha e^2_\alpha \Big[(P^- - {P_\perp^2 \over P^+}) 
		\overline{V}_{1\alpha} + \xi^- \overline{V}_{2\alpha} \Big ]
		\nonumber \\
	&=& {P^+\over 4\pi } \Bigg({2x\over Q}\Bigg)^2 \int d\eta 
		e^{-i\eta x} \sum_\alpha e^2_\alpha \langle PS| 
		\overline{\psi}_\alpha (\xi^-) \nonumber \\
	& & ~~~~~~~~~~~~~~~~~~~~~~~~~  \times 
		\Big(\gamma^--{P_\perp^2 \over (P^+)^2}
                       \gamma^+ \Big)\psi_\alpha 
		(0) - h.c. |PS \rangle \, .  \label{fl}
\end{eqnarray}
Here the first equality may be reduced to the same expression obtained 
by the collinear expansion in the Feynman diagrammatic method up to the 
order twist-four \cite{3EPF83}. But it is obtained directly here in the 
leading order in the $1/q^-$ expansion without involving the concept 
of twist expansion.  
The polarized structure functions come out to be, 
\begin{eqnarray}
	g_1(x,Q^2) &=& {1\over 8\pi} \int d\eta e^{-i\eta x} \sum_\alpha
		e^2_\alpha \Big(\overline{A}_{1\alpha} + {1 \over 2}
           P^+ \xi^- \overline{A}_{2\alpha} \Big) \label{g10} \\
	&=& {1\over 8 \pi S^+} \int d\eta e^{-i\eta x} \sum_\alpha
		e^2_\alpha \langle P S| \overline{\psi}_\alpha (\xi^-)
		\gamma^+ \gamma_5 \psi_\alpha(0) + \overline{\psi}(0) 
		\gamma^+\gamma_5 \psi(\xi^-) |PS \rangle , \label{g1}\\
	g_T(x,Q^2) &=& {1 \over 8\pi} \int d\eta e^{-i\eta x} \sum_\alpha
		e^2_\alpha \overline{A}_{1\alpha} 
		\label{gt0}\\
	&=& {1\over 8\pi S^i_T} \int d\eta 
		e^{-i\eta x} \sum_\alpha e^2_\alpha \langle PS|\overline{
		\psi}_\alpha(\xi^-) \Big(\gamma^i -{P^i\over P^+}
		\gamma^+ \Big)\gamma_5 \psi_\alpha(0) +~ h.c. |PS \rangle 
		\, . \label{gt}
\end{eqnarray}

Thus we have obtained the structure functions as the Fourier transform of
the various matrix elements of bilocal currents, eqs.(\ref{3f2+}-\ref{gt})
being the main results of this section. 
The above results are derived without recourse to perturbation theory,
and also without the use of concept of collinear and massless partons.
They are also the most general expressions for the leading contribution
(in the $1/q^-$ expansion, not the leading contribution in terms of 
twists) to the deep inelastic structure functions in which the target
is in an arbitrary Lorentz frame. Some of these expressions have not 
ever been obtained in previous works. Note that the above expressions of the
structure functions make sense only in the $A^+=0$ gauge, which we have
used. Otherwise, bilocal expressions should always involve a path-ordered
exponential to ensure the gauge invariance. Since the bilocality here is
in the longitudinal direction, only in $A^+=0$ gauge the exponential
factor is one and we have the simple form. 

Notice from the above expressions of the structure functions 
which are derived within the
framework of light front QCD, it appears that the scaling is exact as was
predicted earlier from two seemingly disconnected works. In the pre-QCD era 
Bjorken predicted scaling using current algebra approach and taking infinite
momentum limit, which in our approach is built in. On the other hand,
scaling was obtained through the parton model where partons were treated 
as non-interacting particles. Now we can clearly see the connection, since the
interaction is buried in the bilocal operators and unless we pull them
apart, their form looks the same whether we consider the free theory or
the  interacting one. Also
the hadronic state, between which these bilocal operators are sandwiched,
has the substructure, resolution of which depends on the energy of the probe.
More and more substructures are resolved as we increase $Q^2$ for the probe
and the structure functions become $Q^2$-dependent.  
In this work, our effort is to unmask this 
$Q^2$-dependence which is  
hidden in the hadronic bound states $|PS \rangle$ by  
describing them  in terms of multi-parton wave functions, as will be 
discussed in the next Chapter. 

\newpage
\setcounter{section}{0}
\renewcommand{\thesection}{4.\arabic{section}}
\setcounter{equation}{0}
\renewcommand{\theequation}{4.\arabic{equation}}
{\flushleft \huge\bf {Chapter 4}}
\vskip 1cm
In the previous Chapter, we have seen how various DIS structure functions
are related to the  Fourier transform of  matrix elements of the bilocal
vector and axial vector currents sandwiched between the target states. As
they stand, it seems that these results are similar to what is obtainable
using light front current algebra approach 
where form of the current commutators  are
{\it always assumed}  and the scaling 
appears to be  exact. This is
due to the fact that, as mentioned earlier,  the interactions are buried
in the bilocal operators. Also, the target state substructures, which depends
on the energy scale of the probe, are not manifest in these relations.
Obviously, what is important for DIS structure functions are the matrix
elements of the bilocal operators renormalized at the physical energy scale
$Q$ of the probe. Since there is no reference of the underlying dynamics and
the necessary renormalization in the current algebra approach, many of the
current algebra predictions are invalidated in QCD perturbation theory.
The only exceptions are the sum rules protected by certain conservation
laws. In this Chapter,  we shall see how this renormalization procedure can
be carried out expanding the target state in terms of multi-parton
wave-functions and using the old-fashioned perturbation theory appropriate
to light-front QCD. First we shall see in Sec.4.1  
how partonic interpretation becomes
apparent by introducing multi-parton wave-functions, where the partons are
not necessarily collinearly moving or massless as postulated originally. 
Next we shall propose there a new
factorization scheme which enables us (for certain structure functions) to
separate the soft and hard parts of the dynamics.  It shows how 
nonperturbative and perturbative aspects of
the dynamics can
be dealt with within the same light-front Hamiltonian framework as well as
the importance of the dressed parton structure functions, which we calculate
explicitly  next in Sec.4.2.    
Then in Sec.4.3 we shall calculate  the $F_2$-structure
function for a meson-like target starting from the scratch, 
which shows
among other things how factorization (we introduced) is realized. Lastly, we 
shall discuss in Sec.4.4  
the physical interpretation of the various structure functions in
the light of sum rules they satisfy.
\section{Unravelling the Complexities of structure functions}
\subsection{Multi-parton wave functions}
In our formulation, the structure functions
are proportional to the simple hadronic matrix elements of the bilocal
currents that are separated only in the longitudinal direction. 
In this formulation, no time evolution or propagation is explicitly 
involved in the matrix elements. Hence, unlike the OPE or the
perturbative field theory descriptions of parton model, all the 
perturbative and nonperturbative  dynamics here are completely carried 
by the structure of the target's bound state. 
To unravel this dynamics we first
expand the target state 
in terms of Fock-states introducing multi-parton wave-functions. 
As mentioned earlier, this expansion, which is closer to the real 
physical picture probed in the 
experiments, is meaningful due to the triviality
of light-front vacuum. 
Thus, the bound state of a hadron on light-front is given as
\begin{equation}
        |PS \rangle = \sum_{n,\lambda_i} \int' dx_i d^2\kappa_{\bot i} 
		 | n, x_iP^+,x_iP_{\bot}+ 
		\kappa_{\bot i}, \lambda_i \rangle \Phi^S_n 
		(x_i,\kappa_{\bot i}, \lambda_i) \, , \label{lfwf} 
\end{equation}
where $n$ represents $n$ constituents contained in the Fock state 
$|n, x_i P^+, x_i P_{\bot} + \kappa_{\bot i}, \lambda_i \rangle$, 
$\lambda_i$ is the helicity of the i-th constituent, $\int'$ denotes 
the integral over the space:
\begin{equation}  \label{lfspc}
        \sum_i x_i = 1, ~~ {\rm and} ~~~ \sum_i \kappa_{\bot i} = 0
\end{equation}
while $x_i$ is the fraction of the total longitudinal momentum 
carried by the $i$-th constituent, and $\kappa_{\bot i}$ is its relative 
transverse   momentum with respect to the center of mass frame:
\begin{equation}
        x_i = { p_i^+ \over P^+}~~, ~~~ \kappa_{i\bot} = p_{i\bot} - x_i 
P_{\bot} \end{equation}
with $p_i^+, p_{i\bot}$ being the longitudinal and transverse momenta
of the $i$-th constituent. $\Phi^S_n (x_i,\kappa_{\bot i},\lambda_i)$ 
is the amplitude of the Fock state $| n, x_iP^+,x_iP_{\bot}+ \kappa_{\bot 
i},\lambda_i \rangle $, i.e., the {\it multi-parton wave-function},
which is boost invariant and satisfy the normalization condition: 
\begin{equation}
        \sum_{n,\lambda_i} \int' dx_i d^2\kappa_{\bot i} 
		|\Phi^S_n (x_i,\kappa_{\bot i},\lambda_i)|^2 = 1,
\end{equation}
and is, in principle, determined from the light-front bound state 
equation (long-hand version of $H|PS\rangle=E|PS\rangle$ in light-front),
\begin{equation}
        \Big(M^2 - \sum_{i=1}^n { \kappa_{i\bot}^2 + m_i^2 \over x_i} 
		\Big) \left[\begin{array}{c} \Phi^S_{qqq} \\
                \Phi^S_{qqqg} \\ \vdots \end{array} \right]
                  = \left[ \begin{array}{ccc} \langle qqq
                | H_{int} | qqq \rangle & \langle qqq | H_{int}
                | qqqg \rangle & \cdots \\ \langle qqq g
                | H_{int} | qqq \rangle & \cdots & ~~  \\ \vdots &
                \ddots & ~~ \end{array} \right] \left[\begin{array}{c}
                \Phi^S_{qqq} \\ \Phi^S_{qqqg} \\ \vdots \end{array}
                \right] . \label{lfbef}
\end{equation}
Here $H_{int}$ is the interaction part of the light-front QCD 
Hamiltonian given in Chapter 2. 
Of course, solving this infinite dimensional coupled equation is a huge task
and we shall shortly see how it can, at least, be made feasible by
introducing a new factorization scheme. 
Thus, the complexities of the structure functions carried by 
hadronic bound states are now translated into the language of 
multi-parton wave functions on the light-front, rather than 
composite operators in OPE.

Explicitly, let us look at the structure function $F_2(x,Q^2)$ as obtained
in the previous Chapter:
\begin{eqnarray}
	{F_2
	(x,Q^2)\over x} &=& {1\over 4\pi} \int d\eta e^{-i\eta x} 
		 \overline{V}_{1} \nonumber  
		 \\
	&=& {1\over 4\pi P^+} \int d\eta e^{-i\eta x}  
		 \langle PS| \overline{\psi} (\xi^-) 
		\gamma^+ \psi (0) - \overline{\psi} (0)
		\gamma^+ \psi (\xi^-) |PS \rangle_p,  \label{f2+P}\\
	&=& {1\over 4\pi P^i_\perp} \int d\eta e^{-i\eta x}  
		 \langle PS| \overline{\psi} (\xi^-) 
		\gamma^i \psi (0) - \overline{\psi} (0)
		\gamma^i \psi (\xi^-) |PS \rangle_p,  \label{f2i}
\end{eqnarray}
and for
the illustration purpose, 
consider  only the relation involving the plus component
(usually called the ``good" component),
\begin{equation}
	\overline{\psi}(\xi^-) \gamma^+ \psi(0) = 2 \psi_+^\dagger(\xi^-) 
	\psi_+(0) \, .
\label{plpl}
\end{equation}	
This has no explicit dynamical dependence, and has the lowest mass 
dimension in light-front 
(a twist-two operator in OPE language). The corresponding 
matrix element has straightforward parton interpretation. It can be easily
shown, using the Fock-expansion of the state $|PS\rangle$ as well as  
the dynamical fields $\psi^+$ present,  
that on the light-front eq.(\ref{f2+P}) reduces  to the sum of expectation
values of 
various quark (parton) number operators, which immediately leads to the
fact that $F_2$ is proportional to the parton density distributions
$q_\alpha (x,Q^2)$. 
\begin{eqnarray}
	{F_2(x,Q^2) \over x} &=& \sum_\alpha e^2_\alpha q_\alpha
		(x,Q^2) \, , \label{f2part} \\
	q_\alpha (x,Q^2) &=&  \int d^2 k_\bot 
		\langle PS | \sum_\lambda b_\alpha ^\dagger(x,k_\perp,\lambda) 
		b_\alpha (x,k_\perp,\lambda) | PS \rangle  \nonumber \\
	&=& \int d^2 \kappa_\bot \sum_{n,\lambda_i} \int'' dx_i d^2 
		\kappa_{\bot i} |\Phi^S_{n, \alpha} (x, x_i, 
		\kappa_{\bot i},\lambda_i)|^2 \, , \label{partd} 
\end{eqnarray}
where the $Q^2$-dependence is 
carried by the multi-parton wave functions with the active parton 
renormalized at the scale $Q^2$, $\int''$ 
denotes the integral in the right-hand-side over the space of  
eq.(\ref{lfspc}) except for the active parton $(x, \kappa_\bot)=
(k^+/P^+, p_\bot - xP_\bot)$. (Here we have omitted the antiquark
contributions for simplicity, see Sec.4.3 for the complete description). 
With this consideration it is 
straightforward to derive the logarithmic corrections which is  
same as that obtained in the QCD improved parton model or in the 
OPE, as will be shown later in this Chapter. In this case, all the three 
descriptions are almost the same. 
The only difference here in our framework is that  
the perturbative QCD dynamics is transferred from the composite 
operators in OPE language to the scale-dependent multi-parton wave-functions 
on the light-front, which enables us to describe the nonperturbative 
dynamics in the same framework, as we discuss next in the 
context of new factorization scheme that we are
going to introduce.

\subsection{A scheme for the evaluation of soft and hard
contributions to deep inelastic structure functions}
As shown in the previous Chapter, all the derivations and discussions
of the deep inelastic structure functions in the
$1/q^-$ expansion are rigorously carried out within light-front 
QCD and without recourse to perturbation theory. The remaining 
problem is how to evaluate various matrix elements of the bilocal 
currents. These matrix elements contain both hard and soft 
quark-gluon dynamics. As we have just seen, 
all the hard and soft dynamics probed 
through the structure functions are completely carried by the 
target's bound state in the present formulation. This is the
main advantage of this formalism that allows us to explore the
perturbative and nonperturbative contributions to the structure
functions in the same framework. Let us first introduce in detail the new
factorization scheme for such an exploration. 

We have already seen  how 
the hadronic bound state is  formally expressed in
terms of Fock space expansion on the light-front (see  eq.(\ref{lfwf})),
and it is determined in principle by the light-front bound state
equation given by eq.(\ref{lfbef}). However, the difficulty in determining
the wave-functions by solving eq.(\ref{lfbef}) is that the QCD Hamiltonian
contains more than one energy scale. At different energy scales,
QCD Hamiltonian can exhibit different aspects of the dynamics.
So, let us roughly divide the quark and gluon dynamics into
two energy domains, namely, high energy and low energy. In the 
high energy domain, the dynamics is controlled by the renormalized
QCD Hamiltonian with all the constituents carrying momenta
greater than a scale $\mu_{\rm fact}$ ($\approx 1 GeV$),  
which we call the factorization energy scale. This high 
energy QCD Hamiltonian describes all the hard dynamics of 
quarks and gluons and determines the hard contributions to the
structure functions which can be calculated in the perturbation theory. 
In the low energy domain, the effective QCD Hamiltonian is still 
{\it unknown} but such a low energy QCD Hamiltonian should fairly
determine the low energy structure of the hadrons and is
responsible for the soft contributions to the structure 
functions. 

Schematically, we may write the QCD Hamiltonian on the light-front
for DIS as
\begin{equation}
	H^{LF}_{QCD} = \left\{ \begin{array}{ll}
	 H^H_{QCD} \equiv \int_{k_{i\bot}^2 \geq \mu_{\rm fact}^2} 
		dk_i^+ d^2k_{i\bot} {\cal H}^C_{QCD} (k_i)~&~~{\rm for~hard~
		contributions} \, , \\
		~&~ \\
	 H^M_{QCD} \equiv \int dk_i^+ d^2k_{i\bot} {\cal H}^C_{QCD} 
		(k_i) &~~ {\rm for~mixed~hard~and~soft~modes}\, , \\
		~&~ \\
	 H^L_{QCD} \equiv \int_{k_{i\bot}^2 < \mu_{\rm fact}^2 } 
		dk_i^+ d^2k_{i\bot} {\cal H}^L_{QCD} (k_i) &~~{\rm for~
		soft~contributions}\, , 
			\end{array} \right.
\label{seprn}
\end{equation}
where $H^H_{QCD}$ represents the canonical light-front QCD Hamiltonian 
(with density ${\cal H}^C_{QCD}$ as given earlier) 
in which the 
transverse momenta of all the quarks and 
gluons are restricted to be $\mu^2_{\rm fact} < k_\bot^2 < Q^2$ (i.e., 
hard partons), and $H^L_{QCD}$ denotes a low energy effective 
light-front Hamiltonian in which all the constituents have the 
transverse momentum $k_\bot^2 < \mu^2_{\rm fact}$ (soft partons).  
In addition, we also introduce a 
Hamiltonian $H^M_{QCD}$ which depends only on the interaction part and
which mixes the hard and soft partons. 
The source of such a mixing term can easily be traced out as follows. If one
tries to separate all the momenta integrals in the light-front 
Hamiltonian
into two parts, one ends up with a term involving canonical Hamiltonian
density where all the momenta integrals
are restricted in the high momenta domain and another term where all are
restricted in the low momenta  domain; plus there will be all kinds of
mixing term where some of the integrals are in one domain while the others are
not. But, we should keep in mind that the above argument is naive. 
It is so because even though this way of 
separating the integral gives right $H^H_{QCD}$, and $H^M_{QCD}$ is one of the
mixed term which becomes effective according to the process, $H^L_{QCD}$ is
still a different object. This low energy Hamiltonian is, in principle, 
obtained by integrating 
out all the modes with $k_\bot^2 > \mu_{\rm fact}^2$ from the canonical 
light-front QCD Hamiltonian. This also explains the presence of the 
superscript  ${\cal C}$ 
(for canonical) in Hamiltonian density  
while writing down $H^H_{QCD}$ and $H^M_{QCD}$ only, as in eq.(\ref{seprn}). 
Writing the light-front QCD Hamiltonian in such three parts will 
make the discussion of the perturbative and nonperturbative QCD 
contributions to DIS structure functions much more transparent, as we
will see next.

Now, the target bound state can be expressed by
\begin{eqnarray}  \label{pss}
	| PS \rangle = U_h | PS, \mu_{\rm fact}^2 \rangle \, ,
\end{eqnarray}
with
\begin{eqnarray}
	&& U_h = T^+ \exp \Bigg\{ -{i \over 2} \int_{-\infty}^0 dx^+ 
		(H^H_{QCD}+H^M_{QCD}) \Bigg\}	\, , \label{pqcd} \\ 
 && H^L_{QCD} |PS, \mu_{\rm fact}^2 \rangle = {P_\bot^2 + M^2 \over P^+}
		| PS, \mu_{\rm fact}^2 \rangle  \, .	\label{npqcd} 
\end{eqnarray}
In eq.(\ref{pqcd}), $H^H$ and $H^M$ contain the interaction parts only and
the mixed Hamiltonian $H^M_{QCD}$ is active only
in the extreme right in the time-ordered expansion of the evolution operator
$U_h$ (as explained later). 
In other words, 
the hard and the soft dynamics in the bound states are determined 
separately by $H^H_{QCD}$ and $H^L_{QCD}$ but these two contributions 
are connected by $H^M_{QCD}$ through the action of  
$U_h$ on the state $|PS,\mu^2_{\rm fact} \rangle$.
On the other hand,  the soft dynamics, contained in $|PS,
\mu^2_{\rm fact} \rangle$, must be solved nonperturbatively from 
eq.(\ref{npqcd}), and the key point to solve eq.(\ref{npqcd}) 
is to find the low energy effective Hamiltonian $H^L_{QCD}$. 
A practical procedure to find $H^L_{QCD}$ on the 
light-front may be the use of similarity renormalization group 
approach plus a weak-coupling treatment developed recently
\cite{4Wilson94,4zhang97,4Perry97}. Indeed, a major effort on the study of
light-front QCD is underway at present\cite{4review}.

To see how the perturbative and nonperturbative QCD contributions
can be separately evaluated in the present formalism and how these
two contributions are connected by $H^M_{QCD}$, we substitute 
eqs.(\ref{pss}-\ref{npqcd}) into the expressions of structure 
functions. Denote the structure functions simply 
by $F_i \equiv: \{ F_L, F_2, g_1, g_T \}$,
\begin{eqnarray}
	F_i(x,Q^2) \sim \int d\eta e^{-i \eta x} \sum_\alpha e_\alpha^2
		\langle PS| \overline{\psi}_\alpha (\xi^-) \Gamma_i 
		\psi_\alpha (0) \pm h.c. | PS \rangle \, ,
\end{eqnarray}
where $\Gamma_i$ involves the Dirac $\gamma$-matrices (see the expression
for structure functions derived  
in the previous Chapter). After putting complete set of states in
appropriate places, it follows that
\begin{eqnarray}
        F_i(x,Q^2) = \int d\eta e^{-i\eta x}
                && \sum_\alpha e^2_\alpha \sum_{n_1,n_2} \langle 
		P S, \mu^2_{fact}| n_1\rangle \langle n_2| PS,
		\mu^2_{fact} \rangle \nonumber \\
        && \times \langle n_1 | U_h^{-1} \Big[\overline{\psi}_\alpha
		(\xi^-) \Gamma_i \psi_\alpha (0) \pm h.c \Big]  
		U_h | n_2 \rangle \, , \label{gfock} 
\end{eqnarray}
where $|n_1\rangle, |n_2 \rangle$ are a complete set of quark and 
gluon Fock states with momentum $k_i^2 \leq \mu^2_{\rm fact}$ only as
dictated by orthonormality of the states.
This is indeed the generalized factorization theorem in the light-front
Hamiltonian formulation. The hard contribution is described by the
matrix element,
\begin{equation}  \label{hardc}
	\langle n_1 | U_h^{-1} \Big[\overline{\psi}_\alpha(\xi^-) 
	\Gamma_i \psi_\alpha (0) \pm h.c \Big] U_h | n_2 \rangle \, ,
\end{equation} 
which can be evaluated in the light-front time-ordered perturbation theory
\cite{4zhang93}. The physical picture corresponds to the multi-parton
forward scattering amplitude with all the internal partons carrying a 
momentum
with the transverse component $k_{\bot}$: $\mu^2_{\rm fact} \leq k^2_\bot 
\leq Q^2 $ and the longitudinal momentum fraction $y$: $ x \leq y \leq 1$.  
The soft contribution is characterized by the overlap of the multi-parton 
wave functions in different Fock states: 
\begin{equation}  \label{softc}
	\langle P S, \mu^2_{fact}| n_1\rangle \langle n_2| PS, 
		\mu^2_{fact} \rangle \, , 
\end{equation}
which contains all the quantum correlations and interference effects 
of multi-parton (quarks and gluons) dynamics in the low energy 
domain with $k^2_\bot < \mu^2_{\rm fact}$.  Since we assume all the {\it 
internal} partons in the relevant DIS processes to 
carry high momenta $\mu^2_{\rm fact} \leq k_\bot^2 \leq Q^2$, the 
mixed Hamiltonian $H^M_{QCD}$ 
in the time-ordered expansion of $U_h$ in eq.(\ref{hardc}) 
has the contribution only when it comes at the 
extreme left or extreme right in the expansion, where it can act on
the state containing soft parton producing nonvanishing result. In effect, it
picks up a soft parton from $|n_1 \rangle$ and puts it into a
high energy state. It
is this effect that connects the hard contribution of 
eq.(\ref{hardc}) to the soft contribution in eq.(\ref{softc}).

The simple parton picture in deep inelastic processes corresponds to 
the case when $|n_1\rangle =|n_2 \rangle$ in eq.(\ref{gfock}) with only 
one parton in $|n_1 \rangle$ actively participating in the high energy
process, all others being spectators. This immediately leads to 
\begin{eqnarray}
	F_i(x,Q^2) \sim \sum_\alpha e^2_\alpha \int_x^1 dy {\cal P}_{pp',i} 
		(y,x,{Q^2 \over\mu_{\rm fact}^2}) q_{\alpha i} 
		(y,\mu_{\rm fact}^2) \, ,
\end{eqnarray}
where the hard scattering coefficient ${\cal P}_{pp',i}$ is determined by 
\begin{equation}
        {\cal P}_{pp',i}(y,x,{Q^2\over \mu_{\rm fact}^2}) \simeq \int d\eta 
		e^{-i\eta x} \langle y, k_\bot, s| U_h^{-1} \Big[ 
		\overline{\psi}_\alpha (\xi^-) \Gamma_i \psi_\alpha (0) 
		\mp h.c. \Big] U_h | y, k_\bot, s \rangle  . \label{sqm} 
\end{equation}
Here we have denoted $| y, k_\bot, s \rangle$ ($y=k^+/P^+$) as
the active parton state.  Eq.(\ref{sqm}) means that we have 
suppressed all references to the spectators in the states $| n_1 
\rangle $. The hard scattering coefficient is directly related to the 
so-called 
splitting function whose physical 
interpretation is the probability to find a daughter parton $p'$
in the active parent parton $p$.
The quantity $q_{\alpha i}(y, \mu^2_{\rm fact})$, usually 
called the parton distribution function, is given by 
\begin{equation}
	q_{\alpha i} (y, \mu^2_{\rm fact}) = \sum_n | \langle PS, 
		\mu^2_{\rm fact} | n \rangle |^2 \, ,
\end{equation}
where $n$ runs over all the Fock states containing the active 
parton with momentum fraction $y$.  
Theoretically, the parton distributions are 
determined by solving eq.(\ref{npqcd}). Physically, they
contain only the quantum correlations of 
multi-parton dynamics but no quantum interference effects. 
Example of such distribution functions is given by 
eqs.(\ref{f2part}-\ref{partd}) for ${F_2(x)/x}$ (except the transverse
momenta are now restricted to be soft), 
which manifestly exhibits the simple parton picture.

The above discussions indeed constitute a presentation of factorization 
scheme in the light-front Hamiltonian formulation. The leading
hard contributions to the structure functions are given by the 
the hard scattering coefficient ${\cal P}_{pp', i}(y,x,{Q^2\over 
\mu_{\rm fact}^2})$ and a detail calculation of ${\cal P}_{pp', i}$
based on the light-front time-ordered perturbative expansion of the 
multi-parton wave functions will be presented later in this Chapter.  
The evaluation of soft contribution to 
the structure functions given by $q_{\alpha i}(x, \mu_{\rm fact}^2)$,
however, 
remains for future investigations of nonperturbative light-front 
QCD approaches to the hadronic bound states.  
Thus, a unified treatment of both perturbative and nonperturbative
aspects of deep inelastic structure functions in the
same framework may emerge which permits one to overcome the obstacles
in dealing with the nonperturbative QCD dynamics in contrast to the OPE 
methods and field theoretical parton model approaches.  
To put this factorization scheme on a stronger basis, in the next section, 
we further investigate how
this factorization scheme can be realized perturbatively.

\subsection{Factorization: A Perturbative Analysis}
In this section we show in detail, how the factorization picture 
emerges in a perturbative analysis carried
over to all orders in the case where the bilocal operator involved
does not change the
particle number. The analysis leads to the concept of the structure function
of a dressed parton in DIS.

To explicitly demonstrate the factorization picture on the light-front,
we consider the $F_2$ structure function as a specific example here.
For simplicity we drop reference to the flavor and take $e_\alpha=1$, then
\begin{eqnarray}
{F_2(x,Q^2) \over x} && = { 1 \over 4 \pi} \int d \xi^- e^{-{ i \over 2} P^+
\xi^- x}  \langle PS  \mid
\Big [ (\psi^+)^\dagger(\xi^-) \psi^+(0) - 
(\psi^+)^\dagger(0) \psi^+
(\xi^-) \Big ] \mid PS \rangle. \nonumber \\
&& 
\end{eqnarray}
From the discussion of the last section,  we have
\begin{eqnarray}
{F_2(x,Q^2) \over x} && = q(x,Q^2)  = { 1 \over 4 \pi} \int d \xi^- 
e^{ - { i \over 2} P^+ \xi^- x}  \sum_{n_1,n_2}
\langle PS \mu^2 \mid n_1 \rangle
\langle n_2 \mid PS \mu^2 \rangle \nonumber \\
&& ~~~~\langle n_1 \mid U_h^{-1}
\Big [ (\psi^+)^\dagger(\xi^-) \psi^+(0) - 
(\psi^+)^\dagger(0) \psi^+
(\xi^-) \Big ] 
U_h \mid n_2 \rangle \, , \label{exp2}
\end{eqnarray}
where 
\begin{eqnarray}
U_h && =T^+ \exp\Big\{ -{i\over 2}\int_{-\infty}^0 dx^+ 
\tilde{P}^-_{int}(x^+)\Big\} \nonumber\\
&& = 1 - {i\over 2}\int_{-\infty}^0 dx^+ 
\tilde{P}^-_{int}(x^+) + (-{i\over 2})^2 \int_{-\infty}^0 dx_1^+ 
\tilde{P}^-_{int}(x_1^+)\int_{-\infty}^{x_1^+} dx_2^+ 
\tilde{P}^-_{int}(x_2^+) + \, ....
\label{uhi}
\end{eqnarray}
and $\tilde{P}_{int}^- \equiv P_{int}^{-H} 
+ P_{int}^{-M}$
is denoted as the hard and mixed light-front 
interaction Hamiltonian.

Let us use eq.(\ref{uhi}) in the expression eq.(\ref{exp2}) and 
consider the first few terms in the order-by-order expansion in QCD coupling
constant.
The lowest order term ($\sim g_s^0$) 
yields the low energy non-perturbative distribution
function 
\begin{eqnarray}
	q^{(0)}(x,Q^2) &=& q(x,\mu^2) = \sum_n |\langle PS, \mu^2_{\rm fact}
		| n\rangle |^2 \nonumber \\
	&=& \sum_s \int^\mu d^2 k^\perp \langle PS \mu^2 \mid 
		b_s^\dagger(y P^+, k^\perp) b_s(yP^+, k^\perp) \mid  
		PS \mu^2 \rangle \, .
\end{eqnarray}
The terms linear in the coupling constant ($\sim g_s^1$) is of the form
\begin{eqnarray}
q^{(1)}(x,Q^2) && =  { i \over 2} \sum_{nmp} \langle PS \mu^2
\mid n \rangle \int _{- \infty}^0 dx_1^+ \langle n \mid
{\tilde P}^{-}_{int}(x_1^+) 
\mid m \rangle  \langle m \mid {\cal O}  
  \mid p \rangle \langle p \mid P S \mu^2 \rangle. \nonumber\\
&&   
\label{linq}
\end{eqnarray}
Here we have put complete set of states in appropriate places and denoted 
\begin{eqnarray}
{\cal O} = { 1 \over 4 \pi}
\int d \xi^- e^{ - { i \over 2} P^+ \xi^- x}
\Big [ (\psi^+)^\dagger(\xi^-) \psi^+(0) - 
(\psi^+)^\dagger(0) \psi^+
(\xi^-) \Big ].
\end{eqnarray}
Since the 
plus component of the bilocal operator
conserves particle number on the light-front, eq.(\ref{linq}) will have
non-vanishing contribution only when $\mid m \rangle$ and $\mid p \rangle$
contain equal number of particle in the same momentum range (soft). Even if
there can be possibility of having equal number of particles in these two
states, the momentum range for individual particles will be different, since 
$\mid m \rangle$ is produced after the action of ${\tilde
P}^{-}_{int}(x_1^+)$ and should contain at least one particle with high
momentum. Therefore, by orthogonality of the states, $q^{(1)}(x,Q^2)$
vanishes.

Next we consider the second order
contribution given by, 
\begin{eqnarray}
q^{(2)}(x,Q^2) && =  { 1 \over 4} \sum_{nmpk} \langle PS \mu^2
\mid n \rangle \int _{- \infty}^0 dx_1^+ \langle n \mid
{\tilde P}^{-}_{int}(x_1^+) 
\mid m \rangle  \langle m \mid {\cal O}  
  \mid p \rangle \nonumber \\
&& ~~~~~~~~ \langle p \mid \int _{- \infty}^0 dx_2^+
{\tilde P}^{-}_{int}(x_1^+) \mid k \rangle \langle k \mid P S \mu^2 \rangle.  
\end{eqnarray}
Here we have not considered the contributions where 
intermediate states involve vanishing energy denominators. These
contributions are most conveniently included by introducing the wave
function renormalization constant associated with the parton active in 
the high energy process.
Using,
\begin{eqnarray}
P^-_{int}(x^+) = e^{{i \over 2} P_{free}^- (x^+)} P^-_{int}(0)
e^{-{i \over 2} P_{free}^- (x^+)},
\end{eqnarray}
we perform the $x_1^+$ and $x_2^+$ integrations producing energy
denominators and the resulting expression is 
\begin{eqnarray}
q^{(2)}(x,Q^2) && =  \sum_{nmpk}  
{ 1 \over P^-_{0n} - P^-_{0m} }{ 1 \over P^-_{0k} - P^-_{0p} } 
\langle n \mid {\tilde P}^{-}_{int} (0) \mid m \rangle 
\nonumber \\
&& ~~~~~~~ \langle m \mid {\cal O} \mid p \rangle \langle p \mid 
{\tilde P}^{-}_{int}(0) \mid k
\rangle 
\langle PS \mu^2
\mid n \rangle
\langle k \mid P S \mu^2 \rangle, 
\end{eqnarray}
where the $P^-_{0n}$ denotes the light-front energy of the state
$|n\rangle$ and so on. Note that 
the states $ \mid n \rangle$, and $ \mid k \rangle$ are forced to be low
energy states with $ (k^\perp)^2  < \mu^2$ for otherwise the overlap 
$\langle k \mid P S \mu^2 \rangle$ will be zero.  
We can restrict the states $
\mid m \rangle$, $ \mid p \rangle$ to be high energy states with $
(k^\perp)^2 > \mu^2$. The bilocal operator $ {\cal O}$ picks an active
parton in a high energy state whose longitudinal momentum is forced to be $
x P^+$. Further we need to keep only terms in ${\tilde P}^{-}_{int} $ 
which cause
transitions involving the active parton. (Transitions involving spectators
lead to wave function renormalization of spectator states which are 
cancelled by the renormalization process as shown explicitly
later in this Chapter.)

As we shall see shortly, to order $\alpha_s$ 
(i.e., considering only $q^{(0)}$ and $q^{(2)}$), 
a straightforward evaluation  leads to, 
\begin{eqnarray}
q(x,Q^2) = {\cal N}~ \Big \{ q(x, \mu^2) + { \alpha_s 
\over 2 \pi} C_f \ln{Q^2 \over \mu^2}
\int_x^1 { dy \over y} P(x/y) q(y, \mu^2) \Big \}
\end{eqnarray}
where ${\cal N}$ is the wave function renormalization constant 
of the active parton and $P$ is the splitting function. 
Including the contribution from the wave function renormalization
constant to the same order ($\alpha_s$), we get the factorized form,
\begin{eqnarray}
q(x,Q^2) = \int dy ~{\cal P} (x,Q^2; y, \mu^2) ~ q(y, \mu^2),
\end{eqnarray}
where the hard scattering coefficient 
\begin{eqnarray}
{\cal P} (x,Q^2; y, \mu^2)  =  \delta (x-y) + {
\alpha_s \over 2 \pi} C_f \ln{Q^2 \over \mu^2} \int_0^1 dz \delta(zy-x)
{\tilde P}(z)
\end{eqnarray}
with $ {\tilde P}(x) = P(x) - \delta(1-x) \int_0^1 dy P(y)$.

We note that the above analysis can be carried over to all orders in
perturbation theory with the result
\begin{eqnarray}
{\cal P}(x,Q^2; y, \mu^2) &&= \langle yP^+, k^\perp, s \mid U_h^{-1} 
	{\cal O} U_h \mid yP^+, k^\perp, s \rangle \nonumber \\
&& = \langle yP^+, k^\perp, s; (dressed) \mid {\cal O}
      \mid yP^+, k^\perp, s; (dressed)\rangle, \label{hsc}
\end{eqnarray}

In evaluating the above expression, 
only in the interaction Hamiltonians in the extreme
left and extreme right of the time ordered product we need to keep mixture
of soft and hard partons. This is governed by $P^{-M}_{int}$. They are  
needed to cause the  transition of a soft parton to a hard parton. In the 
rest of the interaction Hamiltonians occurring in the chain, the partons 
are restricted to be hard, i.e., they are determined by $P^{-H}_{int}$ only. 
For the leading logarithmic evolution we are discussing here, they appear 
ordered in transverse momentum.

\section{Unpolarized Dressed parton structure functions}
Now we turn our attention to the calculations of
hard scattering coefficients, ${\cal P}(x,Q^2; y, \mu^2)$,
given by Eq.~(\ref{hsc}).  If we set $k=P$, then $y=1$ and the hard
scattering coefficients just become the structure functions of dressed quark
and gluon targets in DIS, 
\begin{equation}
        f_i^p(x,Q^2) = { 1 \over 4 \pi}\int d\eta
                e^{-i\eta x} ~_p\langle ks | \Big[ \overline{\psi} 
		(\xi^-) \Gamma_i \psi (0) \mp h.c. \Big]
                | ks \rangle_p  \, . \label{sqm1}
\end{equation}
As a matter of fact, we can compute the perturbative QCD
correction to the hadronic structure functions by calculating 
the structure functions of the dressed quarks and gluons.

Here we only outline  the necessary tools for calculating the 
perturbative contribution to the structure functions first (for details of
two-component formalism of light-front field theory, see 
\cite{4zhang93}).
In old-fashioned light-front perturbation theory, the dressed quark
or gluon states can be expanded as follows: 
\begin{eqnarray}
	\mid Ps \rangle_p && = U_h | Ps\rangle = \sqrt{\cal N} \Big 
		\{ \mid Ps \rangle + \sum_{n} \mid n \rangle { 
		\langle n \mid P^{-M}_{int} \mid 
		Ps \rangle  \over (P^- - P_n^-) } \nonumber \\
		&& ~~~~~ +  \sum_{mn} \mid m \rangle{ \langle m \mid 
	P^{-H}_{int} \mid n \rangle \langle n \mid P^{-M}_{int} \mid Ps
	\rangle \over(P^- - P_m^- ) (P^- - P_n^-)} + ... \Big \} \label{exp1}
\end{eqnarray}
where $|Ps \rangle$, the bare single particle state, and $ \mid n 
\rangle$, the two-particle state, $\mid m \rangle$, the three-particle 
state, etc., are eigenstates of the free Hamiltonian. 
Introducing the multi-parton amplitudes (wave functions),
\begin{eqnarray}
	\Phi_n && = { \langle n \mid P^{-M}_{int} \mid Ps \rangle \over 
		(P^- - P_n^- )}, \nonumber \\
	\Phi_m && = \sum_n{ \langle m \mid P^{-H}_{int} \mid n \rangle 
             \langle n \mid P^{-M}_{int} \mid Ps \rangle \over (P^- - 
		P_m^- ) (P^- - P_n^-)}, 
\label{phmn}
\end{eqnarray}
the expansion in Eq. (\ref{exp1}) takes the form
\begin{eqnarray}
	\mid P s \rangle_p = \sqrt{{\cal N}} \Big \{ \mid Ps \rangle + 
		\sum_n \Phi_n \mid n \rangle + \sum_m \Phi_m \mid m 
		\rangle + ... \Big \} \, .
\label{focke}
\end{eqnarray} 

In the above expressions, $P^{-M}_{int}$ and $P^{-H}_{int}$ are the 
interaction parts of the canonical light-front QCD Hamiltonian as given
earlier, but
the former contains the mixed soft and hard partons and the latter
only has hard partons.  Notice that the amplitudes in eq.(\ref{phmn}) 
are given in terms of the vertex function and the energy denominators which
are the main ingredients of the old-fashioned perturbation theory. Also notice
the fact that amplitudes for the Fock-states containing more particles involve
more interaction Hamiltonian. Thus, in the perturbative calculations in the
high energy domain, we can
truncate the expansion in eq.(\ref{focke}) reliably depending on the 
desired order of the the calculation (more on this later). 
With this background we now proceed to calculate the $F_2$ structure function
for dressed quark and gluon targets in the perturbative region. Let us
mention beforehand that the calculations are straightforward and we have
omitted the details whenever it comes to putting some expression into
another as will be mentioned and thereby calculating the matrix elements by
using standard commutation relations for the creation-annihilation
operators; other details are provided as much as is necessary.  
\subsection{Dressed quark structure function}
The $F_2$ structure function of a dressed quark is given by, 
\begin{eqnarray}
	{F_{2q}^{|q\rangle}
	(x,Q^2)\over x} &=& {1\over 4\pi} \int d\eta e^{-i\eta x} 
		 \overline{V}_{1} \nonumber  
		 \\
	&=& {1\over 4\pi P^+} \int d\eta e^{-i\eta x}  
		 {}_{q}\langle ks| \overline{\psi} (\xi^-) 
		\gamma^+ \psi (0) - \overline{\psi} (0)
		\gamma^+ \psi (\xi^-) |ks \rangle_q.  \label{f2+}
\end{eqnarray}
Here we are using the relation involving the plus component of the bilocal
current only. Note, we have added an extra subscript $q$ above  
($g$ in the equation to follow) in $F_2$ to
remind ourselves that it measures the quark (gluon) distribution 
in the target state $|q\rangle$.   
For later purpose, we also 
introduce the gluon structure function (which measures the gluon distribution
in the target) as defined in
Ref.\cite{4pQCD},  
\begin{eqnarray}
	F_{2g}^{|q\rangle}
	(x,Q^2) = { 1 \over 4 \pi P^+} \int d\eta e^{- {i \eta x }} 
		{}_g\langle k \lambda \mid (-)F^{+ \nu a} 
  		(\xi^-)F^{+a}_{~~\nu}(0) + (\xi \leftrightarrow 0)\mid 
		k \lambda \rangle_g \, .
\label{f2Gq}
\end{eqnarray}
Here, the superscript $|q\rangle$ 
in $F_2$ (in both the equations above) denotes the
the target to be a dressed quark. 

In view of the discussion above, explicitly we can expand the dressed quark 
state in terms of bare states of quark, quark 
plus gluon, quark plus two gluons, etc, 
\begin{eqnarray}
	\mid P \sigma \rangle_q = && \sqrt{{\cal N}_q} \Bigg \{ 
		b^\dagger(P,\sigma) \mid 0 \rangle \nonumber \\
	&& ~ + \sum_{\sigma_1,\lambda_2} \int {dk_1^+ d^2 k_1^\perp \over 
		\sqrt{2 (2 \pi)^3 k_1^+}} \int {dk_2^+ d^2 k_2^\perp 
		\over \sqrt{2 (2 \pi)^3 k_2^+}} \psi_2(P,\sigma  
		\mid k_1, \sigma_1; k_2 , \lambda_2)   \nonumber \\
	&& ~~~~~~~~~~~~~~~~~~~ \times \sqrt{2 (2 \pi)^3 P^+} 
		\delta^3(P-k_1-k_2) b^\dagger(k_1,\sigma_1)
		a^\dagger(k_2,\lambda_2) \mid 0 \rangle \Bigg \}, \label{4dsqs}
\end{eqnarray}
where we have truncated the expansion at the two-particle level.
The factor  ${\cal N}_q$ is the wave function renormalization constant 
for the quark and the function $\psi_2(P,\sigma \mid k_1 \sigma_1, k_2 
\lambda_2)$ is the probability amplitude to find a bare quark with 
momentum $k_1$ and helicity $\sigma_1$ and a bare gluon 
with momentum $k_2$ and helicity $\lambda_2$ in the dressed quark.

Let us introduce the Jacobi momenta ($x_i, \kappa_i^\perp$) 
\begin{eqnarray}
k_i^+ = x_i P^+, \,  k_i^\perp = \kappa_i^\perp + x_i P^\perp 
\end{eqnarray} 
so that 
\begin{eqnarray} 
\sum_i x_i = 1, \sum_i \kappa_i^\perp =0.
\end{eqnarray}
The amplitude $ \psi_2$ is related to the corresponding boost invariant
amplitude $ \Phi_2$ as
\begin{eqnarray}
\sqrt{P^+} \psi_2(k_i^+, k_i^\perp) = &&  \Phi_2(x_i,
\kappa_i^\perp).
\end{eqnarray}

Using the 
notation $x=x_1, \kappa_1 = \kappa$ and using the facts $x_1+x_2=1, 
\kappa_1+\kappa_2=0$, we have (see Appendix C for details) 
\begin{eqnarray}
	&& \Phi_2^{s_1, \rho_2}(x,\kappa^\perp; 1-x, - \kappa^\perp) 
		= { 1 \over \Big[ M^2 - {m^2 +(\kappa^\perp)^2 \over x } -
		{(\kappa^\perp)^2 \over 1-x} \Big] } \nonumber \\
	&&~~~~~~~~~\times   { g \over \sqrt{2 (2 \pi)^3}} T^a 
		{1\over \sqrt {1-x}}~
		\chi^\dagger_{s_1} \Big[ - 2 {\kappa^\perp \over 1-x} - 
		{\sigma^\perp.\kappa^\perp  -i m \over x} \sigma^\perp - 
		\sigma^\perp i m \Big] \chi_{\sigma} .{(\epsilon^\perp_{
		\rho_2})}^*, 
\nonumber\\
\label{qap} 
\end{eqnarray} 
where $M$ and $m$ are the masses of dressed quark and bare quark
respectively.

Evaluating the expression in eq.(\ref{f2+}) explicitly, noting that in the 
present case
the contribution from the second term (which has non-vanishing contribution
if anti-quarks are present, see later) in this expression is zero,
we get the quark structure function of the dressed quark
\begin{eqnarray}
	{F_{2q}^{|q\rangle}(x,Q^2) \over x} 
	&=& {\cal N}_q \Big \{ \delta (1-x) 
		\nonumber \\
	&& + \sum_{\sigma_1, \lambda_2} \int dx_2 \int d^2 \kappa_1^\perp
	\int d^2 \kappa_2^\perp \delta (1-x-x_2) \nonumber \\
	&& ~~~~~~~\times \delta^2(\kappa_1^\perp+\kappa_2^\perp) 
		\mid \Phi_2^{\sigma_1, \lambda_2}(x,\kappa_1^\perp; x_2,
	\kappa_2^\perp) \mid^2 \Big \}. \label{f2plus}
\end{eqnarray}
The above equation is just a special case of eq.(\ref{partd}) 
with the target being a
dressed quark which is truncated at the two particle level. 
It makes manifest the parton interpretation of the
quark distribution function, namely, the quark distribution function of a
dressed quark is the
incoherent sum of probability densities 
to find a bare parton (quark) with longitudinal
momentum fraction $x$ in various multi-particle Fock states of the dressed 
quark. Since we have computed the distribution function in field theory,
there are also significant
differences from the traditional parton model\cite{4Feynman72}. Most important
difference is the fact that the partons in field theory have transverse
momenta ranging from zero to infinity. Whether the structure function scales
or not now depends on the ultraviolet behaviour of the multi-parton
wave functions. By analyzing various interactions, one easily finds that in
super renormalizable interactions, the transverse momentum integrals 
converge in the ultraviolet and the structure function scales, 
whereas in renormalizable
interactions, the transverse momentum integrals diverge in the ultraviolet
which in turn leads to scaling violations in the structure function. 

Using eq.(\ref{qap}) and taking the bare and dressed quarks to be massless
$M=m=0$,
we arrive at
\begin{eqnarray} 
	&& \sum_{\sigma_1,\lambda_2} \int d^2 \kappa^\perp \mid
		\Phi_2^{\sigma_1,\lambda_2}(x,\kappa^\perp, 1-x, 
		-\kappa^\perp) \mid^2 \nonumber \\
	&&~~~~~~~~~~~~~~~= {g^2 \over (2 \pi)^3}  C_f { 1 + x^2 \over 1-x} 
	\int d^2 \kappa^\perp { 1 \over (\kappa^\perp)^2}\, , \label{qdens}
\end{eqnarray}    
where $C_f = {N^2 -1 \over 2N}$. Recalling that 
$\mid \Phi_2(x, \kappa^\perp) \mid^2$ is the probability density to
find a quark with momentum fraction $x$ and relative transverse momentum
$\kappa^\perp$ in a parent quark, we define the probability density to find
a quark with momentum fraction $x$ inside a parent quark as
the splitting function 
\begin{eqnarray}
	P_{qq}(x)= C_f {1 + x^2 \over 1-x}.
\label{pqqa}
\end{eqnarray}

The transverse momentum integral in eq.(\ref{qdens}) is divergent at 
both limits of integration.  We regulate the lower limit by $\mu$ and the 
upper limit by $Q$. Thus we arrive at, 
\begin{eqnarray}
	{F_{2q}^{|q\rangle}(x,Q^2) \over x} = {\cal N}_q  \Big [ \delta (1-x) \, 
		+ \, {\alpha_s \over 2 \pi}C_f{1+x^2 \over 1-x} \ln{Q^2 
		\over \mu^2} \Big].
\label{nqcon}
\end{eqnarray}
Note that $\mu$ has to be large enough so that perturbative calculation is
not invalidated. 
The normalization condition $\langle PS|P^\prime S^\prime\rangle =
2(2\pi)^3P^+\delta^3(P-P^\prime)\delta_{SS^\prime}$
in this case reads 
\begin{eqnarray}
	{\cal N}_q \Big[ 1 + {\alpha_s \over 2 \pi} C_f
		\int dx{ 1 + x^2 \over 1-x}
		\ln{Q^2 \over \mu^2} \Big ] = 1.
\end{eqnarray}
Within the present approximation (valid only up to $\alpha_s$), we can write
\begin{eqnarray}
	{\cal N}_q = 1 - {\alpha_s \over 2 \pi} C_f \int dx{ 1 + 
		x^2 \over 1-x} \ln{Q^2 \over \mu^2} .
\label{nqval}
\end{eqnarray}
In the second term we recognize the familiar expression of wave function 
correction of the state $n$ in old fashioned perturbation theory, 
namely,  $ \sum'_m {\mid \langle m \mid V \mid n \rangle \mid^2 \over 
(E_n -E_m)^2}$.

Thus, putting eq.(\ref{nqval}) in eq.(\ref{nqcon}), we get to order $\alpha_s$,
\begin{eqnarray}
	{F_{2q}^{|q\rangle}(x,Q^2) \over x} 
	= \delta(1-x) + {\alpha_s \over 2 \pi}  
		\ln{Q^2 \over \mu^2} ~C_f~\Big[ {1+x^2 \over 1-x} - 
		\delta(1-x) \int dy {1+y^2 \over 1-y} \Big]. \label{onel}
\end{eqnarray}
Note that eq.(\ref{onel}) can also be written as
\begin{eqnarray}
	{F_{2q}^{|q\rangle}(x,Q^2) \over x} = 
	\delta(1-x) + {\alpha_s \over 2 \pi} 
		C_f \ln{Q^2 \over \mu^2}\Big[ {1 +x^2 \over (1-x)_{+}} + 
		{3 \over 2} \delta(1-x) \Big], \label{f2qq}
\end{eqnarray}
which is a more familiar expression. Note that, 
by construction, $\mid \Phi_2(x, 
\kappa^\perp) \mid^2$ is a probability density. However, this function is 
singular as $x \rightarrow 1$ (gluon longitudinal momentum fraction 
approaching zero). To get a finite probability density we have to 
introduce a cutoff $\epsilon$  ($x_{gluon} > \epsilon)$, for example. In 
a physical cross section, this $\epsilon $ cannot appear and here we have 
an explicit example of this cancellation. 
Note that the function 
${\tilde P}_{qq} =  C_f{1 +x^2 \over (1-x)_{+}} + {3 \over 2} \delta(1-x) $
does not have the probabilistic interpretation since it includes contribution
from virtual gluon emission. This is immediately transparent from the
relation
\begin{eqnarray}
	\int dx {\tilde P}_{qq}(x) = 0.
\end{eqnarray}
 We also note that the divergence arising from small transverse momentum
(the familiar mass singularity) cannot be handled properly in the present 
calculation. This is to be contrasted
with the calculation of the physical hadron structure function 
where the mass singularities can be properly absorbed into the
non-perturbative part of the structure function. On the other hand, $\ln Q^2$
dependence indicates towards the logarithmic scaling violation of $F_2$ 
structure function. Of course, this is not the end of the story since in our
leading order calculation $\alpha_s$ is small but constant. In reality,
$\alpha_s\sim 1/\ln Q^2$ and one has to consider all $(\alpha_s \ln Q^2)^n$ 
terms as in LLA. Our interest here does not permit such a discussion.

In view of eq.(\ref{pqqa}), 
the probability density to find a gluon with momentum fraction 
$x$ inside a parent quark is defined as the splitting function 
\begin{eqnarray}
	P_{Gq}(x) = C_f {1 + (1-x)^2 \over x},  
\end{eqnarray}
which is obtained by replacing $x$ with $(1-x)$ in $P_{qq}$, since in the
two particle
Fock-state, rest of the longitudinal momentum fraction is carried by the
associated gluon. This can also be obtained directly 
from the gluon distribution $F_{2g}^{|q\rangle}$ 
as given in eq.(\ref{f2Gq}) in
the dressed quark state. A similar calculation as above gives us 
\begin{eqnarray}
	F_{2g}^{|q\rangle}&& = 
	{ \alpha_s \over 2 \pi} \ln{Q^2 \over \mu^2} C_f x 
                {1 +(1-x)^2 \over x} \label{f2qg}\nonumber\\   
		&& = { \alpha_s \over 2 \pi} \ln{Q^2 \over \mu^2} C_f x
			P_{Gq}(x).
\end{eqnarray}
and serves as a clarification.
It is easy to check that
\begin{eqnarray}
	\int_0^1 dx ~x \Big [ {\tilde P}_{qq}(x) + P_{Gq}(x) \Big ] =0.
\end{eqnarray}

Our discussion and results for $F_{2q}^{|q\rangle}$ 
follows here from eq.(\ref{f2+}) (or, to be more precise eq.(\ref{f2+P})), 
which involves only the plus component of the bilocal current matrix
element. Eq.(\ref{f2i})  suggests that  
one should obtain the same information
regarding $F_{2q}^{|q\rangle}$ 
from  the   matrix element of the transverse components of bilocal current 
as well. This is not obvious from the operator structure of the transverse
component 
\begin{equation}
	\overline{\psi}(\xi^-) \gamma^i \psi(0) = 
	\overline{\psi}_-(\xi^-) \gamma^i_\bot 
		\psi_+(0) + \overline{\psi}_+(\xi^-) \gamma^i_\bot \psi_-(0) 
	\, ,
\end{equation}	
which explicitly depends on the interaction in QCD. However, it
is shown explicitly\cite{4Hari97} with a similar calculation as
above that although the operator
structures are different for different components of the bilocal current,  
$F_{2q}^{|q\rangle}$ comes out to be the same and have simple partonic
picture, contrary to the popular notion\cite{4jafst}. Thus, the equivalence
of eq.(\ref{f2+P}) and eq.(\ref{f2i}), which we found for the first time in
our work, has been established.
\subsection{Dressed gluon structure function}
Now we consider the calculation of structure functions 
where the target state is a dressed gluon. 
The dressed gluon state can be expanded as
\begin{eqnarray}
	\mid P \lambda \rangle_g =&& {\sqrt{\cal N}_g} \Big \{
		a^{\dagger}(P, \lambda) \mid 0 \rangle  \nonumber \\
	&& + \sum_{\sigma_1 \sigma_2} \int 
		{ dk_1^+ d^2 k_1^\perp \over \sqrt{2 (2 \pi)^3 k_1^+}}
		{ dk_2^+ d^2 k_2^\perp \over \sqrt{2 (2 \pi)^3 k_2^+}}
		\sqrt{2 (2 \pi)^3 P^+} \delta^3(P-k_1-k_2) \nonumber \\
	&& ~~~~ \psi_{2(q {\bar q})}(P, \lambda \mid k_1 \sigma_1, k_2 
		\sigma_2) b^\dagger(k_1 \sigma_1) d^\dagger(k_2, \sigma_2)  
		\mid 0 \rangle \nonumber \\
	&& + { 1 \over 2} \sum_{\lambda_1 \lambda_2} \int 
		{ dk_1^+ d^2 k_1^\perp \over \sqrt{2 (2 \pi)^3 k_1^+}}
		{ dk_2^+ d^2 k_2^\perp \over \sqrt{2 (2 \pi)^3 k_2^+}}
		\sqrt{2 (2 \pi)^3 P^+} \delta^3(P-k_1-k_2) \nonumber \\
	&& ~~~~ \psi_{2(gg)}(P, \lambda \mid k_1 \lambda_1, k_2 \lambda_2)
	a^\dagger(k_1 \lambda_1) a^\dagger(k_2, \lambda_2)  \mid 0 \rangle 
		\Big \}. \label{stateg}
\end{eqnarray}
where two-particle sector now contains $q\bar q$-states as well as two-gluon
states.
Note the symmetry factor ${ 1 \over 2}$ (in the last line) for the state
having identical bosons.

As before we introduce the boost invariant amplitudes
\begin{eqnarray}
	\sqrt{P^+}\psi_{2(q {\bar q})}(k_i^+, k_i^\perp) &=&  \Phi_{2(q
		{\bar q})} (x_i,\kappa_i^\perp), \nonumber \\
	\sqrt{P^+}\psi_{2(gg)}(k_i^+, k_i^\perp) &=&  \Phi_{2(gg)} (x_i,
		\kappa_i^\perp).
\end{eqnarray}
In terms of energy denominators and relevant vertex functions, 
the $ q {\bar q}$ wave function of the dressed gluon is given by
\begin{eqnarray}
	& & \Phi_2^{s_1, s_2}(x,\kappa^\perp; 1-x, - \kappa^\perp) 
		= { 1 \over \Big[ M^2 - {m^2 +(\kappa^\perp)^2 \over 
		x (1-x)}  \Big] } \nonumber \\ 
	& & ~~~~~~~\times   { g \over \sqrt{2 (2 \pi)^3}} T^a 
		\chi^\dagger_{s_1} \Big[ {\sigma^\perp.\kappa^\perp 
		\over x} \sigma^\perp - \sigma^\perp
		{\sigma^\perp .\kappa^\perp \over 1-x}- i { m \over x 
		(1-x)} \sigma^\perp \Big] \chi_{-s_{2}} 
		{(\epsilon^\perp_{\rho_2})}^*, \label{gap} 
\end{eqnarray} 
where $M$ and $m$ are the masses of dressed gluon and the bare quark
respectively. Henceforth, we shall work with $M=m=0$. 
The $gg$ wave function of the dressed gluon state is given by
\begin{eqnarray}
	\Phi_{2(gg)}(x, \kappa^\perp) =&& { g \over \sqrt{2 (2 \pi)^3}} 
		2 i f^{abc} { x(1-x) \over (\kappa^\perp)^2} { 1 \over 
		\sqrt{x}} { 1 \over \sqrt{1-x}} \nonumber \\
	&& ~~~ \epsilon^j_{\lambda_1} \epsilon^l_{\lambda_2}
		(\epsilon^i_{\lambda})^*\Big [ - \kappa^i \delta_{lj} + 
		{ \kappa^j \over x} \delta_{il} + {\kappa^l
		\over 1-x} \delta_{ij} \Big ],
\end{eqnarray}

Now explicit calculation of gluon structure function 
$F_{2g}$ given in eq.(\ref{f2Gq}) (but now 
for the dressed gluon target) gives the
following. 
The contribution from the first term in eq.(\ref{stateg}) comes out to be 
\begin{eqnarray}
	F_{2g}^{|g\rangle(1)} = {\cal N}_g \delta(1-x).
\end{eqnarray}
Since the the operator in $F_{2g}$ contains the creation and
annihilation operator ($a$, $a^{\dagger}$) for gluons only and 
the $q {\bar q}$ component of the state contains those of quarks, 
the corresponding contribution to structure function 
is a disconnected one which we omit.
The contribution 
from the $gg$ component of the dressed gluon state comes out to be 
\begin{eqnarray}
	F_{2g}^{|g\rangle(3)}= {\cal N}_g 
		{\alpha_s \over 2 \pi} \ln{Q^2 \over \mu^2} 2 N 
		\Big [ { x \over 1-x} + { 1-x \over x} + x(1-x) \Big ]x.
\end{eqnarray}
We define the probability density to a find a gluon with momentum fraction
$x$ in the dressed gluon, $P_{GG}(x)$ by 
\begin{eqnarray}
	P_{GG}(x) = 2 N \Big [ {x \over 1-x} + {1-x \over x} + 
		x (1-x) \Big ].
\end{eqnarray}
Collecting the above contributions together, we have,
\begin{eqnarray}
	F_{2g}^{|g\rangle}(x,Q^2) = 
		 {\cal N}_g \Big [ \delta(1-x) + {\alpha_s 
		\over 2 \pi} \ln{Q^2 \over \mu^2} 2N 	
		[{ x \over 1-x} + { 1-x \over x} + x (1-x)]x  \Big].
\label{wnf2g}
\end{eqnarray} 

The coefficient $ {\cal N}_g$ is determined from the longitudinal 
momentum sum rule for the dressed gluon target, namely, we require,
\begin{eqnarray}
	\int_0^1 dx F_2^{|g\rangle}(x) = \int_0^1 dx \Big [ 
		F_{2g}^{|g\rangle}(x) 
		+ F_{2q}^{|g\rangle}(x) \Big ] =
		{ 1 \over 2 (P^+)^2} ~{}_g\langle P 
		\mid \theta^{++}(0) \mid P \rangle_g =1 \label{normg}.
\end{eqnarray}
Thus we need to evaluate
\begin{eqnarray}
{ 1 \over 2 (P^+)^2} ~ {}_g\langle P \mid \theta^{++}_q(0) \mid P \rangle_g.
\end{eqnarray}
Explicit evaluation leads to 
\begin{eqnarray}
{ 1 \over 2 (P^+)^2} {}_g\langle P \mid \theta^{++}_q(0) \mid P \rangle_g =
{\alpha_s \over 2 \pi} \ln{Q^2 \over \mu^2} { 1 \over 2} \int dx \Big [ x^2 +
(1-x)^2 \Big] {\cal N}_g.
\end{eqnarray}
We define the probability density to find a quark with momentum fraction $x$
in a dressed gluon as the splitting function $P_{qG}(x)$: 
\begin{eqnarray}
P_{qG}(x) = {1 \over 2} \Big [ x^2 + (1-x)^2 \Big ]. 
\end{eqnarray}

From eq.(\ref{normg}) we arrive at 
\begin{eqnarray}
	{\cal N}_g \Big [ 1 + {\alpha_s \over 2 \pi}\ln{Q^2 \over \mu^2} 
		\int dx \Big \{  [x^2 + (1-x)^2 ]+ 2N [ { x \over 1-x} 
		+ { 1-x \over x} + x (1-x)]x \Big\} \Big] =1.
\end{eqnarray}
Thus to order $ \alpha_s$, we have 
\begin{eqnarray}
	{\cal N}_g =1 - {\alpha_s \over 2 \pi} \ln{Q^2 \over \mu^2} 
		\int dx \Big \{  [x^2 + (1-x)^2 ]+ 2N 
	[ { x \over 1-x} + { 1-x \over x} + x (1-x)]x \Big\}.
\end{eqnarray}
Performing the $x$-integration in ${\cal N}_g$ and putting it 
back into eq.(\ref{wnf2g}), we get 
dressed gluon structure function to order $\alpha_s$,
\begin{eqnarray}
	F_{2g}^{|g\rangle}x,Q^2) &&= \delta(1-x) + {\alpha_s \over 2 \pi} 
		\ln{Q^2 \over \mu^2} \nonumber \\
	&&~~~\Big \{ 2N \Big [ [{x \over (1-x)_+} + { 1-x \over x} + 
		x (1-x)]x + { 11 \over 12}\delta(1-x) \Big ] - { 1 \over 3} 
		\delta (1-x) \Big \}. 
\end{eqnarray}
Including the end point ($ x \rightarrow 1 $) contributions, we define,
\begin{eqnarray}
	{\tilde P}_{GG}(x) = 2N \Big \{ \Big [ { x \over (1-x)_+} 
		+ { 1-x \over x} + x (1-x) \Big ] + { 11 \over 12} 
		\delta(1-x) \Big \} - { 1 \over 3} \delta(1-x).
\end{eqnarray}
To the best of our knowledge, this is the first time gluon
splitting function has been calculated using multi-parton 
wave-functions. There exist some discussions in the literature 
regarding the calculation of DIS splitting functions using the 
language of multi-parton wave-functions mainly due to Lepage and 
Brodsky \cite{4Brodsky}. But for the gluon splitting function, 
they have simply quoted the result from Altarelli-Parisi paper 
\cite{4AP}. It is easy to verify that
\begin{eqnarray}
	\int_0^1 dx ~x  \Big [ 2P_{qG}(x) + {\tilde P}_{GG}(x) \Big ] =0.
\end{eqnarray}  

Note that in the last two subsections we have presented the calculations of
$F_2$ structure function for a dressed parton (quark or gluon) target.
These calculations can also be extended for longitudinal structure function
$F_L$ as well as the polarized structure functions. In fact, some of these
extensions
have already been performed giving rise to interesting new results. For
details see some of the works done by us and our collaborators
\cite{4rajk},\cite{4zhang96}. Our
discussion here together with these references mentioned, constitute a
complete presentation of how to perform the perturbative calculations (to the
leading order) in a simple and straightforward way within the Hamiltonian
framework of 
light front QCD. 
\section{Structure Function of hadron: Parton picture, Scale evolution and
Factorization}
In the previous sections in this Chapter, we have seen how 
the nonperturbative contribution to the structure functions and the 
scaling violations from the perturbative QCD corrections can be unified and 
treated in the same framework in our formalism. This is done by 
studying how the 
factorization of soft and hard part of the structure function 
can be realized, leading to the introduction of dressed parton structure
functions and through the perturbative calculation of them. 
In this section, we shall 
address the issues associated with scaling  violations in the structure 
function of the ``meson-like" bound state {\it without} invoking the proposed
factorization. This gives us the opportunity to explicitly demonstrate the
validity of the factorization outlined before. 
\subsection{Parton picture}
Let us first discuss the emergence of parton picture for the structure
function of a composite state.
We expand the state $\mid P \rangle $ for $ q {\bar q}$ bound state in 
terms of the Fock components $q {\bar q}$, $q {\bar q}g$, ... as follows.
\begin{eqnarray}
\mid P \rangle = && \sum_{\sigma_1, \sigma_2} 
\int {dk_1^+ d^2 k_1^\perp \over \sqrt{2 (2 \pi)^3 k_1^+}} 
\int {dk_2^+ d^2 k_2^\perp \over \sqrt{2 (2 \pi)^3 k_2^+}} 
\nonumber \\
&& \psi_2(P \mid k_1, \sigma_1; k_2, \sigma_2) \sqrt{2 ((2 \pi)^3 P^+}
\delta^3(P-k_1-k_2) b^\dagger(k_1, \sigma_1) d^\dagger(k_2,\sigma_2) \mid 0
\rangle \nonumber \\
&& + \sum_{\sigma_1,\sigma_2,\lambda_3} 
\int {dk_1^+ d^2 k_1^\perp \over \sqrt{2 (2 \pi)^3 k_1^+}} 
\int {dk_2^+ d^2 k_2^\perp \over \sqrt{2 (2 \pi)^3 k_2^+}} 
\int {dk_3^+ d^2 k_3^\perp \over \sqrt{2 (2 \pi)^3 k_3^+}} 
\nonumber \\
&& \psi_3(P \mid k_1, \sigma_1; k_2, \sigma_2; k_3, \lambda_3)
\sqrt{2 (2 \pi)^3 P^+} \delta^3(P-k_1 -k_2 -k_3) 
\nonumber \\
&&~~~~~b^\dagger(k_1 ,\sigma_1)
d^\dagger (k_2, \sigma_2) a^\dagger(k_3, \lambda_3) \mid 0 \rangle \nonumber
\\
&& + \, \, \, ... \, \, \, . \label{meson1}
\end{eqnarray}
Here $\psi_2$ is the probability amplitude to find a quark and an antiquark
in the meson, $\psi_3$ is the probability amplitude to find a quark,
antiquark and a gluon in the meson etc. 

We introduce the boost invariant amplitudes as before, 
\begin{eqnarray}
	\sqrt{P^+}\psi_2(k_i^+, k_i^\perp) = && 
 		\Phi_2 (x_i, \kappa_i^\perp), \nonumber \\
	P^+\psi_3(k_i^+, k_i^\perp) = &&  \Phi_3(x_i, \kappa_i^\perp),
\end{eqnarray}
and so on. Notice that the $P^+$-dependences of $\psi_2$ and $\psi_3$ are
different.
Now we evaluate the expression in eq.(\ref{f2+P}) explicitly 
with the ``meson-like target" given in eq.(\ref{meson1}).
The contribution from the first term (from the quark) in eq.(\ref{f2+P}),
in terms of $\Phi$s, comes out to be the following.
\begin{eqnarray}
	{F_2^q(x) \over x} = && \sum_{\sigma_1,\sigma_2} \int dx_2 \int
	d^2\kappa_1^\perp \int d^2 \kappa_2^\perp \delta (1 - x -x_2)
	\delta^2(\kappa_1 +  \kappa_2) \mid \Phi_2^{\sigma_1,
	\sigma_2}(x,\kappa_1^\perp; x_2 \kappa_2^\perp) \mid^2 \nonumber \\
&& + \sum_{\sigma_1, \sigma_2, \lambda_3} \int dx_2 \int dx_3 \int d^2
	\kappa_1^\perp \int d^2 \kappa_2^\perp \int d^2 \kappa_3^\perp 
	\delta(1 -x -x_2 -x_3) \delta^2(\kappa_1 + \kappa_2 + \kappa_3)
	\nonumber \\
&& ~~~~~~~~~~~~~~~~~~~~ \mid \Phi_3^{\sigma_1, \sigma_2, \lambda_3}(x,
\kappa_1^\perp; x_2, \kappa_2^\perp; x_3, \kappa_3^\perp) \mid^2 + ... ~~ .
\label{exact}
\end{eqnarray}
Again, the partonic interpretation of the $F_2$ structure function 
is manifest in this expression. Using different techniques and 
approximations, the same result has also been obtained by Brodsky 
and Lepage \cite{4Brodsky}.

Contributions to the structure function from the second term (from the
anti-quark) in 
Eq. (\ref{f2+P}) is given by,
\begin{eqnarray}
	{F_2^{\bar q}(x) \over x} = && \sum_{\sigma_1,\sigma_2} \int dx_1 \int
	d^2\kappa_1^\perp \int d^2 \kappa_2^\perp \delta (1 - x_1 -x)
	\delta^2(\kappa_1 +  \kappa_2) \mid \Phi_2^{\sigma_1,
	\sigma_2}(x_1,\kappa_1^\perp; x, \kappa_2^\perp) \mid^2 \nonumber \\
&& + \sum_{\sigma_1, \sigma_2, \lambda_3} \int dx_1 \int dx_3 \int d^2
	\kappa_1^\perp \int d^2 \kappa_2^\perp \int d^2 \kappa_3^\perp 
	\delta(1 -x_1 -x -x_3) \delta^2(\kappa_1 + \kappa_2 + \kappa_3)
\nonumber \\
&& ~~~~~~~~~~~~~~~~~~~~ \mid \Phi_3^{\sigma_1, \sigma_2, \lambda_3}(x_1,
\kappa_1^\perp; x, \kappa_2^\perp; x_3, \kappa_3^\perp) \mid^2 + ... ~~ .
\label{extbr}
\end{eqnarray}
Notice that eq.(\ref{exact}) and eq.(\ref{extbr}) are the special case of
eq.(\ref{partd}) with Fock expansion truncated at the three particle level. 
The normalization condition guarantees that
\begin{eqnarray}
\int dx \big[ {F_2^{q}(x) \over x} + {F_2^{\bar q} (x) \over x} \big ] = 2
\end{eqnarray}
which reflects the fact that there are two valence particles in the meson.
Since the bilocal current component ${\bar {\cal J}}^+$ involves only
fermions explicitly, we appear to have missed the contributions from the
gluon constituents altogether. Gluonic contribution to the structure
function $F_2$ is most easily calculated by studying the hadron
expectation value of the conserved longitudinal momentum operator $P^+$. 

From the normalization condition, it is clear that the valence distribution
receives contribution from the amplitudes  $\Phi_2$, $\Phi_3$, ...  at any
scale $\mu$. This has interesting phenomenological implications. In the
model for the meson with only a quark-antiquark pair of equal mass, 
the valence distribution function will peak at $x = {1 \over 2}$. If there
are more than just the two particles in the system, the resulting valence
distribution will no longer be symmetric about $ x = {1 \over 2}$
 as a simple
consequence of longitudinal momentum conservation and 
the peak shifts to the lower value of $x$.

The eq.(\ref{exact}) as it stands is useful only when the bound state
solution in QCD is known in terms of the multi-parton wave-functions. The
wave-functions, as they stand,
span both the perturbative and non-perturbative sectors of the theory.
Great progress in the understanding of QCD in the high energy sector is made
in the past by separating the soft (non-perturbative) and hard
(perturbative) regions of QCD via the machinery of factorization. It is of
interest to see under what circumstances 
a factorization occurs in the formal result of eq.(\ref{exact}) 
and a perturbative picture of scaling violations emerges
finally. We shall explicitly address this issue in the following section
where we consider only the ${F_2^q\over x}$ as in eq(\ref{exact}), 
which is sufficient to explain
all the relevant issues. All the arguments can be copied for anti-quark
contribution ${F_2^{\bar q}\over x}$. 

\subsection{Perturbative picture of scaling violations in a bound state}

To address the issue of scaling violations in the structure function of 
the ``meson-like" bound state, it is convenient to separate the momentum
space into low-energy and high-energy sectors. Such a separation has been
introduced in the past in the study of renormalization  of bound
state equations \cite{4Yukawa} in light-front field theory. The two sectors
are formally defined by introducing cutoff factors in the momentum space
integrals. How to cutoff the momentum integrals in a sensible and convenient
way in light-front theory is a subject under active research at the 
present time. Complications arise because of the possibility of large energy
divergences from both small $k^+$ and large $k^\perp$ regions.
In the following we investigate only the effects of logarithmic
divergences arising from
large transverse momenta, ignoring the 
subtleties arising from both small $x$
($x \rightarrow 0$)
and large $x$ ($x \rightarrow 1$) regions and subsequently use simple
transverse momentum cutoff. For complications arising from $x
\rightarrow 1$ region see Ref. \cite{4Brodsky}. 
     
\subsubsection{Scale separation}
We define the soft region to be $\kappa^\perp < \mu$ and the hard
region to be $ \mu < \kappa^\perp < \Lambda$, where
$\mu$ serves as a factorization scale which separates soft and hard regions.
Since it is an intermediate scale introduced artificially purely for
convenience, physical structure function should be independent of $\mu$.
The multi-parton amplitude $\Phi_2$ is a function of a single relative
transverse momentum $\kappa_1^\perp$ and we define
\begin{eqnarray}
 \Phi_2 = \left \{ \begin{array}{c} \Phi_2^{s}, 
~~ 0 < \kappa_1^\perp < \mu, \\
     \Phi_2^{h}, ~~~ \mu < \kappa_1^\perp < \Lambda. \end{array} \right.
\label{sfthd1}
\end{eqnarray}  
The amplitude $\Phi_3$ is a function of two relative momenta,
$\kappa_1^\perp$ and $\kappa_2^\perp$ and  we define
\begin{eqnarray}
	\Phi_3 = \left \{ \begin{array} {ll} 
	\Phi_3^{ss}, ~~~ & 0 < \kappa_1^\perp, \kappa_2^\perp < \mu \\
	\Phi_3^{sh}, ~ & 0 < \kappa_1^\perp < \mu, ~ \mu < \kappa_2^\perp < 
		\Lambda \\ 
	\Phi_3^{hs}, ~ & \mu < \kappa_1^\perp < \Lambda, ~ 0 <
		\kappa_2^\perp < \mu \\
	\Phi_3^{hh}, ~ & \mu < \kappa_1^\perp, \kappa_2^\perp <\Lambda .
		\end{array} \right.
\end{eqnarray}
Let us consider the quark distribution function $q(x) = {F_2(x) \over x}$
defined in eq.(\ref{exact}). In presence of the ultraviolet cutoff
$\Lambda$, $q(x)$ depends on $\Lambda$ and schematically we have,
\begin{eqnarray}
	q(x,\Lambda^2) = \sum \int_0^{\Lambda}\mid \Phi_2 \mid^2 + \sum 
		\int_0^\Lambda \int_0^\Lambda \mid \Phi_3 \mid^2.
\label{sfthd2}
\end{eqnarray}
For convenience, we write,
\begin{eqnarray}
	q(x,\Lambda^2) =q_2(x,\Lambda^2) + q_3(x,\Lambda^2).
\end{eqnarray}
where the subscripts $2$ and $3$ denotes the two-particle and three-particle
contributions respectively. Thus, schematically we have from
eq.(\ref{sfthd1}-\ref{sfthd2}),
\begin{eqnarray}
	q(x,\Lambda^2) = && q(x,\mu^2) + \sum \int_\mu^\Lambda \mid 
		\Phi_2^h \mid^2 \nonumber \\
	&&+ \sum \int_0^\mu \int_\mu^\Lambda \mid \Phi_3^{sh} \mid^2 + 
		\sum \int_\mu^\Lambda\int_0^\mu \mid \Phi_3^{hs} 
		\mid^2 \nonumber \\
	&& + \sum \int_\mu^\Lambda \int_\mu^\Lambda \mid \Phi_3^{hh} \mid^2,
\label{qmesh}
\end{eqnarray}
where we have defined $q(x,\mu^2)=\int^\mu_0 \mid\Phi_2^s\mid^2 +
\int^\mu_0 \int^\mu_0 \mid\Phi_2^{ss}\mid^2$. 
Now we investigate the contributions from the amplitudes $\Phi_3^{sh}$ and
$\Phi_3^{hs}$ to order $\alpha_s$ in the following.       

\subsubsection{Dressing with one gluon}
Here we consider the dressing of bare $q\bar q$-state with a single gluon,
i. e., wish to truncate the Fock-state expansion after three-particle state. 
We substitute the Fock expansion eq.(\ref{meson1}) in the bound state
eq.(\ref{lfbef})  and
make projection on a three particle state $b^\dagger (k_1,\sigma_1)
d^\dagger(k_2, \sigma_2) a^\dagger(k_3, \sigma_3) \mid 0 \rangle $ from the
left. In terms of the amplitudes $\Phi_2$, $\Phi_3$, we get,
\begin{eqnarray}
\Phi_3^{\sigma_1 \sigma_2 \lambda_3}(x, \kappa_1; x_2, \kappa_2;
1-x-x_2, \kappa_3) = {\cal M}_1 + {\cal M}_2,
\label{m1pm2}
\end{eqnarray}
where the amplitudes
\begin{eqnarray}
	{\cal M}_1 = && { 1 \over E} (-) { g \over \sqrt{2 (2 \pi)^3}} T^a
		{ 1 \over \sqrt{1 - x - x_2}} ~V_1~ \Phi_2^{\sigma_1' 
		\sigma_2}(1-x_2, -\kappa_2^\perp; x_2,\kappa_2^\perp) 
\end{eqnarray}
and
\begin{eqnarray}
	{\cal M}_2 = && { 1 \over E} { g \over \sqrt{2 (2 \pi)^3}} T^a
		{ 1 \over \sqrt{1 - x - x_2}} ~V_2~ \Phi_2^{\sigma_1 
		\sigma_2'}(x,\kappa_1^\perp;1-x,-\kappa_1^\perp) ,
\end{eqnarray}
with the energy denominator
\begin{eqnarray}
	E= \big[ M^2  - {m^2 + (\kappa_1^\perp)^2 \over x} -
		{m^2 + (\kappa_2^\perp)^2 \over x_2} - {(\kappa_3^\perp)^2 
		\over 1 - x -x_2} \big ],
\end{eqnarray}  
and the vertex functions are given by     
\begin{eqnarray}
	V_1=\chi_{\sigma_1}^\dagger \sum_{\sigma_1'}\big [ { 2 
	\kappa_3^\perp \over 1 - x -x_2} - { (\sigma^\perp. \kappa_1^\perp
	- i m) \over x} \sigma^\perp + \sigma^\perp {(\sigma^\perp. 
	\kappa_2^\perp -im) \over 1-x_2} \big] \chi_{\sigma_1'}. 
		(\epsilon^\perp_{\lambda_1})^*
\end{eqnarray}
and
\begin{eqnarray}
	V_2=\chi_{-\sigma_2}^\dagger \sum_{\sigma_2'}
		\big [ { 2 \kappa_3^\perp \over 1 - x -x_2} - \sigma^\perp
		{ (\sigma^\perp. \kappa_2^\perp
		- i m) \over x_2}  +  {(\sigma^\perp. \kappa_1^\perp -
		im) \over 1-x} \sigma^\perp 
		\big] \chi_{-\sigma_2'}. (\epsilon^\perp_{\lambda_1})^*.
\end{eqnarray}
Physically eq(\ref{m1pm2}) shows the relation between $\Phi_2$ and $\Phi_3$,
i.e., how a three particle state goes to two particle state, 
where ${\cal M}_1$ 
is obtained when the hard gluon attaches to the quark and ${\cal M}_2$ is
obtained when it attaches to the antiquark.
\subsubsection{Perturbative analysis}
Now we consider the various possible contributions to the wave-function 
$\Phi_3$ for different ranges of momenta as
discussed earlier.
For $\kappa_1^\perp$ hard and $\kappa_2^\perp$ soft (i. e., we
concentrate on $ \Phi^{hs}_3$), we can safely assume 
$\kappa_1^\perp+\kappa_2^\perp \approx \kappa_1^\perp$. 
Now, with this consideration,  
the contribution
from ${\cal M}_1$ to $\Phi_3$ in eq.(\ref{m1pm2}) 
can be simplified and reduces to 
\begin{eqnarray}
\Phi_{3,1}^{\sigma_1,\sigma_2,\Lambda_3} (x,\kappa_1^\perp;x_2,\kappa_2^\perp;
1-x-x_2,-\kappa_2^\perp)= && - { g \over \sqrt{2 (2 \pi)^3}} T^a {x
\sqrt{1-x-x_2} \over 1 - x_2} {1 \over (\kappa_1^\perp)^2}
\nonumber \\ 
&& ~~ \chi^\dagger_{\sigma_1}\sum_{\sigma_{1}'}\big [ { 2 \kappa_1^\perp
\over 1-x-x_2}+ {\sigma^\perp. \kappa_1^\perp \over x} \sigma^\perp
\big] \chi_{\sigma_1}'
. (\epsilon^\perp_{\lambda_1})^*
\nonumber \\
&&~~~~~\Phi_2^{\sigma_1',\sigma_2}(1-x_2,-\kappa_2^\perp; x_2, \kappa_2^\perp).
\end{eqnarray}
We see that  the multiple
transverse momentum integral over $\kappa_1$ and $\kappa_2$    
of $\Phi_3$ factorizes into two independent integrals ($\kappa_2$ only
occurs in the argument of $\Phi_2$), while the longitudinal
momentum fraction integral over $x_2$ (relabelled as $y$ in the following 
equation) does not and becomes a convolution.
Thus, the contribution from ${\cal M}_1$ to the structure function is
(see eq.(\ref{qmesh}))
\begin{eqnarray}
\sum \int^\Lambda_\mu \int^\mu_0 
\mid \Phi_{3,1}^{hs} \mid^2 = {\alpha_s \over 2 \pi} C_f
\ln{\Lambda^2 \over \mu^2} \int_x^1 {dy \over y} P_{qq}({x \over y})
q_2(y,\mu^2), 
\label{phs3}
\end{eqnarray}
where
\begin{eqnarray}
P_{qq}\left({x \over y}\right) = { 1 + ({x \over y})^2 \over 1 - {x \over y}}.
\end{eqnarray}

On the other hand, for the same configuration 
($\kappa_1^\perp$ hard, $\kappa_2^\perp$ soft) 
contribution from ${\cal M}_2$ does not factorize and the asymptotic
behaviour of the integrand critically depends on the asymptotic behaviour of
the two-particle wave function  $\Phi_2$. To determine this behaviour,
we have to analyze the bound state equation which shows that for large
transverse momentum $\Phi_2 (\kappa^\perp) \approx { 1 \over
(\kappa^\perp)^2}$. Thus contribution from ${\cal M}_2$ for scale evolution
is suppressed by the bound state wave function. Analysis of the interference
terms between ${\cal M}_1$ and $ {\cal M}_2$ (while calculating 
$\mid \Phi_3\mid^2$, see eq.(\ref{m1pm2})) shows that their
contribution also is suppressed by the bound state wave function. 
Same is true for $\Phi_3^{hh}$. 
 
For the other configuration $\kappa_1^\perp$ soft and $\kappa_2^\perp$ 
(i. e., now concentrating on 
$\Phi_3^{sh}$, which is the only thing left in eq.(\ref{qmesh}) 
to be considered),
contributions from ${\cal M}_1$ and the interference terms are suppressed by
the wave function. Contribution from ${\cal M}_2$ factorizes {\it both} in
transverse and longitudinal space and generate a pure wave function
renormalization contribution:   
\begin{eqnarray}
\sum \int \mid \Phi_{3,2}^{sh} \mid^2 = {\alpha_s \over 2 \pi} C_f 
\ln{\Lambda^2 \over \mu^2} \int_0^1 dy {1 + y^2 \over 1-y}q_2(x,\mu^2).
\label{psh3}
\end{eqnarray} 
Notice that the soft distribution $q_2(x,\mu^2)$ is a function of $x$ and
hence the $y$-integral is not a convolution.

Thus, we see that even though the multi-parton contributions to the structure
function involve both coherent and incoherent phenomena, in the hard region
coherent effects are suppressed by the wave function and we are left with
calculable incoherent contributions.

\subsubsection{Corrections from normalization condition}
In the dressed quark calculation, we have seen that the singularity that
arises as $x
\rightarrow 1 $ from real gluon emission is cancelled by the correction from
the normalization of the state (virtual gluon emission contribution
from wave function
renormalization). In the meson bound state calculation, so far we have
studied the effects of a hard real gluon emission. In this section we study
the corrections arising from the normalization condition of the quark
distribution in the composite bound state.

Collecting all the terms arising from the hard gluon emission contributing
to the quark
distribution function, we have (see eq.(\ref{qmesh}), (\ref{phs3}) and
(\ref{psh3})),
\begin{eqnarray}
q(x,\Lambda^2) &&= q_2(x,\mu^2) + q_3(x,\mu^2) \nonumber \\
&& ~~~~~~ + { \alpha_s \over 2 \pi} C_f
\ln{\Lambda^2 \over \mu^2} \int_x^1 {dy \over y} P_{qq} ({x \over y}) q_2(y,
\mu^2) \nonumber \\
&& ~~~~~~~ + {\alpha_s \over 2 \pi} C_f \ln{\Lambda^2 \over \mu^2}
q_2(x,\mu^2) \int dy P(y).
\end{eqnarray}
We also have a similar expression for the antiquark distribution function in 
$F_2$.

The normalization condition on the quark distribution function should be
such that there is one valence quark in the bound state at any scale $Q$.
We choose the factorization scale $ \mu = Q_0$.
Let us first set the scale $ \Lambda = Q_0$. Then we have (in the truncated
Fock space)
\begin{eqnarray}
\int_0^1 dx ~ q_2(x,Q_0^2) + \int_0^1 dx ~ q_3(x ,Q_0^2) = 1.
\label{norq0}
\end{eqnarray}
Next set the scale $ \Lambda = Q$. We still require
\begin{eqnarray}
\int_0^1 dx ~ q_2(x,Q^2) + \int_0^1 dx ~ q_3(x,Q^2) = 1.
\label{norq}
\end{eqnarray}

Carrying out the integration explicitly, we arrive at 
\begin{eqnarray}
\int_0^1 dx ~ q_2(x,Q_0^2) \big[ 1 + {2 \alpha_s \over 2 \pi} C_f \ln{Q^2 \over
Q_0^2} \int dy P(y)\big] + \int_0^1 dx ~ q_3(x,Q^2) = 1.
\label{724}
\end{eqnarray}

Thus we face the necessity to $``$renormalize" our quark distribution
function $q_2(x,Q_0^2)$ in such a way that the renormalized one now
represents the true quark distribution at the scale $Q_0$ and contains all
the dynamics up to the scale $Q_0$ only and independent of any higher scale
$Q$. Let us define a renormalized quark 
distribution function 
\begin{eqnarray}
q_2^R (x,Q_0^2) = q_2(x, Q_0^2) \big [1 + 2 {\alpha_s \over 2 \pi} C_f \ln{Q^2
\over Q_0^2} \int_0^1 dy P(y) \big].
\label{rn25}
\end{eqnarray}
We note that the evolution of $q_3$ requires an extra hard gluon which is
not available in the truncated Fock space. Thus in the present approximation 
$ q_3(x,Q^2 ) = q_3(x,Q_0^2) = q_3^R(x,Q_0^2)$, 
so that, to order $\alpha_s$ we get from eq.(\ref{724}),
\begin{eqnarray}
\int_0^1 dx ~q_2^R(x,Q_0^2) + \int_0^1 dx ~ q_3(x,Q_0^2) = 1,
\end{eqnarray}
which is the renormalized version of eq.(\ref{norq0}).
We have, to order $\alpha_s$ from eq.(\ref{rn25}),
\begin{eqnarray}
q_2(x,Q_0^2) = q_2^R(x,Q_0^2)  \big [ 1 - 2 {\alpha_s \over 2 \pi} C_f
\ln{Q^2 \over Q_0^2} \int_0^1 dy P(y) \big]. 
\end{eqnarray}
Collecting all the terms, to order $\alpha_s$, we have the normalized quark
distribution function,
\begin{eqnarray}
q(x,Q^2)&& = q_2^R(x,Q_0^2) \nonumber \\
&& ~~~~~~+ {\alpha_s \over 2 \pi} C_f \ln{Q^2 \over Q_0^2} \int_0^1 dy
~ q_2^R(y, Q_0^2) \int_0^1 dz ~ \delta(zy-x)~ {\tilde P}(z) \nonumber \\
&& ~~~~~~ + q_3(x,Q^2) ,
\label{aplo}
\end{eqnarray}
noindent with $ {\tilde P}(z) = P(z) - \delta(z-1) \int_0^1 dy P(y)$.    
It is also easy to see that the normalization condition eq.(\ref{norq}) is
satisfied with the $q(x,Q^2)$ given in eq.(\ref{aplo}), since as shown
earlier 
$\int_0^1 dz ~ {\tilde P}(z) = 0$.

We see that just as in the dressed quark case, the singularity arising as $
x \rightarrow 1$ from real gluon emission is cancelled in the quark
distribution function once the normalization
condition is properly taken into account. From this derivation
we begin to recognize the emergence of the Altarelli-Parisi 
evolution equation. In fact, this is the solution of Altarelli-Parisi
equation to the leading order in $\alpha_s$, with $\alpha_s$ being constant
and unrenormalized. 

Let us summarize what we have done here in a more physical terms. 
In this section we have carried out an analysis of the scale evolution of
structure functions of a meson-like composite system. We have separated the
parton transverse momenta into soft and hard parts. The three body wave
function which is a function of two relative momenta has soft, hard and
mixed components. The mixed components of the three body wave function which
are functions of soft and hard momenta are responsible for the scale
evolution of the soft part of the structure function in the 
perturbation theory. 

In the analysis with wave functions, there are two contributions to the three
body wave function: One where the gluon is absorbed by the quark and
second where the gluon is absorbed by the anti-quark (spectator). There
appears a non-vanishing contribution  when the hard gluon is
absorbed by the antiquark. This corresponds to the transition caused by the
interaction Hamiltonian when the active parton remains soft, while
a hard spectator makes transition to a soft spectator state. This leads to 
wave function renormalization of the spectator anti-quark but this is
eventually cancelled by the normalization condition as discussed here 
in detail. 
This justifies a posteriori the prescription given in Sec.4.1 that
we need to keep only those terms in $P^{-(H)}$ which cause transitions
involving the active parton. 

In the wave function analysis here, we see that 
there are also contributions that are omitted
a priori in the proposed scheme which lead to factorization in Sec.4.1.
All of these contributions are suppressed by the asymptotic behaviour of the
bound state wave function as we have explicitly shown. In summary, the
detailed analysis carried out with the help of multi-parton wave-functions
in Sec.4.3.B justifies the 
approximations made in Sec.4.1 which lead to the
emergence of factorization to all orders in perturbation theory and to the
simple scale evolution picture.   

\section{Physical interpretation of the structure functions from sum rules}
In this last section, we shall explore the physical meaning of the 
deep inelastic structure functions in our framework of light-front 
QCD. The physical meaning of the structure functions can be easily  
understood from the sum rules they obey. 
These sum rules generally arise 
from the existence of conservation laws. Here we consider the 
sum rules for $F_2$ in unpolarized deep inelastic scattering in detail which
is an attempt to show that the structure functions (in general) can be 
connected to the
matrix elements of some physical light front operators and thereby, making
their physical meaning obvious. We first consider the energy-momentum tensor
in QCD, spatial integration of what gives  rise to various physical operators
which are connected to the unpolarized structure functions.
\subsection{Energy-momentum tensor in QCD}

The symmetric, gauge-invariant energy-momentum tensor in QCD is
given by
\begin{eqnarray}
	\theta^{\mu \nu} = && { 1\over 2} \overline{\psi} i \big [ 
		\gamma^\mu D^\nu +\gamma^\nu D^\mu \big ] \psi 
		-F^{\mu \lambda a} F^{\nu}_{~~\lambda a} + {1 \over 4} 
		g^{\mu \nu} (F_{\lambda \sigma a } )^2 \nonumber \\
	&& -g^{\mu \nu} \overline{\psi} \left ( i \gamma^\lambda 
		D_\lambda -m  \right ) \psi.
\end{eqnarray}
The last term vanishes using the equation of motion.
{\it Formally,} we split the energy momentum tensor into a 
$``$fermionic" part $\theta^{\mu \nu}_{q}$ and a $``$gauge bosonic" 
part $ \theta^{\mu \nu}_g$:
\begin{eqnarray}
	\theta^{\mu \nu}_q = { 1 \over 2} \overline{\psi} i \Big [ 
		\gamma^\mu D^\nu +\gamma^\nu D^\mu \Big ] \psi,
\end{eqnarray}
and 
\begin{eqnarray}
	\theta^{\mu \nu}_g =  -F^{\mu \lambda a} F^{\nu}_{~~ \lambda a } + 
		{1 \over 4} g^{\mu \nu}(F_{\lambda \sigma a })^2 ,
\end{eqnarray}
with $ F^{\nu}_{~~\lambda a} = \partial^\nu A_{\lambda a} - \partial_\lambda
A^\nu_a + g f_{abc} A^\nu_b A_{\lambda c} $.
To be consistent with the study of deep inelastic structure function 
which is formulated in $A^+=0$ gauge, we shall work in the same gauge. 

The fermionic and gauge bosonic part of the longitudinal 
momentum densities are given respectively by,
\begin{eqnarray}
	\theta^{++}_{q} = i \overline{\psi} \gamma^+ \partial^+ \psi. 
		\label{denp} 
\end{eqnarray}
\begin{eqnarray}
	\theta^{++}_g = - F^{+ \lambda} F^{+}_{~~\lambda} = \partial^+ A^i 
		\partial^+ A^i.
\end{eqnarray}
Thus the total longitudinal momentum density,
\begin{eqnarray}
	\theta^{++} =  i {\bar \psi} \gamma^+ \partial^+ \psi +
		\partial^+ A^i \partial^+ A^i,
\end{eqnarray}
and is free of interactions at the operator level itself. 

Next consider the transverse momentum density
\begin{eqnarray}
	\theta^{+i}_q = { 1 \over 2} {\bar \psi} i \Big [ \gamma^+ D^i 
		+ \gamma^i D^+ \Big ] \psi = \theta^{+i}_{q-1} + 
		\theta^{+i}_{q-2},
\label{th+iq}
\end{eqnarray}
with
\begin{eqnarray}
	\theta^{+i}_{q-1} = { 1 \over 2} {\bar \psi} i \gamma^+ D^i \psi 
		~~ {\rm and} ~~ \theta^{+i}_{q-2} = { 1 \over 2} 
		{\bar \psi} i \gamma^i \partial^+ \psi. 
\end{eqnarray}
which depend explicitly on the interaction (see later). 
Now, the longitudinal and transverse 
momentum operators are obtained by spatial integration as 
\begin{eqnarray}
	P^{+,i} = { 1 \over 2} \int dx^- d^2 x_\perp \theta^{++,i}.
\label{ppls}
\end{eqnarray}

Similarly we can take out the $(+-)$ component of 
the energy-momentum tensor whose
spatial integration gives the Hamiltonian and can be used to find the
physical meaning of $F_L$ \cite{4rajk}. 
Here we concentrate on $F_2$ and work out the
connection of it with $\theta^{++}$ in the following.

\subsection{Longitudinal momentum sum rule}
The content of the momentum sum rule is known for a long time.
The sum rule simply says that if we add up the longitudinal 
momentum fractions carried by all the 
quarks, antiquarks, and the gluons 
in the nucleon, we should get one.
Here we shall rederive it in our framework, paving the way  for finding
connections between the structure functions, in general, and various
physical light-front operators. 
From the expression of $F_2$ in terms of the plus component
of the bilocal current matrix element given in Eq.(\ref{f2+P}), we have, 
\begin{equation}
F_2= {x \over 4 \pi P^+} \int d \eta 
		e^{-i \eta x} \sum_\alpha
		e^2_\alpha \langle P \mid \big [\overline{\psi}_\alpha
		(\xi^-) \gamma^+ \psi_\alpha (0) - \overline{\psi}_\alpha(0) 
		\gamma^+ \psi_\alpha (\xi^-) \big ] \mid P \rangle 
\end{equation}
Now for the first term in the integral, we use translational invariance, 
\begin{equation}
\overline{\psi}_\alpha
		(\xi^-) \gamma^+ \psi_\alpha (0) = 
\overline{\psi}_\alpha
		(0) \gamma^+ \psi_\alpha (-\xi^-) ,
\nonumber
\end{equation}
and change the variable $\xi^-\rightarrow -\xi^-$, to obtain,
\begin{eqnarray}
F_2 && = {x \over 4 \pi P^+} \int d \eta 
		\big[e^{i \eta x}-e^{-i \eta x}\big] \sum_\alpha
		e^2_\alpha \langle P \mid 
		\overline{\psi}_\alpha(0) 
		\gamma^+ \psi_\alpha (\xi^-) \mid P \rangle 
\nonumber\\
&& = {1 \over 4 \pi P^+} \int d \eta 
		{\partial \over i\partial\eta}
   		\big[e^{i \eta x}+e^{-i \eta x}\big] \sum_\alpha
		e^2_\alpha \langle P \mid 
		\overline{\psi}_\alpha(0) 
		\gamma^+ \psi_\alpha (\xi^-) \mid P \rangle .
\label{f2cros}
\end{eqnarray}
Notice that we have replaced $x$ by ${\partial\over i\partial\eta}$, which
changes the sign of the second exponential. Now, we 
integrate over all possible values of $x$, so that,
\begin{eqnarray}
\int^\infty_{-\infty} dx F_2(x) && = {1 \over 4 \pi P^+} \int d \eta 
		{\partial \over i\partial\eta}\Big(\int^\infty_{-\infty} dx 
   		\big[e^{i \eta x}+e^{-i \eta x}\big]\Big) \sum_\alpha
		e^2_\alpha \langle P \mid 
		\overline{\psi}_\alpha(0) 
		\gamma^+ \psi_\alpha (\xi^-) \mid P \rangle 
\nonumber\\
&& = {1 \over 4 \pi P^+} \int d \eta 
		{\partial \over i\partial\eta}\Big( 2(2\pi)\delta(\eta)\Big)
		\sum_\alpha
		e^2_\alpha \langle P \mid 
		\overline{\psi}_\alpha(0) 
		\gamma^+ \psi_\alpha (\xi^-) \mid P \rangle 
\nonumber\\
&& = {1 \over ( P^+)^2}  
		\sum_\alpha
		e^2_\alpha \langle P \mid 
		\overline{\psi}_\alpha(0) 
		\gamma^+ \partial^+\psi_\alpha (0) \mid P \rangle .
\end{eqnarray}
Last step is obtained first by partial integration and ignoring the surface
term assuming the fields to vanish at $\xi^-\rightarrow\pm\infty$; and then
doing the $\eta$-integration by using $\delta(\eta)$. Note that we have used
${\partial \over \partial\eta}= {1\over P^+}\partial^+$. Now, using the
crossing symmetry
$$F_2(x)=F_2(-x)$$
in LHS and the fact 
that physical structure function is nonvanishing only for $x$
lying between $0$ to $1$, we finally obtain the sum rule
\begin{eqnarray}
	\int_0^1 dx F_2(x) = {1 \over 2 (P^+)^2} \sum_\alpha 
		e^2_\alpha \langle P \mid \theta^{++}_{q\alpha} \mid P 
		\rangle \, . \label{f2pone}
\end{eqnarray}
The above sum rule shows that $F_2$ is connected to the fermionic part of
the light-front
longitudinal momentum density $\theta_q^{++}$.

Similarly, we can define the $``$gluon structure function" \cite{4pQCD} as
mentioned earlier, 
\begin{eqnarray}
	F_{2g}(x) = { 1 \over 4 \pi P^+} \int d\eta e^{- {i \eta x }} 
		\langle P \mid (-)F^{+ \nu a} (\xi^-)F^{+a}_{~~\nu}(0)
		+(\xi^- \leftrightarrow 0) \mid 
		P \rangle \, ,
\end{eqnarray}
where the second term obtained by interchanging the longitudinal coordinates
$(\xi^-\leftrightarrow 0)$, ensures the crossing symmetry $F_{2g}(x)
=F_{2g}(-x)$.
Now a similar exercise as in the case of $F_2$ gives,
\begin{eqnarray}
	\int_0^1 dx F_{2g}(x) = { 1 \over 2 (P^+)^2} 
		\langle P \mid \theta^{++}_g \mid P \rangle \, .
\label{fsumg}
\end{eqnarray}
This shows the connection between $F_{2g}$ and the gluonic part of the
$\theta^{++}$, similar to the quark case.
Now, if we assume $e_\alpha=1$, the physical picture of the $F_2$ structure
function functions become even more clear, since the sum of eq.(\ref{f2pone})
and eq.(\ref{fsumg}) then is nothing but the statement of the longitudinal  
momentum sum rule.
\begin{eqnarray}
	\int_0^1 dx \big [ F_2 + F_{2g} \big ] && =
  	{ 1 \over 2 (P^+)^2} \Big [
		\langle P \mid \theta^{++}_q \mid P \rangle +
		\langle P \mid \theta^{++}_G \mid P \rangle \Big]  
\nonumber\\
	&&= { 1 \over 2 (P^+)^2} 
		\langle P \mid \theta^{++} \mid P \rangle = 1 \, ,
\end{eqnarray} 
since from eq.(\ref{ppls}) it follows 
$\langle P \mid \theta^{++} \mid P \rangle = 2
P^+P^+$, where we have used  the normalization of state 
$ \langle P  \mid P^\prime \rangle = 2(2\pi)^3P^+\delta^3(P-P^\prime)$.

Similarly, starting from Eq.(\ref{f2i}) 
in terms of the transverse component of bilocal current
matrix element and following exactly the same procedure used earlier, we
obtain, 
\begin{eqnarray}
	\int^1_0 dx F_2(x,Q^2) 
	&=& {1\over P^+P^i_\bot} \sum_\alpha e^2_\alpha 
		\langle PS | \overline{\psi}(0) \gamma^i 
	i\partial^+\psi_\alpha (0) 
		|PS \rangle  \nonumber \\
	&=& {1\over P^+P^i_\bot} \sum_\alpha e^2_\alpha 
		\langle PS | \theta_{F\alpha}^{+i} |PS \rangle \, .
\label{f2tone}
\end{eqnarray}
Notice that $\theta_{F\alpha}^{+i}$ does not appear to be the same as 
$\theta_{q\alpha}^{+i}$ as defined in eq.(\ref{th+iq}).
However, explicit demonstration (see Chapter 6) shows that
\begin{equation}
\overline{\psi} \gamma^i iD^+\psi = \overline{\psi} \gamma^i i\partial^+\psi
= \overline{\psi} \gamma^+ iD^i\psi
\end{equation}
up to certain surface terms which do not contribute to the transverse momenta
operator $P^i$ as given in
eq.(\ref{ppls}). Thus, if we are interested only in $P^i$, 
$\theta_{F\alpha}^{+i}$ is indeed equivalent to $\theta_{q\alpha}^{+i}$ and
the sum rule in eq.(\ref{f2tone}) suggests that the integral of $F_2$ over
all $x$ measures the expectation value of the fermionic part of transverse
momentum density  in the
target state. From eq.(\ref{f2pone}) and eq.(\ref{f2tone}) we see that 
\begin{equation}
{1\over P^+}\langle P\mid \theta_q^{++}\mid P \rangle = 
{1\over P^i}\langle P\mid \theta_q^{+i}\mid P \rangle,
\end{equation} 
which shows  the underlying covariance of the theory at the level of 
matrix elements. The
demonstration here shows that (as also is suggested from the explicit
calculation in Ref.\cite{4Hari97}) interaction dependence of the transverse
bilocal current matrix element is spurious and both plus and transverse 
components contain
the same information and have same partonic interpretation.
It clearly shows that drawing conclusions by
looking at the operator structure is quite misleading, as is the case with 
transverse component of bilocal current 
operators that are twist three in the working definition of twist introduced
by Jaffe\cite{4jafst}. 

Since $F_2$ involves quark charges in specific combinations, 
it does not give the direct test of the above momentum sum rule. To
test the sum rule experimentally, one can combine the 
data for both the  electron-proton and electron-neutron 
deep-inelastic scattering and assume that the sea is flavor symmetric, 
then
\begin{eqnarray}
 	\int dx \Big[ F_2^{ep}(x) + F_2^{en}(x) \Big] &=& {5\over 9}
		{1\over (P^+)^2} \sum_\alpha \langle PS 
		| \theta_{F\alpha}^{++} |PS \rangle \nonumber \\
	&=& {5\over 9} {1\over P^+P^i_\bot} \sum_\alpha  
		\langle PS | \theta_{F\alpha}^{+i} |PS \rangle \, .
\end{eqnarray}
This shows that ${9\over 5} \int dx \Big[ F_2^{ep}(x) + F_2^{en}(x) 
\Big]$ is the total longitudinal momentum fraction carried by the all the 
quarks in proton and neutron.
If the quarks carry all the momentum, then we expect that
\begin{equation}
	\int dx \Big[ F_2^{ep}(x) + F_2^{en}(x) \Big] = {5\over 9} \, .
\end{equation}
Experimental data shows that the above integral is $0.28$. In other
words, as is well-known, there are gluons and sea-quarks, and 
half of the momentum in hadrons are carried by them.
 
Similar physical interpretation for the structure functions $F_L$ and $g_1$
are already worked out by showing their connections with the matrix elements
of appropriate
physical light-front operators (for details we refer to the original works
\cite{4rajk},\cite{4zhang96}).
Thus, in the light-front Hamiltonian formulation, physical
interpretation of the DIS structure functions are most conveniently obtained
as they can be shown to be connected with some physical light-front
operators via the sum rule.


\newpage
\setcounter{section}{0}
\renewcommand{\thesection}{5.\arabic{section}}
\setcounter{equation}{0}
\renewcommand{\theequation}{5.\arabic{equation}}

{\flushleft\huge\bf {Chapter 5}}
\vskip 1cm

After obtaining satisfactory results in the leading order calculation in
bare coupling as presented in the previous Chapter, it is natural to extend
the dressed quark structure function to the next higher order, namely, 
to the fourth order $ O(g^4)$. 
However, a complete fourth-order calculation is always very involved
one\cite{5doks}\cite{5lsmith}. 
In light-front, it is particularly so for the
renormalization becoming more complicated due to lack of covariance. 
Instead, we recall 
that in the dressed parton structure function calculations to the
lowest non-trivial order, we truncated the Fock-expansion of the
dressed parton state at the two-particle level and obtained 
$$
{F^{|q\rangle}_{2q}(x)\over x}= 
\delta(1-x) + {\alpha_s\over 2\pi}C_f\ln {Q^2\over\mu^2}
{\tilde P}_{qq}(x).
$$
Also recall that the coupling constant $\alpha_s$ is a bare parameter in our
theory. 
For the purpose of
fourth-order extension, we need to {\it include} three-particle sector. For
example, in the dressed quark state, one  includes 
states where a bare quark
is associated with two gluons or a pair of quark and anti-quark. To this
order, we expect to obtain (from our experience in LLA)
$$
{F^{|q\rangle}_{2q}(x)\over x}= 
\delta(1-x) + {\alpha_s(Q^2)\over 2\pi}C_f\ln {Q^2\over\mu^2}
{\tilde P}_{qq}(x) 
+ {1\over 2} \left({\alpha_s\over 2\pi}\right)^2
\left(\ln {Q^2\over \mu^2}\right)^2
\int_x^1 {dy\over y} C_f{\tilde P}_{qq}(y) C_f{\tilde P}_{qq}({x\over y}).
$$
Notice that, in the leading order term, we  
have the renormalized running coupling 
in QCD (which, strictly speaking, contains fourth order term in bare
coupling), while the last term corresponds to the well known ladder
contribution involving two rungs where the coupling is still a bare
parameter and needs further higher order calculation for its running.   
Thus, the coupling constant 
renormalization being a part this calculation, one
should have a complete understanding of it in light-front theory as a first
step towards 
working out the fourth order extension of the structure function and we
shall do that in this Chapter.   

Now, the coupling constant renormalization as required for the fourth order
extension of structure function calculation is also very important from a
different viewpoint as we discuss now and we shall put major emphasis on
that in our discussion. 
In light-front field theory, at present, many higher order 
calculations need to be
performed using time ($x^+$) ordered perturbative techniques in 
order to overcome several conceptual and practical problems
\cite{5wi94,5bpp}. 
Investigations have revealed a very complex structure 
for the coefficient functions (accompanying the divergences)
which emerge at the end of notoriously long and tedious calculations
of individual  
time ($x^+$) ordered diagrams which are  higher order in the coupling.
It appears that
almost no guidance is available to look for possible mistakes in these
structures.
In contrast, in covariant 
field theory, the
structures accompanying the divergences are quite simple.
The complexity of the former is due to the fact that power
counting is different on the light-front \cite{5wi94}. In the latter case, 
simplicity of the structure is due to the underlying Lorentz
symmetry (rotational and boost invariance) which can be maintained at every stage
of the calculation.
Since the light-front formalism do possess some kinematical 
symmetries, it is
worthwhile to investigate whether they
can provide some
constraint on the possible structure of coefficient functions 
for individual $x^+$ ordered diagrams. 

Two of the most important kinematic symmetries in light-front field theory  
which
are relevant especially for phenomenological concerns are the longitudinal and
transverse boost symmetries. As we discussed in Chapter 2, 
the longitudinal boost symmetry is a scale
symmetry on the light-front whereas transverse boost symmetry is simply
Gallilean symmetry in two dimensions in non-relativistic 
dynamics \cite{5gref}. 
The implications of the Gallilean symmetry for the structure of the
interaction vertices resulting from the light-front 
Hamiltonian at tree level are known. 
For example,
the symmetry can be utilized \cite{5ln} to reduce 
the number of free parameters in an interaction Hamiltonian constructed 
at tree level purely from light-front power counting. 
The implications of this symmetry beyond tree level is not well-understood.

Incidentally, we
mention that previous calculation \cite{5perry} 
of vertex corrections have employed the four-component representation 
of Brodsky and
Lepage \cite{5bl} and only the final answers after summing different time
orderings have been presented which is of no use to us. 
A calculation \cite{5zh93c} (more suitable to our purpose) of the vertex
correction using the
two-component representation\cite{5zh93b,5bks} have studied only the two 
{\it specific cases} of 
helicity-flip part of the vertex (proportional to quark mass) and zero momentum
($q^{+,\perp}=0$) limit for the gluon. 
But for the structure function calculation, it is necessary that the
particles connected to the vertex have generic momenta, which makes the
calculation complicated exactly in a way as mentioned earlier. 
Therefore, the calculation of vertex correction with {\it general
kinematics} 
gives us the opportunity to investigate 
whether and how the Gallilean
symmetry manifests itself beyond tree level 
and whether the symmetry can
provide some guidance in understanding the complex structure of coefficient
functions accompanying the divergences. As we shall see here, 
explicit calculation reveals that the accompanying structures are
either proportional to the canonical vertex or independent of the total
transverse momentum 
and thereby exhibiting transverse boost invariance. 
Since, these 
processes are relevant for the calculation of
asymptotic freedom in light-front QCD we also present the 
$\beta$-function calculation for the most {\it general kinematics} in the
two-component formalism. 

We reemphasize the fact that our motivation here is to study 
the role played by Gallilean boost symmetry in
ensuring the correctness of the structure of coefficient functions appearing
in the calculation beyond tree level for each $x^+$ ordered diagram
separately. Such nontrivial checks are going to be extremely helpful 
in any higher order calculation using old-fashioned time-ordered
perturbation theory used in LFQCD, as is the case in extending 
the calculations of structure functions presented earlier    
to the fourth order in coupling. 

\section{Transverse Boost Symmetry: Canonical Considerations}
In light-front theory, the generators of transverse boost are given by
\begin{eqnarray}
E^i ~= ~M^{+i} ~= ~{1 \over 2} ~\int dx^- d^2 x^\perp ~\Big [ ~x^i ~
\theta^{++} ~ - x^+ ~ \theta^{+i} ~ \Big ]
\end{eqnarray}
where $\theta^{\mu \nu}$ is the symmetric energy-momentum tensor.
As discussed in Chapter 2, the generators $E^i$ leave $x^+=0$ invariant 
and hence are kinematic
operators. We recall that the motion in the transverse plane generated  by
$E^i$ and $J^3$ are Gallilean in nature. This is ensured by the fact that the 
commutation relations satisfied by them, namely, 
\begin{eqnarray}
\Big [~ E^i,E^j ~\Big ] ~=~ 0~,~~~
\Big [ J^3,E^i \Big ] ~=~ i \epsilon^{ij} E^j~, 
\end{eqnarray}
where $\epsilon^{ij}$ the two dimensional antisymmetric tensor,
ressemble the corresponding Gallilean generators in
the non-relativistic dynamics.
Also the commutation relation
\begin{equation}
\Big [ ~ E^i, P^j ~ \Big ] ~ = ~ -i \delta^{ij} P^+~,
\end{equation}
shows that the light-front longitudinal momentum plays the role of 
non-relativistic mass, which occurs instead of $P^+$ 
in the corresponding commutation
relation in the non-relativistic case. 
Thus the generators $E^i$s act just like Gallilean boosts in 
the transverse plane, familiar from non-relativistic dynamics. 

In light-front theory involving fermions and gauge bosons, the interaction
vertices have a nontrivial structure (see ${\cal V}_1$ below).
Gallilean symmetry implies that the interaction vertices in the
theory (in momentum space) are independent of the total transverse momentum
in the problem. Since the issues associated with Gallilean invariance are
most transparent in the
two-component representation (which we have always been using in this 
dissertation), it is most convenient to 
use this representation of light-front QCD in
contrast to the more familiar four-component representation \cite{5bl}.

The canonical quark-gluon vertex in our notation is (see Fig. 1)   
\vskip .5cm
\begin{center}
\begin{picture}(300,100)(0,0)
\Text(160,-8)[]{{\footnotesize 
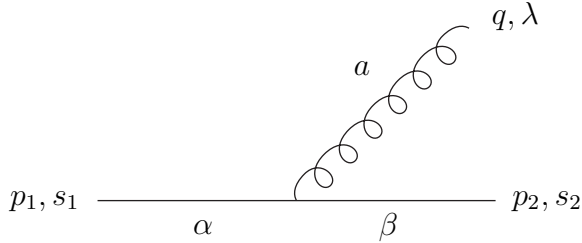
Fig.1 The canonical quark-gluon vertex in light-front QCD.}}
\SetOffset(100,0)
\Text(-40,35)[]{$p_1, s_1$}
\Text(150,35)[]{$p_2, s_2$}
\Text(138,105)[]{$q, \lambda$}
\Text(20,25)[]{$\alpha$}
\Text(90,25)[]{$\beta$}
\Text(80,85)[]{$a$}
\Gluon(55,35)(120,100){5}{6}
\Line(-20,35)(130,35) 
\end{picture}
\end{center}
\vskip .2cm
\begin{eqnarray}
{\cal V}_1 ~= &&~g ~T^a ~ \sqrt{p_1^+~p_2^+}~ \chi^\dagger_{s_{2}}~
 \Big [ - 2 {q^\perp \over q^+}
+ \sigma^\perp {\sigma^\perp . p_1^\perp \over p_1^+} + {\sigma^\perp
.p_2^\perp \over p_2^+} \sigma^\perp  +im \Big( {1 \over p_1^+} - {1 \over
p_2^+} \Big) \Big ] ~\chi_{s_{1}} ~ .~
(\epsilon^\perp_\lambda)^*. \nonumber  \\ \label{canon}
&&
\end{eqnarray}
Since the mass term (helicity-flip interaction) is irrelevant for the
Gallilean invariance, we drop it in the following.
Note that the canonical vertices and energy denominators in
$x^+$ ordered diagrams in the two-component representation 
are manifestly invariant under Gallilean boost. 
(See  Appendix D for an explicit example at the one loop level).

\section{One loop calculations}
 
In the {\it massless} limit, the helicity-flip contribution 
vanishes and the canonical vertex has the
structure
\begin{eqnarray}
{\cal V}_1 ~= ~g ~T^a ~ \sqrt{p_1^+~p_2^+}~ \chi^\dagger_{s_{2}}~
 \Big [ - 2 {q^\perp \over q^+}
+ \sigma^\perp {\sigma^\perp . p_1^\perp \over p_1^+} + {\sigma^\perp
.p_2^\perp \over p_2^+} \sigma^\perp   \Big ] ~\chi_{s_{1}} ~ .~
(\epsilon^\perp_\lambda)^*. 
\end{eqnarray}
In this section we consider corrections to this vertex at one loop level
in LF Hamiltonian perturbation theory.
Specifically we consider the corrections arising from quark-gluon
vertex and the three-gluon vertex. Note that the corrections arising from
instantaneous vertices in the theory do not contribute to the divergent
structure of the vertex for zero quark mass at one loop level and hence will
not be considered here at all.

In order to perform the calculations beyond tree level, we need to
regulate the loop momenta. 
How to introduce regulators in light-front theory is, at present, an active 
subject of research \cite{5ann}. One may (1) introduce cutoffs on the 
sum of light-front energies (the so-called boost invariant cut-off), or 
(2) choose to cutoff energy
differences at vertices (which emerge naturally in similarity renormalization
perturbation theory) or (3) simply cutoff single particle momenta. 
We employ the third choice for the regulators, namely, we put cutoffs, such
that $ k_i^+ > \epsilon$,
$ \mu < k_i^\perp < \Lambda$, which is 
simple to implement but obviously violate both longitudinal
and transverse boost invariance.  
Since the vertices and energy denominators are explicitly invariant under
the Gallilean boost (see Appendix D), the violation of the symmetry can
occur only through the explicit appearance of total transverse momentum 
$P^\perp$ in the limits
of integration. From power counting, the vertex corrections at one loop
level are only logarithmically divergent in the transverse plane. 
Hence we expect the coefficient functions accompanying the logarithms to
still exhibit the symmetry.

Consider the one loop corrections to the vertex involving two
quark-gluon vertices. There are two time-ordering contributions shown in
Figs. 2(a) and 2(b). 
\vskip .6cm
\begin{center}
\begin{picture}(300,100)(0,0)
\Text(150,-8)[]{{\footnotesize 
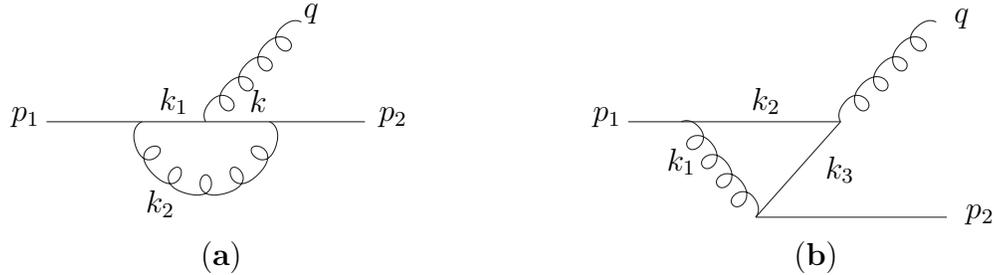
Fig.2 Contribution to the quark-gluon vertex from contributions involving
two quark-gluon vertex.}}
\SetScale{.8}
\SetOffset(0,35)
\Text(-24,30)[]{$p_1$}
\Text(115,30)[]{$p_2$}
\Text(84,69)[]{$q$}
\Text(32,36)[]{$k_1$}
\Text(27,-3)[]{$k_2$}
\Text(64,36)[]{$k$}
\Text(50,-23)[]{({\bf a})}
\GlueArc(55,35)(30,180,360){5}{5}
\Gluon(55,35)(100,82){5}{4}
\Line(-20,35)(130,35) 
\SetOffset(220,35)
\Text(-24,30)[]{$p_1$}
\Text(117,-7)[]{$p_2$}
\Text(110,67)[]{$q$}
\Text(4,13)[]{$k_1$}
\Text(36,36)[]{$k_2$}
\Text(64,10)[]{$k_3$}
\Text(55,-23)[]{({\bf b})}
\Gluon(80,35)(125,82){5}{4}
\Line(-20,35)(80,35) 
\Line(80,35)(40,-10)
\Line(40,-10)(130,-10)
\Gluon(5,35)(40,-10){5}{4}
\end{picture}
\end{center}
\vskip .3 cm
The contribution from Fig. 2(a) is 
\begin{eqnarray}
{\cal V}_{2a} ~=&&~{g^3 \over 2 (2 \pi)^3}~ T^b T^a T^b ~ \sqrt{p_1^+~p_2^+} 
\int_\epsilon^{p_2^+-\epsilon} dk^+ \int d^2 k^\perp 
~ \theta(\Lambda - \mid k^\perp \mid) \nonumber \\
&& ~~~~~~{ 1 \over k_3^+}~ 
{ 1 \over p_1^- - k_1^- - k_3^-} ~{ 1 \over p_1^- - q^- - k^- - k_3^-}
\nonumber \\
&& ~~~~~~ \sum_{\sigma_1 \sigma_2 \lambda_1}
~\chi^\dagger_{s_{2}}~ 
\Big [ -2  {k_3^\perp \over k_3^+} + \sigma^\perp {\sigma^\perp .k^\perp \over
k^+} + {\sigma^\perp . p_2^\perp \over p_2^+} \sigma^\perp \Big ]~ 
\chi_{\sigma_1} ~. ~\epsilon^\perp_{\lambda_{1}} ~ \nonumber \\
&& ~~~~~~~~~\chi^\dagger_{\sigma_1} ~
\Big [ -2  {q^\perp \over q^+} + \sigma^\perp {\sigma^\perp .k_1^\perp \over
k_1^+} + {\sigma^\perp . k^\perp \over k^+} \sigma^\perp \Big ] ~
~\chi_{\sigma_{2}} ~. ~(\epsilon^\perp_{\lambda})^* 
\nonumber \\
&&~~~~~~~~~\chi^\dagger_{\sigma_{2}} 
~\Big [ -2  {k_3^\perp \over k_3^+} + \sigma^\perp {\sigma^\perp .p_1^\perp 
\over p_1^+} + {\sigma^\perp . k_1^\perp \over k_1^+} \sigma^\perp \Big ]~ 
\chi_{s_1} ~. ~(\epsilon^\perp_{\lambda_{1}})^*. \label{eq2a}
\end{eqnarray}
Here $ k_1^{(+,\perp)} = q^{(+,\perp)} + k^{(+,\perp)}$ and $ k_3^{(+,\perp)}
 =
p_2^{(+,\perp)} - k^{(+,\perp)}$. 

The calculation is cumbersome but straightforward. Here we mention the
important points that one should keep in mind while actually performing the
calculations. First of all we notice that, there are two sources of
divergences: one coming from the lower limit of integration over
longitudinal momenta $k^+$ and the other coming from the high value of
transverse momenta $k^\perp$. The $k^+$-integration for various terms 
yields either finite or logarithmically 
divergent contribution. 
The leading divergence coming from the transverse momenta integration 
is 
only logarithmic. But to obtain the correct structure of the co-efficient 
functions accompanying the log-divergence, one has to be careful. 
We briefly discuss the general procedure that can be pursued in evaluating
the expression in eq.(\ref{eq2a}) as well as for the other diagrams in 
Fig.2
and Fig.3.
Notice that all the
vertex functions involved in eq.(\ref{eq2a}) depend on the integration variable $k$ and 
are linear functions of $k^\perp$. 
Similarly, both the energy denominators are
quadratic functions of $k^\perp$. Thus, the general structure of the
transverse momenta integration becomes 
\begin{equation}
\int dk^+ d^2k^\perp { (A_1^\perp\cdot k^\perp)(k^\perp)^2
+B_1(k^\perp)^2
+(A_2^\perp\cdot k^\perp) +B_2\over [(k^\perp)^2 +(a_1^\perp\cdot k^\perp) 
+b_1]
[(k^\perp)^2 +(a_2^\perp\cdot k^\perp) +b_2]}\, ,
\end{equation}
where the coefficients $A_i^\perp$s and  $B_i$s as well as  
$a_i^\perp$s and  $b_i$s  
are independent of $\kappa^\perp$ and 
completely determined by the external particles'
quantum numbers (momenta, helicity etc.) {\it and}  
the longitudinal momenta $k^+$. 
To determine large $k^\perp$
behavior we expand the energy denominators as
\begin{equation}
{1\over [(k^\perp)^2 
+(a_1^\perp\cdot k^\perp) +b_1]} \sim {1\over (k^\perp)^2}\left[
1- {a_1^\perp\cdot k^\perp\over (k^\perp)^2} -{b_1\over
(k^\perp)^2}\right]\, . 
\end{equation}
Putting this back into the general expression and dropping the terms which
vanish either due to large values of $k^\perp$ or being an odd function of
$k^\perp$, we get
\begin{eqnarray}
&&\int dk^+d^2k^\perp {1\over (k^\perp)^4}\left[  
B_1~(k^\perp)^2 -(A_1^\perp\cdot k^\perp)(k^\perp)^2\left\{
{a_1^\perp\cdot k^\perp\over (k^\perp)^2} 
+{a_2^\perp\cdot k^\perp\over (k^\perp)^2}\right\}
\right]\, 
\nonumber\\
&&~~~= \int dk^+\big( B_1 -C_1-C_2\big)\int_\mu^\Lambda {d^2k^\perp\over
(k^\perp)^2}\, ,\nonumber\\
&&~~~= \int dk^+ \big( B_1 -C_1-C_2\big)~ (2\pi)\ln {\Lambda\over\mu}\, ,
\end{eqnarray}
where we have used 
\begin{equation}
(A_1^\perp\cdot k^\perp) (a_i^\perp\cdot k^\perp) = C_i~ (k^\perp)^2\, .
\end{equation}
Now $k^+$ dependence of $\big( B_1 -C_1-C_2\big)$ is such that the
integration over $k^+$ yields
either finite or logarithmically divergent contribution, i.e.,
\begin{equation}
\int_\epsilon dk^+ \big( B_1 -C_1-C_2\big)= A+B\ln{p_1^+\over \epsilon} 
+C\ln{p_2^+\over \epsilon} +D\ln{q^+\over \epsilon}\, ,
\end{equation}
where $p_1^+$, $p_2^+$ and $q^+$ are the longitudinal momenta of the
external particles (see the Fig.2(a)) and $\epsilon$ is a small cut-off used
for the lower limit of $k^+$-integration. Here 
$A$, $B$ etc. are known once $B_1$, $C_1$ and $C_2$ are determined
through explicit calculation. For example, we can easily read out $C$ and
$A$ from eq.(\ref{eq2a1}) and eq.(\ref{eq2a2}) respectively where we
presented the contributions from the diagram in Fig.2(a); 
$B$ and $D$ are zero for this
diagram.
Thus, in general, we expect two types of
divergence structures: one involving single logarithm such as 
$\sim\ln{\Lambda\over\mu}$
and the other involving a product of logarithms  such as 
$\sim\ln{\Lambda\over\mu}
\ln{p^+\over\epsilon}$. 

Following the above procedure, after 
a long and tedious calculation, we arrive at two types of divergent
contributions, as mentioned, {\it viz},  
one containing  product of logarithms and other containing a
single logarithm. We have taken $\mu$ and $\Lambda$ to be much larger 
than the external momentum 
scales in the problem.
Divergent contributions that contain products of logarithms:
\begin{eqnarray}
{\cal V}_{2a}^{I}~=&&~g ~T^a ~ 
\sqrt{p_1^+~p_2^+} ~\chi^\dagger_{s_{2}}~  
\Big [ - 2 {q^\perp \over q^+} +  \sigma^\perp { \sigma^\perp .p_1^\perp \over
 p_1^+} +  { \sigma^\perp . p_2^\perp \over p_2^+} \sigma^\perp \Big
]~\chi_{s_{1}}~. ~(\epsilon^\perp_\lambda)^*~ \nonumber \\
&&~~~~~~~~ {g^2 \over 8 \pi^2} ~ \Big( -{1 \over 2}~ C_A ~+ ~C_f \Big)~
\ln{\Lambda \over \mu} ~  4 ~\ln{p_2^+ \over \epsilon}.\label{eq2a1}
\end{eqnarray}
Since the coefficient of the divergent factor is proportional to the
canonical vertex, the transverse boost invariance of the above result is manifest. 
Divergent contributions that contain single logarithm are 
\begin{eqnarray}
{\cal V}_{2a}^{II}~ = && ~g ~ T^a~
\sqrt{p_1^+~p_2^+} ~\chi^\dagger_{s_{2}}
 ~\Big [ 
6 {q^\perp \over q^+} - 6 {p_1^\perp \over p_1^+}
 - {\sigma^\perp . p_2^\perp \over p_1^+}\sigma^\perp
+ {p_2^+ \over p_1^+} {\sigma^\perp . p_1^\perp \over p_1^+} \sigma^\perp 
 \Big]~\chi_{s_{1}}~. ~(\epsilon^\perp_\lambda)^*~ \nonumber \\ 
&& ~~~~~~~~~{ g^2 \over 8 \pi^2}~ \Big ( -{1 \over 2}~ C_A~ + ~C_f \Big )~
 \ln{\Lambda \over \mu} \label{eq2a2}.
\end{eqnarray}
In this case the coefficient of the divergent factor involving transverse
momenta is not proportional to the canonical vertex. However, in terms of
the internal momenta (see Appendix D), the quantity inside the square
bracket can be rewritten as
\begin{eqnarray}  
\Big [ 
6 {q^\perp \over q^+} - 6 {p_1^\perp \over p_1^+}
 - {\sigma^\perp . p_2^\perp \over p_1^+}\sigma^\perp
+ {p_2^+ \over p_1^+} {\sigma^\perp . p_1^\perp \over p_1^+} \sigma^\perp 
 \Big] \nonumber \\
=-{1 \over P^+} \Big [{ 6 \kappa_1^\perp \over 1-x} + \sigma^\perp
.\kappa_1^\perp \sigma^\perp \Big ],
\end{eqnarray}
which satisfies the constraint from  Gallilean invariance, namely,
independent of the total transverse momenta $P^\perp$.

Contribution from Fig. 2(b) is    
\begin{eqnarray}
{\cal V}_{2b} ~= &&~
(-) ~{g^3 \over 2 (2 \pi)^3} ~ T^b T^a T^b ~ \sqrt{p_1^+~p_2^+}
~\int_\epsilon^{q^+-\epsilon} dk_2^+ ~
\int d^2 k_2^\perp ~ \theta(\Lambda - \mid k^\perp \mid)~ { 1 \over p_1^+ - k_2^+}
~{ 1 \over p_1^- - k_1^- - k_2^-} ~\nonumber \\
&& ~~~~~{ 1 \over p_1^- - k_2^- - k_3^- - p_2^-}~ 
\sum_{\sigma_1 \sigma_2 \lambda_1}~ 
\chi^\dagger_{s_{2}} 
~\Big [ -2  {k_1^\perp \over k_1^+} + \sigma^\perp {\sigma^\perp .k_3^\perp \over
k_3^+} + {\sigma^\perp . p_2^\perp \over p_2^+} \sigma^\perp \Big ] 
~\chi_{\sigma_{1}} ~.~ \epsilon^\perp_{\lambda_{1}} ~ \nonumber \\
&& ~~~~~~~~~\chi^\dagger_{\sigma_1} 
~\Big [ -2  {q^\perp \over q^+} + \sigma^\perp {\sigma^\perp .k_2^\perp \over
k_2^+} + {\sigma^\perp . k_3^\perp \over k_3^+} \sigma^\perp \Big ] 
~\chi_{\sigma_{2}} ~. ~(\epsilon^\perp_{\lambda})^*  \nonumber \\
&& ~~~~~~~~~\chi^\dagger_{\sigma_{2}} ~
\Big [ -2  {k_1^\perp \over k_1^+} + \sigma^\perp {\sigma^\perp .p_1^\perp 
\over p_1^+} + {\sigma^\perp . k_2^\perp \over k_2^+} \sigma^\perp \Big ] 
~\chi_{s_1} ~. ~(\epsilon^\perp_{\lambda_{1}})^*. \label{eq2b}
\end{eqnarray}
Here $ k_1^{(+,\perp)} = p_1^{(+,\perp)} - k_2^{(+,\perp)}$ and 
$ k_3^{(+,\perp)} =
q^{(+,\perp)} - k_2^{(+,\perp)}$. The overall 
negative sign arises from the anti
symmetry property of fermionic states. Note that this negative sign is missing
from Eq. (A8) of Ref.\cite{5zh93c}.

As in the previous case, explicit evaluation leads to terms containing
two types of divergences.
Divergent contributions that contain products of logarithms are
\begin{eqnarray}
{\cal V}_{2b}^{I}~ = &&~ g ~ T^a 
~ \sqrt{p_1^+} ~\sqrt{p_2^+} ~\chi^\dagger_{s_{2}}~ 
 \Big [ - 2 {q^\perp \over q^+} +  \sigma^\perp { \sigma^\perp .p_1^\perp 
\over
 p_1^+} +  { \sigma^\perp . p_2^\perp \over p_2^+} \sigma^\perp \Big ]~
\chi_{s_{1}}~. ~(\epsilon^\perp_\lambda)^*~ \nonumber \\
&& ~~~~~~~ {g^2 \over 8\pi^2}~ \Big ( -{1 \over 2}~C_A ~+ ~C_f \Big )~
 \ln{\Lambda \over \mu} ~  2 ~ \ln{p_1^+ \over p_2^+} . \label{eq2b1}
\end{eqnarray}
Again, the transverse boost invariance of this result is manifest since the
contribution is proportional to the canonical vertex.
Divergent contributions that contain single logarithm  are 
\begin{eqnarray}
{\cal V}_{2b}^{II}~ = &&~(-) ~g ~T^a
~ \sqrt{p_1^+~p_2^+} ~\chi^\dagger_{s_{2}}~  
\Big [ 
 3 \sigma^\perp {
\sigma^\perp.p_1^\perp \over p_1^+} 
+ 3 {\sigma^\perp .p_2^\perp \over p_2^+} \sigma^\perp \nonumber \\
&&- 6 {p_1^\perp \over p_1^+} - {\sigma^\perp . p_2^\perp \over p_1^+}
\sigma^\perp
+ {p_2^+ \over p_1^+} {\sigma^\perp . p_1^\perp \over p_1^+} \sigma^\perp
 \Big]~ 
\chi_{s_{1}}~.~ (\epsilon^\perp_\lambda)^* \nonumber \\
&& { g^2 \over 8 \pi^2} ~ 
\Big ( -{1 \over 2}~C_A ~+~ C_f \Big )~\ln{\Lambda \over \mu}. \label{eq2b2}
\end{eqnarray}
The transverse boost symmetry of the terms inside the square bracket is not
manifest but becomes explicit once we express the result in terms of the
internal momenta. Alternatively, 
by subtracting and adding the term  $ - 6 {q^\perp \over q^+}$ to these
terms we can rewrite the terms inside the square bracket as the canonical
term plus the terms contained in the square bracket in eq.(\ref{eq2a2})
which again shows the boost invariance of the result in eq.(\ref{eq2b2}).  
\vskip .2cm
\begin{center}
\begin{picture}(300,100)(0,0)
\Text(150,-8)[]{{\footnotesize 
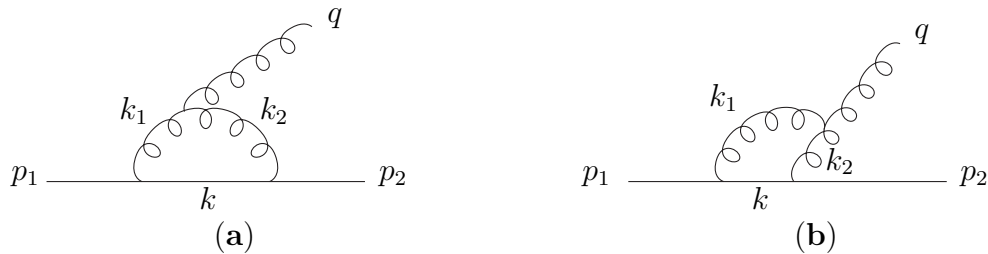
Fig.3 Contribution to the quark-gluon vertex from contributions involving
one quark-gluon and}}
\Text(150,-18)[]{{\footnotesize  one three-gluon vertex.}}
\SetScale{.8}
\Text(-24,30)[]{$p_1$}
\Text(115,30)[]{$p_2$}
\Text(93,90)[]{$q$}
\Text(17,55)[]{$k_1$}
\Text(70,55)[]{$k_2$}
\Text(45,22)[]{$k$}
\Text(55,7)[]{({\bf a})}
\GlueArc(55,35)(30,0,180){5}{5}
\Gluon(45,68)(105,108){5}{4}
\Line(-20,35)(130,35) 
\SetOffset(220,0)
\Text(-28,30)[]{$p_1$}
\Text(115,30)[]{$p_2$}
\Text(95,84)[]{$q$}
\Text(20,60)[]{$k_1$}
\Text(64,36)[]{$k_2$}
\Text(34,22)[]{$k$}
\Text(55,7)[]{({\bf b})}
\GlueArc(55,35)(30,55,180){5}{4}
\Gluon(58,35)(108,100){5}{5}
\Line(-20,35)(130,35) 
\end{picture}
\end{center}

\vskip .6cm
Consider, next, the one loop contributions to the quark-gluon vertex
involving one quark-gluon vertex and one three gluon vertex. There are two
time ordering contributions shown in Figs. 3(a) and 3(b).
The contribution from Fig. 3(a) is
\begin{eqnarray}
{\cal V}_{3a}~=&&~ {g^3 \over 2 (2 \pi)^3}~ (-i f^{abc} T^b T^c)
 ~ \sqrt{p_1^+~p_2^+}
~ \int_\epsilon^{p_2^+-\epsilon} dk^+
~ \int d^2 k^\perp ~ \theta(\Lambda - \mid k^\perp \mid)
~{ 1 \over k_1^+} ~{1 \over k_2^+} ~{ 1 \over p_1^- - k_1^- - k^-}
\nonumber \\
&& ~~~~~{ 1 \over
p_1^- - q^- - k_2^- - k^-}~
~ \sum_{\sigma_1, \lambda_{1}, \lambda_{2}} 
~\chi^\dagger_{s_{2}} ~\Big [ -2 { k_2^\perp \over k_2^+} + \sigma^\perp
{\sigma^\perp . k^\perp \over k^+} + {\sigma^\perp. p_2^\perp \over p_2^+}
\sigma^\perp \Big ] ~\chi_{\sigma_1} ~.~ \epsilon^\perp_{\lambda_{2}}~ 
\nonumber \\
&&~~~~~~~~ \chi^\dagger_{\sigma_1}~ \Big [ -2 { k_1^\perp \over k_1^+} + \sigma^\perp
{\sigma^\perp . p_1^\perp \over p_1^+} + {\sigma^\perp. k^\perp \over k^+}
\sigma^\perp \Big ]~ \chi_{s_{1}} ~. ~(\epsilon^\perp_{\lambda_{1}})^* 
\nonumber \\
&&~~~~~~\epsilon^j_{\lambda_{1}} ~(\epsilon^i_{\lambda})^*~
(\epsilon^l_{\lambda_{2}})^* ~\Bigg [ \Big [ (k_1^i + k_2^i) - {q^i \over
q^+} (k_1^+ + k_2^+) \Big ] \delta_{lj}
- \Big [ (k_1^l + q^l) - {k_2^l \over k_2^+} (k_1^+ + q^+) \Big ] \delta_{ij}
\nonumber \\
&&~~~~~~~~+  \Big [ (q^j - k_2^j) - {k_1^j \over k_1^+} (q^+ - k_2^+) \Big ]
\delta_{il} \Bigg ] .  \label{eq3a}
\end{eqnarray}
Here $ k_1^{(+,\perp)} = p_1^{(+,\perp)} - k^{(+,\perp)}$ and 
$ k_2^{(+,\perp)} =
p_2^{(+,\perp)} - k^{(+,\perp)}$.
Divergent contributions that contain products of logarithms are
\begin{eqnarray}
{\cal V}_{3a}^{I}~=&& ~g ~T^a~  ~\sqrt{p_1^+~p_2^+} ~\chi^\dagger_{s_{2}}~ 
\Big [ 
- 2 {q^\perp \over q^+} + \sigma^\perp {\sigma^\perp.p_1^\perp \over p_1^+}
+ {\sigma^\perp. p_2^\perp \over p_2^+} \sigma^\perp 
 \Big ] ~
\chi_{s_{1}}~. ~(\epsilon^\perp_\lambda)^* 
\nonumber \\
&& ~~~~~~~{g^2 \over 8 \pi^2}  ~ {1 \over 2}~ C_A 
~ \ln{\Lambda \over \mu} 
~2 ~\ln{p_1^+ p_2^+ \over q^+
\epsilon}.\label{eq3a1}
\end{eqnarray}
The boost invariance of this result is again clear.
Divergent contributions that contain single logarithm are
\begin{eqnarray}
{\cal V}_{3a}^{II}~= &&~g ~T^a ~  \sqrt{p_1^+~p_2^+} ~\chi^\dagger_{s_{2}}~ 
\Big [ 6 {q^\perp \over q^+} - 6 {p_1^\perp \over p_1^+} + {\sigma^\perp.
p_2^\perp \over p_1^+} \sigma^\perp - {p_2^+ \over p_1^+} {\sigma^\perp .
p_1^\perp \over p_1^+} \sigma^\perp
 \Big ] ~
\chi_{s_{1}}~. ~(\epsilon^\perp_\lambda)^* \nonumber \\
&& ~~~~~~~~~~ {g^2 \over 8 \pi^2}~{1 \over 2}~ C_A~ \ln{\Lambda \over \mu}.
\label{eq3a2}
\end{eqnarray}
Expressing the terms inside the square bracket in terms of the internal
momenta we get $ - { 1 \over P^+} \Big [ {6 \kappa_1^\perp \over 1-x} -
\sigma^\perp. \kappa_1^\perp \sigma^\perp \Big ]$ which makes boost invariance
explicit.

The contribution from Fig. 3(b) is
\begin{eqnarray}
{\cal V}_{3b}~=&&~ {g^3 \over 2 (2 \pi)^3} (-i f^{abc}T^b T^c)~
\sqrt{p_1^+~p_2^+} ~
\int_\epsilon^{q^+ - \epsilon} dk_1^+~
\int ~d^2 k_1^\perp ~ \theta(\Lambda - \mid k^\perp \mid )
{ 1 \over k_1^+} ~{1 \over k_2^+}~ { 1 \over p_1^- - k_1^- - k^-}~
\nonumber \\
&& { 1 \over
p_1^- - k_1^- - k_2^- - p_2^-}~ \sum_{\sigma_1 \lambda_1 \lambda_2} 
\chi^\dagger_{s_2} ~ \Big [ -2 { k_2^\perp \over k_2^+} + \sigma^\perp
{\sigma^\perp . k^\perp \over k^+} + {\sigma^\perp. p_2^\perp \over p_2^+}
\sigma^\perp \Big ]~ \chi_{\sigma_1} ~. ~(\epsilon^\perp_{\lambda_{2}})^*~
\nonumber \\
&& ~~~~~ 
\chi^\dagger_{\sigma_1} ~\Big [ -2 { k_1^\perp \over k_1^+} + \sigma^\perp
{\sigma^\perp . p_1^\perp \over p_1^+} + {\sigma^\perp. k^\perp \over k^+}
\sigma^\perp \Big ] ~\chi_{s_{1}} ~.~ (\epsilon^\perp_{\lambda_{1}})^* \nonumber \\
&& ~~~\epsilon^j_{\lambda_{1}} ~(\epsilon^i_{\lambda})^*~
\epsilon^l_{\lambda_{2}}~ \Bigg [ \Big [ (k_1^i - k_2^i) - {q^i \over
q^+} (k_1^+ - k_2^+) \Big ] \delta_{lj} 
- \Big [ (k_1^l + q^l) - {k_2^l \over k_2^+} (k_1^+ + q^+) \Big ] \delta_{ij}
\nonumber \\
&& ~~~~~~~+  \Big [ (q^j + k_2^j) - {k_1^j \over k_1^+} (q^+ + k_2^+) \Big ]
\delta_{il} \Bigg ] .  \label{eq3b}
\end{eqnarray}
Here $ k^{(+,\perp)} = p_1^{(+,\perp)} - k_1^{(+,\perp)}$ and 
$ k_2^{(+,\perp)} =
q^{(+,\perp)} - k_1^{(+,\perp)}$.

Divergent contributions that contain products of logarithms are
\begin{eqnarray}
{\cal V}_{3b}^{I}~=&&~g ~ T^a ~
  \sqrt{p_1^+~p_2^+} ~\chi^\dagger_{s_{2}}~ 
\Big [ 
- 2 {q^\perp \over q^+} + \sigma^\perp {\sigma^\perp.p_1^\perp \over p_1^+}
+ {\sigma^\perp. p_2^\perp \over p_2^+} \sigma^\perp 
 \Big ] ~
\chi_{s_{1}}~. ~(\epsilon^\perp_\lambda)^* ~ \nonumber \\
&& ~~~~~~{g^2 \over  8 \pi^2} ~{1 \over 2} C_A ~   \ln{\Lambda \over \mu} ~ 
6~ \ln{q^+ \over \epsilon} \label{eq3b1}
\end{eqnarray}
which is manifestly boost invariant.
Divergent contributions that contain single logarithm are
\begin{eqnarray}
{\cal V}_{3b}^{II}~=&&~ g ~ T^a~
   \sqrt{p_1^+}~
\sqrt{p_2^+}~ \chi^\dagger_{s_{2}}~ 
\Big [ - 3 \sigma^\perp {\sigma^\perp. p_1^\perp \over p_1^+} - 3
{\sigma^\perp. p_2^\perp \over p_2^+}\sigma^\perp \nonumber \\
&& ~~~+ 6 {p_1^\perp \over p_1^+}
+ {p_2^+ \over p_1^+}~ {\sigma^\perp. p_1^\perp \over p_1^+} \sigma^\perp
 - {\sigma^\perp.
p_2^\perp \over p_1^+} \sigma^\perp  
 \Big ] ~
\chi_{s_{1}}~. ~(\epsilon^\perp_\lambda)^* 
 {g^2 \over 8 \pi^2}~ {1 \over 2}~ C_A ~ \ln{\Lambda \over \mu}.
\label{eq3b2}
\end{eqnarray}
As in the case of eq.(\ref{eq2b2}), an addition and subtraction of
${6q^\perp \over q^+}$ in the square bracket, renders itself as a 
combination of
canonical vertex and the quantity already encountered in eq.(\ref{eq3a2}) and
thereby manifestly boost invariant.
\section{Coupling Constant Renormalization}
For the sake of 
completeness, we present here the results for the other diagrams which are
relevant for the coupling constant renormalization. We also calculate the 
$\beta$-function which exactly matches
with the well known results and therefore extends the results arrived at in
the Ref.\cite{5zh93c}, to the {\it most general kinematics} in the
two-component formalism.

The sum of divergent contributions from Figs. 2(a) and 2(b) is
\begin{eqnarray}
{\cal V}_2~=&&~g T^a~
\sqrt{p_1^+~p_2^+}~ \chi^\dagger_{s_{2}}~ \Big [
-2 {q^\perp \over q^+} + \sigma^\perp {\sigma^\perp. p_1^\perp \over p_1^+}
+ {\sigma^\perp .p_2^\perp \over p_2^+} \sigma^\perp \Big ]~ \chi_{s_{1}}
~.~ (\epsilon^\perp_\lambda)^* ~ \nonumber \\
&& {g^2 \over 8 \pi^2} ~\Big ( -{1 \over 2}~C_A ~+~ C_f \Big )~
\ln{\Lambda \over \mu} ~\Big( 2 ~\ln{p_1^+ p_2^+ \over \epsilon^2} - 3 \Big ),
\end{eqnarray}
where we observe the emergence of the canonical vertex structure.

The sum of divergent contributions from Figs. 3(a) and 3(b) is
\begin{eqnarray}
{\cal V}_3 ~=&&~ g~ T^a~
 ~ \sqrt{p_1^+}~
\sqrt{p_2^+}~ \chi^\dagger_{s_{2}}~ 
\Big [ 
- 2 {q^\perp \over q^+} + \sigma^\perp {\sigma^\perp.p_1^\perp \over p_1^+}
+ {\sigma^\perp. p_2^\perp \over p_2^+} \sigma^\perp 
 \Big ] ~
\chi_{s_{1}}~.~ (\epsilon^\perp_\lambda)^*  \nonumber \\
&& ~~~~~~ {g^2 \over 8 \pi^2} {1 \over 2}~ C_A ~
\ln{\Lambda \over \mu}
~ \Big ( 2~ \ln{p_1^+ p_2^+ \over \epsilon^2}~ +~ 4~ \ln{q^+ \over \epsilon}
~ -~ 3
\Big) ,
\end{eqnarray}
where we again observe the emergence of the canonical vertex.

The diagrams in Figs. 4(a), 4(b), 5(a) and 5(b) 
correspond to the renormalization of the
external quark and gluon legs that are connected to the vertex. Their
contributions are given below.
\begin{eqnarray}
{\cal V}_{4a}=g~ T^a~
 ~ \sqrt{p_1^+ p_2^+}~ \chi^\dagger_{s_{2}}~ 
&& ~~\Big [ 
- 2 {q^\perp \over q^+} + \sigma^\perp {\sigma^\perp.p_1^\perp \over p_1^+}
+ {\sigma^\perp. p_2^\perp \over p_2^+} \sigma^\perp 
 \Big ] ~
\chi_{s_{1}}~.~ (\epsilon^\perp_\lambda)^*  \nonumber \\
&& ~~~~~~ {g^2 \over 4 \pi^2} ~ C_f~
\ln{\Lambda \over \mu}
~ \Big ( {3 \over 2}- 2~\ln{p_1^+  \over \epsilon}~ 
\Big) ,
\end{eqnarray}
\begin{eqnarray}
{\cal V}_{4b}=g~ T^a~
 ~ \sqrt{p_1^+ p_2^+}~ \chi^\dagger_{s_{2}}~ 
&& ~~\Big [ 
- 2 {q^\perp \over q^+} + \sigma^\perp {\sigma^\perp.p_1^\perp \over p_1^+}
+ {\sigma^\perp. p_2^\perp \over p_2^+} \sigma^\perp 
 \Big ] ~
\chi_{s_{1}}~.~ (\epsilon^\perp_\lambda)^*  \nonumber \\
&& ~~~~~~ {g^2 \over 4 \pi^2} ~ C_f~
\ln{\Lambda \over \mu}
~ \Big ( {3 \over 2}- 2~\ln{ p_2^+ \over \epsilon}~ 
\Big) ,
\end{eqnarray}
\begin{eqnarray}
{\cal V}_{5a}=-g~ T^a~
 ~ \sqrt{p_1^+ p_2^+}~ \chi^\dagger_{s_{2}}~ 
&& ~~\Big [ 
- 2 {q^\perp \over q^+} + \sigma^\perp {\sigma^\perp.p_1^\perp \over p_1^+}
+ {\sigma^\perp. p_2^\perp \over p_2^+} \sigma^\perp 
 \Big ] ~
\chi_{s_{1}}~.~ (\epsilon^\perp_\lambda)^*  \nonumber \\
&& ~~~~~~ {g^2 \over 8 \pi^2} {4 \over 3}~ N_fT_f~
\ln{\Lambda \over \mu} ,
\end{eqnarray}

\begin{center}
\begin{picture}(300,100)(0,0)
\Text(150,-5)[]{{\footnotesize 
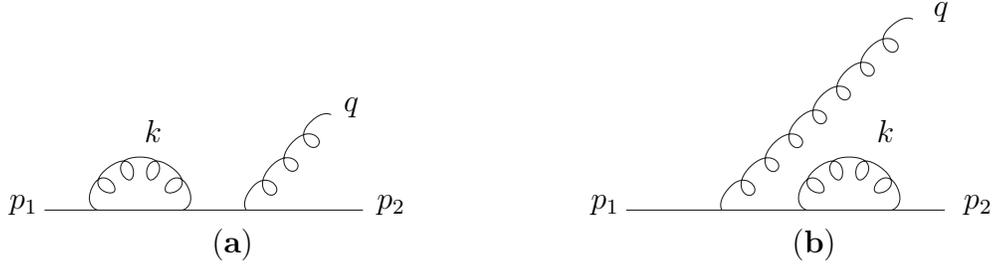
Fig.4 Contribution to the quark-gluon vertex from
quark wave-function renormalization.}}
\SetScale{.8}
\Text(-24,30)[]{$p_1$}
\Text(115,30)[]{$p_2$}
\Text(100,67)[]{$q$}
\Text(25,58)[]{$k$}
\Text(55,15)[]{({\bf a})}
\GlueArc(25,35)(20,0,180){5}{4}
\Gluon(75,35)(115,80){5}{3}
\Line(-20,35)(130,35) 
\SetOffset(220,0)
\Text(-24,30)[]{$p_1$}
\Text(117,30)[]{$p_2$}
\Text(103,103)[]{$q$}
\Text(82,58)[]{$k$}
\Text(55,15)[]{({\bf b})}
\GlueArc(85,35)(20,0,180){5}{4}
\Gluon(25,35)(115,125){5}{7}
\Line(-20,35)(130,35) 
\end{picture}
\end{center}
\vskip .7cm
\begin{center}
\begin{picture}(300,100)(0,0)
\Text(150,-5)[]{{\footnotesize 
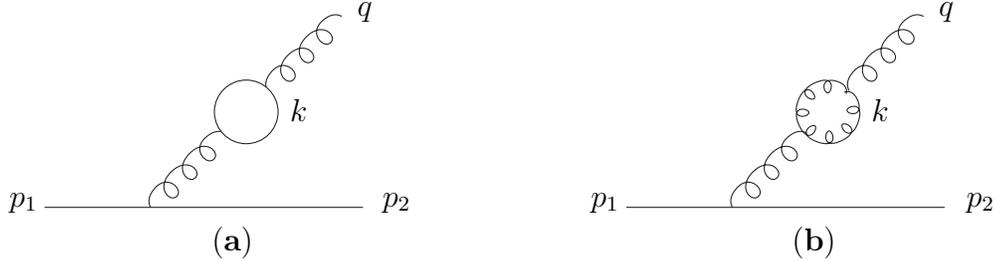
Fig.5 Contribution to the quark-gluon vertex from
gluon wave-function renormalization.}}
\SetScale{.8}
\Text(-24,30)[]{$p_1$}
\Text(117,30)[]{$p_2$}
\Text(105,102)[]{$q$}
\Text(80,65)[]{$k$}
\Text(55,15)[]{({\bf a})}
\Gluon(30,35)(65,70){5}{3}
\Gluon(85,90)(120,125){5}{3}
\BCirc(75,80){15}
\Line(-20,35)(130,35) 
\SetOffset(220,0)
\Text(-24,30)[]{$p_1$}
\Text(118,30)[]{$p_2$}
\Text(105,102)[]{$q$}
\Text(80,65)[]{$k$}
\Text(55,15)[]{({\bf b})}
\Gluon(30,35)(65,70){5}{3}
\Gluon(85,90)(120,125){5}{3}
\GlueArc(75,80)(12,45,410){3}{7}
\Line(-20,35)(130,35) 
\end{picture}
\end{center}
\vskip .3cm
\begin{eqnarray}
{\cal V}_{5b}=g~ T^a~
 ~ \sqrt{p_1^+ p_2^+}~ \chi^\dagger_{s_{2}}~ 
&& ~~\Big [ 
- 2 {q^\perp \over q^+} + \sigma^\perp {\sigma^\perp.p_1^\perp \over p_1^+}
+ {\sigma^\perp. p_2^\perp \over p_2^+} \sigma^\perp 
 \Big ] ~
\chi_{s_{1}}~.~ (\epsilon^\perp_\lambda)^*  \nonumber \\
&& ~~~~~~ {g^2 \over 8 \pi^2} ~ C_A~
\ln{\Lambda \over \mu}
~ \Big ( {11 \over 3}- 4~\ln{q^+  \over \epsilon}~ 
\Big ) .
\end{eqnarray}
 
Now, to evaluate the contributions to the coupling constant, we have to
multiply ${\cal V}_4$ and ${\cal V}_5$ with ${1\over2}$ 
in order to take into account
the proper correction due to the renormalization of initial and final states
\cite{5bd}. Thus adding the contributions we get,
\begin{eqnarray}
\delta {\cal V}_1 =&&~~\big( {1\over2}{\cal V}_4 + {1\over 2}{\cal V}_5 +
{\cal V}_2 +
{\cal V}_3 \big)
\nonumber\\=&&~~g~ T^a~
 ~ \sqrt{p_1^+ p_2^+}~ \chi^\dagger_{s_{2}}~ 
\Big [ 
- 2 {q^\perp \over q^+} + \sigma^\perp {\sigma^\perp.p_1^\perp \over p_1^+}
+ {\sigma^\perp. p_2^\perp \over p_2^+} \sigma^\perp 
 \Big ] ~
\chi_{s_{1}}~.~ (\epsilon^\perp_\lambda)^*  \nonumber \\&&\quad\quad
{g^2\over 8\pi^2}\Big( {11\over6} C_A-{2 \over 3}N_fT_f\Big)~\ln {\Lambda\over
\mu}.
\end{eqnarray}
Note that all the mixed divergences cancel. The correction to the coupling
constant is given by
\begin{eqnarray}
g_R~=~g(1+\delta g)~=~g\Big[~1+ {g^2\over 8\pi^2}\Big( {11\over6} C_A-{2 \over 
3}N_fT_f\Big)\ln{\Lambda\over \mu}~\Big].
\end{eqnarray}
We compute the $\beta$-function as 
\begin{eqnarray}
\beta(g)~&&=~-~{\partial{g_R} \over \partial{\ln\Lambda}}\nonumber\\
&&=~-~{g^3 \over 16\pi^2}\Big({11\over 3}C_A-{4 \over 
3}N_fT_f\Big) ,
\end{eqnarray}
which is well known result to the one-loop order.

Running of the QCD coupling can be depicted as follows. Notice that the
regularized coupling constant depends on the ultraviolate cut-off $\Lambda$
as well as $\mu$ :
\begin{equation}
g_R~=~g_R(\Lambda, \mu)\, .
\end{equation} 
To obtain the renormalized coupling, we have to remove the cutoff dependence
by adding suitable counter term 
and then take $\Lambda\rightarrow\infty$. But we must notice that one energy
scale $\mu$ has crept in the coupling constant and should be kept there
with large enough value 
to ensure the validity of the perturbative calculation.
The counter term is chosen to be the following : 
\begin{equation}
g_{\rm ct}(\Lambda, \mu_r)~ =~ -  
{g^3\over 8\pi^2}\Big( {11\over6} C_A-{2 \over 
3}N_fT_f\Big)\ln{\Lambda\over \mu_r}\, .
\label{gm1m2}
\end{equation}
Notice that the counter term introduces another arbitrary energy 
scale $\mu_r$ which is 
large for obvious reason, but need
not be the same as $\mu$. We call $\mu_r$ the renormalization scale. So,
we get the renormalized coupling constant as, 
\begin{eqnarray}
g_{\rm ren}(\mu,\mu_r)~=&&\lim_{\Lambda\rightarrow\infty}~[g_R(\Lambda,\mu)+
g_{\rm ct}(\Lambda, \mu_r)]~\nonumber\\
=&&g\Big[1 +{g^2\over 8\pi^2}\Big( {11\over6} C_A-
{2 \over 
3}N_fT_f\Big)\ln{\mu_r\over \mu}\Big]\, .\label{rcon}
\end{eqnarray} 
Thus the renormalized coupling constant 
at some scale $\mu$ always depends on the
renormalization scale  as well as its value at that scale obtained by
setting $\mu=\mu_r$ in eq.(\ref{rcon}) as 
the renormalization condition: 
\begin{equation}
g(\mu_r)\equiv g_{\rm ren}(\mu, \mu_r)\mid_{\mu=\mu_r}=g\, . \label{rcond}
\end{equation}
Therefore, from eq.(\ref{rcon}) and  eq.(\ref{rcond}) 
we get the renormalized coupling constant at any perturbatively large scale 
$Q$ as 
\begin{eqnarray}
&&g(Q)\equiv  g_{\rm ren}(Q,\mu_r)=
g(\mu_r)\Big[ 
1+ {g^2(\mu_r)\over 8\pi^2}\Big( {11\over6} C_A-{2 \over 
3}N_fT_f\Big)\ln{\mu_r\over Q}\Big]\nonumber\\
&&~~~~~~~~~~~~~~~~~~={g(\mu_r)\over 
1+ {g^2(\mu_r)\over 8\pi^2}\Big( {11\over6} C_A-{2 \over 
3}N_fT_f\Big)\ln{Q\over \mu_r}}\, ,\label{gr1}
\end{eqnarray}
or, in terms of $\alpha_s(Q^2)={g^2(Q^2)\over 2\pi}$  
we have 
\begin{equation}
\alpha_s(Q^2)={\alpha_s(\mu_r^2)\over 
1+ {\alpha_s(\mu_r^2)\over 4\pi}\Big( {11\over6} C_A-{2 \over 
3}N_fT_f\Big)\ln\left({Q\over \mu_r}\right)^2}\, .\label{alr2} 
\end{equation}
Expressions in eq.(\ref{gr1}) and eq.(\ref{alr2}) are  
the well known results for the running QCD coupling constant.
 
\section{ Discussion }
In this Chapter, we have presented the calculations necessary for 
coupling constant renormalization in light-front QCD with the 
most {\it general
kinematics}, which is the first step towards extending the structure
function calculation to fourth order. This calculation and also other
similar ones    
employing time ($x^+$) ordered perturbative techniques in
light-front theory are known to be straightforward but long and tedious. A
lot of effort has to be invested in the calculation of coefficient functions
accompanying the divergences for individual diagrams.
No clue seems to be there in the intermediate steps 
regarding the correctness of the calculations due
to lack of covariance. Other than actually performing the calculations, 
we tried to investigate in parallel the 
role played by the Gallilean
boost symmetry present in light-front formulation here. 
To the best of our knowledge, this is the {\it first work} to investigate the
utility of Gallilean boost symmetry in determining the correctness of the 
structure of the coefficient functions accompanying the
divergences in light-front perturbation theory beyond the tree level.

In this initial investigation we have employed the simplest choice of regulators 
that cutoff single particle momenta.  
One should note that in addition to possible violations of boost invariance,
such simple minded cutoff procedure could in principle even introduce
non-analyticities in the structure of counter terms (see Sec. VI of Ref.
\cite{5wi94} for an explicit example). However, in the case of vertex
diagrams, we encounter only logarithmic transverse divergences.
Even with finite cutoffs,
violations of transverse boost invariance can appear only inside the 
logarithms and we
expect the symmetry to be 
present in the non-trivial structure of the coefficient functions 
that accompany the 
divergences. We are primarily interested in understanding the complex
structure of these coefficient functions on the basis of Gallilean symmetry.
Incidentally we note that, in contrast, longitudinal boost invariance is a scale 
invariance in
light-front theory. 
The implication of longitudinal boost symmetry for the
coefficient functions is trivial, namely, simple scaling behavior.

Let us summarize our findings. Out of all the  $x^+$-ordered diagrams
relevant for our calculation, four involve wave-function renormalization
correction and have the structure of the canonical vertex. For the remaining
diagrams which correspond to vertex corrections, 
the divergent contributions from each of the
them contain terms that involve (I) product of logarithms 
and (II) single
logarithm. For contributions that belong to (I), we find that 
for each diagram separately, 
the coefficient of the divergent factor is proportional to the canonical
vertex and hence Gallilean boost invariance is manifestly maintained. For
contributions that belong to (II), for each diagram, the coefficient of the
divergent factor is not proportional to the canonical vertex. Nevertheless,
in each case, rewriting the coefficient in terms of the internal momenta 
explicitly shows that the coefficient is independent of the total transverse
momentum $P^\perp$. Hence for the contributions that belong to (II) the
constraint from transverse boost invariance is maintained, even though the
canonical form is not reproduced. 

Our results show that two-dimensional Gallilean invariance which is
manifest at tree level is also exhibited in the coefficient functions
accompanying the divergences in the regulated theory at the one loop level in
the case of quark-gluon vertex in light-front QCD even with a regulator that
violates the symmetry. Since the symmetry is only a part of the complete
Lorentz symmetry, we expect the constraints which follow from the
invariance to be less restrictive. Indeed, our results show
that the structure of
the vertex that satisfies transverse boost invariance is not unique.

Even-though the canonical vertex structure is not reproduced in the
coefficient of the single logarithms, it still has some usefulness in
practical calculations since it obeys constraint from Gallilean boost 
invariance. The coefficient functions accompanying single logarithms are
obtained after isolating the leading double logarithms and they exhibit a
complicated structure. It is quite easy to
make a mistake in the sign  in {\it one} of the terms for individual $x^+$
ordered diagrams. Our calculations show that using the underlying
transverse boost
symmetry one can easily recognize the mistake in the calculation and hence
correct it.

Finally, of course, we  summarized the results for the
complete set of diagrams contributing to the 
coupling constant renormalization
for the massless quark case. We have extracted the $ \beta $-function and
obtained the running QCD coupling renormalized at a particular energy 
scale, both of which
match with the well-known results and therefore extends the results
arrived at previously in the literature to the most general kinematics. 
%
%
Using the
two-component representation \cite{5zh93b} 
we have presented {\it for the
first time} the
results separately for each $x^+$ ordered diagram with arbitrary 
external momenta which is essential to study the renormalization of the
helicity-non flip parts of the vertex. Present calculations together with the
calculations presented in Ref. \cite{5zh93c} explicitly show that linear
divergences of the type ${ 1 \over \epsilon}$ where $\epsilon$ is the cutoff
on longitudinal loop momenta occur in individual time-ordered diagrams only
in radiative corrections to the chiral symmetry breaking part of the
quark gluon vertex. This divergence is a special feature of non-abelian
gauge theory. At one loop level, this divergence cancels with our
choice of regulators when
different time-ordered diagrams are summed up. Since intermediate states
involved are, in general, different in different time ordered diagrams, the
cancellation may no longer be operative once more sophisticated regulators
that explicitly depend on the intermediate states are employed. This needs
to be investigated in detail in the future because of its nontrivial
consequences for the renormalization of chiral symmetry breaking terms in the 
QCD Hamiltonian. 

As far as the structure function calculation to the fourth order goes, one
has to essentially embed the coupling 
constant renormalization calculation in another
loop, making the final calculation a two-loop one. 
Which seems to be  a straightforward extension of the calculations presented
here (which are all one-loop); but one has to pay proper attention to the
intricacies involved in the complete two-loop perturbative calculations. 
And hopefully it would yield the solution of Altarelli-Parisi equation to
the fourth order in coupling  providing greater 
validity of calculating structure functions using our approach. 


\newpage
\setcounter{section}{0}
\renewcommand{\thesection}{6.\arabic{section}}
\setcounter{equation}{0}
\renewcommand{\theequation}{6.\arabic{equation}}

{\flushleft\huge\bf {Chapter 6}}
\vskip 1cm
In this Chapter, we try to investigate the nucleonic spin structure in
the light-front QCD Hamiltonian formulation 
that we have been using throughout to describe DIS structure functions. 
Like the longitudinal momentum, it
is of importance to know how the spin of the nucleon is shared among its 
various constituents. Individual constituents are allowed to have orbital
motions and hence orbital angular momenta which then combine with their 
intrinsic angular momenta (or, spin) to give a net total angular momentum
(or, simply the spin, if the orbital angular momentum quantum number is zero)
for the nucleon. As of now, we can only measure the intrinsic helicity part
carried by the fermionic constituents (namely, the quarks or antiquarks),
through the measurement of  polarized structure function $g_1$. The
orbital helicity parts or even the intrinsic helicity for the gluons are not
known how to be measured in the experiments. Since helicity is the object 
measured in the experiment, we shall only be concerned with the third
component of the angular momentum operator throughout this Chapter. Thus,
as it turns out, the contribution from constituents' orbital 
helicity to the total 
helicity of the nucleon is more of theoretical interest at present. In
fact, it is still theoretically uncertain how the total 
helicity operator (relevant for nucleon), 
as defined in the field theory, separates into a sum of corresponding pieces
describing intrinsic and orbital helicity of the constituents. The problem
seems to lie in the gauge invariance. The quark or gluon fields in a
particular gauge becomes a combination of both by  a gauge transformation. 
So, an operator defined to be, for example,  the quark orbital helicity
operator solely in terms of quark fields in a particular gauge loses its
characteristics in another gauge. Thus, the separation we are interested in
seems to be unrealistic due to lack of gauge invariance. We shall discuss
this problem in detail in Sec.6.1, which hints towards the resolution that
the gauge fixing may be necessary in order to define such operators. There
we shall
derive such a gauge fixed (in $A^+=0$ gauge) light-front operators, which
necessarily does the separation we are looking for. Next, in  Sec.6.2, 
with the help of
these operators we shall define the relevant structure functions, which can
measure the orbital (and intrinsic) helicity contributions of quarks or
gluons to the total helicity of the nucleon. Then  we calculate
these structure functions for a dressed parton target, 
which shows the utility of our definitions by 
obtaining various anomalous dimensions necessary to study their 
$Q^2$-evolution in  Sec.6.3. Validity of our calculations
are checked by showing that the helicity sum rules are satisfied in  Sec.6.4.
Thus, the plausibility of 
defining such structure functions and the calculations
leading to various anomalous dimensions in a transparent way (as will be
discussed here) exhibit one of the major triumph of our formulation
presented in this dissertation. 
\section{Light-front Helicity operator $J^3$ from the manifestly gauge 
invariant energy momentum tensor}
\subsection  {A Brief Review}
Before we show the detailed derivation of the light-front helicity operator
$J^3$, let us first review the situation which necessitates such an 
exercise. 
The well-known {\it proton spin crisis} that 
emerged after the the publication of EMC data on the
measurement of polarized structure function $g_1(x,Q^2)$, 
drew a lot of theoretical interest on how the total helicity of 
the nucleon is distributed among its constituents. 
In the parton model, partons are
assumed to be moving collinearly (i. e., with zero relative transverse 
momenta ${\vec k_T} =0$) and do not have any orbital motion ${\vec L}={\vec
r}\times{\vec k_T}=0$. Thus, the total helicity of the nucleon comes solely 
from the intrinsic helicity of the constituents (quarks and gluons). Now an
integral of $g_1(x,Q^2)$ over all possible $x$ can be shown (see later) to
measure, out of the total nucleonic helicity, only 
the intrinsic helicity part ($\Delta \Sigma$) coming 
from the fermionic constituents.  EMC results 
showed that this $\Delta \Sigma$ contribution was very small, which appeared
to be quite puzzling and became known as the proton spin crisis.
Incidentally, the role of orbital angular 
momentum in deep inelastic scattering was  
first emphasized by Sehgal\cite{lms} and then by Ratcliff\cite{pgr} 
in the context of Altarelli-Parisi equation for real partons in
QCD. But it  should be noted that, in the 
interpretation of EMC
data, orbital helicity parts of the constituents were ignored as
some higher twist contribution, even if the partons in reality might 
have some non-zero
transverse momenta ${\vec k_T} \neq 0$. On the other hand, it is
understood that an anomaly contribution coming from the gluonic sector is
responsible for a cancellation to occur, giving rise to the small measured
value of $\Delta \Sigma$. It also made ambiguous whether to call 
$\Delta \Sigma$ as the intrinsic helicity contribution from quarks and
anti-quarks or the one before the cancellation occurred. Whichever be the
situation, it is now realized that to understand the nucleonic spin structure
properly, one should put more emphasis in studying the helicity sum
rule, which anyway has to be satisfied and free from anomaly due to the 
conservation of angular momentum. 

The total helicity operator $J^3$ acting
on a nucleon state $\mid PS \rangle$ gives,
\begin{equation}
J^3\mid PS \rangle = \pm {1\over 2} \mid PS \rangle\, ,
\end{equation}
depending on the value of $S=\pm{1\over 2}$. Thus, we have,  
\begin{eqnarray}
{ 1 \over {\cal N}} \langle PS \mid \Big [ J^3_{q(i)} + J^3_{q(o)} +
J^3_{g(i)} + J^3_{g(o)} \Big ] \mid PS \rangle = \pm { 1 \over 2}\, ,
\end{eqnarray}
with ${\cal N}= \langle PS\mid PS\rangle$. Here we have assumed for the time
being that the total $J^3$ can be separated into corresponding intrinsic and
orbital helicity operators for quarks and gluons.  It is this separation
which becomes ambiguous due to lack of gauge invariance of the individual
parts. 

Before we proceed further, let us give a brief account of recent
developments in this regard. Jaffe and Manohar\cite{jm} 
first noted that angular momentum operator
constructed from the gauge invariant, symmetric, energy-momentum tensor
fails to display the distinction between intrinsic 
and orbital angular momentum.
They suggested the use of free field theory form which is interaction
independent and can be separated unambiguously into quark and gluon orbital
and spin parts. 
Ji, Tang and Hoodbhoy\cite{jth}, starting from the Jaffe-Manohar choice,
studied the {\it asymptotic fraction}  of the
nucleon spin carried by quarks and gluons at the one loop level. 
They mentioned that this separation is gauge variant and 
supported the choice of light-front gauge and light-front coordinates for
their calculation.
Later on Ji\cite{ji97} introduced a gauge invariant definition of $J_q$ 
and $J_g$ starting from gauge invariant symmetric energy-momentum tensor. 
But these are now interaction dependent contrary to the well-known 
kinematical nature of the  angular
momentum operators (only $J^3$ in light-front theory is so). 
No justification has been given why they
are called angular momentum operators. 

In fact, Singleton and Dzhunushaliev\cite{sd} claim 
to show by explicit calculations  
that the gauge invariant orbital angular momentum operator proposed by
Ji\cite{ji97} do not obey the 
angular momentum algebra and hence do not qualify
as the angular momentum operator.
It seems to be in agreement with the observation made by Chen and Wang\cite{cw} 
regarding the gauge invariant definition (that they do not obey the
angular momentum algebra) and they further put more
emphasis on the matrix elements, which are important as far as the experiment
is concerned. From their argument it appears that 
the gauge dependent 
operators may have gauge
invariant expectation values in the  hadronic eigenstates having definite  
angular momentum.  
There are other works (for example, see Refs. \cite{ovt},
\cite{hjl1}, \cite{bj}, \cite{hjl2}, \cite{mhs}) 
which show that the issue of gauge invariant
separation is still under hot debate. In our case, we always work with the
gauge fixed theory which is free from any of these ambiguity as we discuss
next, in detail.
\subsection{$J^3$ in light-front gauge}
It is well-known that the {\it energy-momentum density} (which gives rise to
Hamiltonian and three-momentum) and the {\it generalized angular momentum 
density} (which gives
rise to angular momentum and boosts) can be expressed in a manifestly
covariant, gauge invariant form. But, in order to define the Poincare 
generators in quantum 
field theory, one has to choose a particular hypersurface over which these
densities are integrated. Thus, if one chooses $x^0=0$ as the surface
over which integration is to be performed, one gets using standard notations, 
$P^\mu = \int d^3x
\theta^{0\mu}$. On the other hand, the choice $x^+=0$ gives $P^\mu = \int
dx^-d^2x^\perp
\theta^{+\mu}$. 
Therefore, the the Poincare generators explicitly depends on the frame of 
reference. It may not be surprizing that it also depends on the 
gauge choice in the case of gauge theory. 
This of course does not imply that the 
theory has lost Lorenz and gauge symmetry. The symmetries are no longer 
manifest, but the physical observables in the theory still obey the consequences 
of the symmetries. 

Poincare generators can be further classified as kinematical (which do not
contain interactions and do not change the quantization surface) and
dynamical (which contain interactions and change the quantization surface).
Which operator is dynamical and which is kinematical of course depends on
the choice of quantization surface. It is well-known that in light-front
field theory, on which our formalism of deep inelastic scattering is based
on,
the generators of boosts and the rotation in the transverse plane
(light-front helicity) are kinematical like three momenta
whereas the generators of rotations
about the two transverse axes are dynamical like the Hamiltonian. 
Thus, the operator in light-front
field theory relevant to the {\it proton spin crisis} is the light-front
helicity operator which belongs to the kinematical subgroup. In light-front
literature, it is customary to construct this operator from the canonical
symmetric energy momentum tensor and one explicitly finds that this operator
is indeed free of interaction and has the same form as in free field
theory\cite{ks}.

In non-Abelian gauge theories like QCD, one should be extra cautious since
such theories are known to exhibit non-trivial topological effects. In this
work,  we restrict our attention to the topologically trivial sector of QCD
which is relevant for DIS. 
In this sector, interactions do not affect kinematical 
generators\cite{wein}. 
In view of the prevailing confusion in the literature as we mentioned,
we provide an explicit demonstration of this
fact in this section in the case of the light-front helicity operator.
      
We start from the manifestly gauge invariant, symmetric energy momentum
tensor in QCD:
\begin{eqnarray}
\Theta^{\mu \nu} && = { i \over 2} {\overline \psi} [ \gamma^\mu D^\nu +
\gamma^\nu D^\mu ]\psi - F^{\mu \lambda a} F^{\nu a} _{~ \lambda} \nonumber \\
&&  ~~ - g^{\mu \nu} \Big \{ -{1 \over 4} (F_{\lambda \sigma a})^2 +
{\overline \psi}(i \gamma^\lambda D_\lambda -m) \psi \Big \},
\end{eqnarray}
where $ i D^\mu = i \partial^\mu + g A^\mu $, 
$ F^{\mu \lambda a} = \partial^\mu  A^{\lambda a} - \partial^\lambda
A^{\mu a} + g f^{abc} A^{\mu b} A^{\lambda c}$ , $ 
F^{\nu a}_{~ \lambda } = \partial^{\nu a}  A_{\lambda} - \partial_{\lambda}
A^{\nu a} + g f^{abc} A^{\nu b} A^c_{\lambda}$.

We define the light-front helicity operator
\begin{eqnarray}
{\cal J}^3 = { 1 \over 2} \int dx^- d^2 x^\perp  [ x^1  \Theta^{+2} - x^2
\Theta^{+1}].
\end{eqnarray}
${\cal J}^3$ is a manifestly gauge invariant operator 
by construction. However, 
it depends explicitly on the interaction through $\Theta^{+i}$ 
and does not appear to be a
kinematical operator at all. Furthermore, 
it is not apparent that ${\cal J}^3$
generates the correct transformations as an angular momentum operator. 
Thus at this stage, we are not justified to call it a helicity operator.

Explicitly, we have,
\begin{eqnarray}
{\cal J}^3 && = { 1 \over 2} \int dx^- d^2 x^\perp \Big \{
x^1  [ { i \over 2} {\overline \psi} (\gamma^+ D^2 + \gamma^2 D^+) \psi 
        - F^{+ \lambda a} F^{2a}_{~\lambda}]  \nonumber \\
&& ~~ - x^2  [ { i \over 2} {\overline \psi} (\gamma^+ D^1 + \gamma^1 D^+) \psi 
        - F^{+ \lambda a} F^{1 a}_{~\lambda}] \Big \} .   
\label{j3in}
\end{eqnarray}
Notice that the last term in $\Theta^{+i}$ does not contribute for
$g^{+i}=0$. 
The fermion field can be decomposed as usual, 
$ \psi^{\pm} = \Lambda^{\pm} \psi$, with $ \Lambda^{\pm} = { 1 \over
4} \gamma^{\mp} \gamma^{\pm}$ and 
we shall work in the gauge $A^+=0$. 
In this gauge, we {\it still have residual gauge freedom} associated with
$x^-$-independent gauge transformations. Note that only $\psi^+$ and $A^i$
are dynamical variables whereas $\psi^-$ and $A^-$ are constrained. 

Let us proceed to calculate $\Theta^{+2}$ explicitly in terms of dynamical
fields. 
We have,
\begin{eqnarray}
{ i \over 2} {\overline \psi} ( \gamma^+ D^2  + \gamma^2 D^+) \psi =
{\psi^+}^\dagger i \partial^2 \psi^+  
+ g {\psi^+}^\dagger T^a \psi^+ A^2_a + { i \over 2} {\overline \psi}
\gamma^2 i \partial^+ \psi,
\end{eqnarray}
where the last term depends on the interaction through the constrained field
$\psi^-$ and the interaction dependence in the second term is explicit. 
We use the constraint equation
\begin{eqnarray}
i \partial^+ \psi^- = \Big [ \alpha^\perp \cdot  ( i \partial^\perp + g
A^\perp) + \gamma^0 m \Big ] \psi^+,
\end{eqnarray}
to eliminate the constraint variable $\psi^-$ as follows.
\begin{eqnarray}
{ i \over 2} {\overline \psi} \gamma^2 \partial^+ \psi && = 
{ i \over 2}  \psi^{+\dagger}\gamma^0\gamma^2(\partial^+ \psi^-) + 
{ i \over 2}  \psi^{-\dagger} \gamma^0 \gamma^2(\partial^+  \psi^+)
\nonumber\\ 
&&={ i \over 2}  \psi^{+\dagger}\alpha^2(\partial^+ \psi^-) - 
{ i \over 2}  (\partial^+\psi^{-\dagger}) \alpha^2  \psi^+ 
+ { i \over 2} \partial^+ \Big ( {\psi^-}^\dagger \alpha_2 \psi^+
\Big )\nonumber\\ 
&&= { 1 \over 2}  \psi^{+\dagger}\alpha^2 \Big [ \alpha^\perp\cdot
(i{\overrightarrow \partial^\perp} + gA^\perp) + \gamma^0 m \Big]\psi^+ + 
{ 1 \over 2}  \psi^{+\dagger} \Big [ \alpha^\perp\cdot
(-i{\overleftarrow\partial^\perp} + gA^\perp) + \gamma^0 m \Big]
\alpha^2\psi^+ \nonumber\\
&& ~~~~~~~~~~~~~\quad\quad\quad
+ { i \over 2} \partial^+ \Big ( {\psi^-}^\dagger 
\alpha_2 \psi^+\Big )\nonumber\\
&& =\Big[ {i\over 2} \psi^{+\dagger} \alpha^2\alpha^j(\partial^j \psi^+) -  
{i\over 2} (\partial^j\psi^{+\dagger}) \alpha^j\alpha^2\psi^+\Big]
+{g\over 2} \psi^{+\dagger}(\alpha^2\alpha^j + \alpha^j\alpha^2)T^a\psi^+
A^{ja}\nonumber\\ 
&& ~~~~~~~~~~~~~\quad\quad\quad
+ { i \over 2} \partial^+ \Big ( {\psi^-}^\dagger 
\alpha_2 \psi^+\Big )\nonumber\\
&& = i {\psi^+}^\dagger
\partial^2 \psi^+ 
+ {1 \over 2} \partial^1({\psi^+}^\dagger \Sigma^3 \psi^+) + g {\psi^+}^\dagger T^a
\psi^+ A^{2a} \nonumber \\
&& ~~~~~~~~~~~~~\quad\quad\quad
+ { i \over 2} \partial^+ \Big ( {\psi^-}^\dagger \alpha_2 \psi^+
\Big ) - { i \over 2} \partial^2 \Big ( {\psi^+}^\dagger \psi^+ \Big ) 
\label{lastln}
\end{eqnarray}
Notice that $\partial^{+,2}$ are like total derivatives 
(but not $\partial^1$), since 
$\Theta^{+2}$ term is
multiplied by $x^1$  and we have used this fact to
switch around $\partial^{+,2}$ producing surface terms (see the last line in
eq.(\ref{lastln})).  
First and second terms in the last line follow from the 
first term in the previous line
for $j=2$ and  $j=1$ respectively with a corresponding surface term. Also note
that the terms involving $m$ has canceled and we have used 
$ \Sigma^3 = i \alpha^2 \alpha^1$.

Now we restrict ourselves to the topologically trivial sector by
requiring that the dynamical fields ($\psi^+$ and $A^i$) vanish at $x^{-,i}
\rightarrow \infty$. The residual gauge freedom and the surface terms 
are no longer present 
and so we drop total derivatives of
$\partial^+$ and $ \partial^2$. 
Thus we obtain, 
\begin{eqnarray}
{ i \over 2} {\overline \psi} (\gamma^+ D^2 + \gamma^2 D^+) \psi = 2 i
{\psi^+}^\dagger  \partial^2 \psi^+  
+ {1 \over 2} \partial^1 ({\psi^+}^\dagger \Sigma^3 \psi^+) + 2 g {\psi^+}^\dagger T^a
\psi^+ A^{2a}.
\label{Th+iqaq}
\end{eqnarray}
Notice that the last term explicitly depends on the interaction and rest of
the expression is the same as one would have got in the free theory. 

Similarly, we can calculate the gluonic contribution to $\Theta^{+i}$. 
In the gauge $A^+=0$, we have, 
\begin{eqnarray}
- F^{+ \lambda a} F^{2 a}_{~ \lambda } &&= -\big( \partial^+A^{\lambda
a}\big) \big( \partial^2 A_\lambda^a -\partial_\lambda A^{2a} + gf^{abc}
A^{2 b} A^c_\lambda \big) \nonumber\\
&& = { 1 \over 2 } \big(\partial^+ A^{-a}\big)\big(\partial^+ 
A^{2a}\big) + \partial^+ A^{ja} (\partial^2 A^{ja} - \partial^j A^{2a}) 
+ g f^{abc} (\partial^+ A^{ja}) A^{2b} A^{jc} \nonumber \\
&& = - { 1 \over 2 } (\partial^+)^2 A^{-a}
A^{2a} + \partial^+ A^{ja} (\partial^2 A^{ja} - \partial^j A^{2a}) 
+ g f^{abc} (\partial^+ A^{ja}) A^{2b} A^{jc} \nonumber \\
&& ~~~~ + { 1 \over 2} \partial^+ \Big (\partial^+ A^{-a}  A^{2a} \Big ),
\end{eqnarray}
where we have done partial integration using $\partial^+$. 
This  still depends on the constrained field $A^{-a}$. 
We have the constraint equation for the elimination of the variable $A^-$,
\begin{eqnarray}
 { 1 \over 2} (\partial^+)^2 A^{-a} = \partial^+ \partial^i A^{ia} + g
f^{abc} A^{ib} \partial^+ A^{ic} + 2 g {\psi^+}^\dagger T^a \psi^+.
\end{eqnarray}
Thus, we obtain, 
\begin{eqnarray}
-F^{+ \lambda a} F^{2a}_{~ \lambda } &&= 
\partial^+ A^{ja} (\partial^2 A^{ja} - \partial^j A^{2a}) +
\big(\partial^jA^{ja}\big)\big(\partial^+A^{2a}\big)- 2g 
{\psi^+}^\dagger T^a \psi^+\nonumber\\
&& ~~~~+ { 1 \over 2} \partial^+ \Big ( \partial^+ A^{-a} A^{2a} \Big )
- \partial^+ \Big ( \partial^i A^{ia} A^{2a} \Big )
\nonumber\\
&& =\partial^i A^{ia} \partial^+ A^{2a} +
\partial^+ A^{ja}(\partial^2 A^{ja} - \partial^j A^{2a}) - 2 g
{\psi^+}^\dagger T^a \psi^+ A^{2a} \nonumber \\
&&~~~~~ + { 1 \over 2} \partial^+ \Big ( \partial^+ A^{-a} A^{2a} \Big )
- \partial^+ \Big ( \partial^j A^{ja} A^{2a} \Big ) \nonumber \\
&& = \partial^+ A^{1a} \partial^2 A^{1a} + \partial^+ A^{2a} \partial^2
A^{2a} + \partial^1 (A^{1a} \partial^+ A^{2a}) - 2 g
{\psi^+}^\dagger T^a \psi^+ A^{2a}\nonumber \\
&& ~~~~+ { 1 \over 2} \partial^+ \Big ( \partial^+ A^{-a} A^{2a} \Big )
- \partial^+ \Big ( \partial^j A^{ja} A^{2a} \Big )
- \partial^+ \Big ( A^{1a} \partial^1 A^{2a} \Big ).
\label{Th+igl}
\end{eqnarray}
Again we have performed appropriate partial integrations. Note that the
terms 
corresponding to the interaction among the gluon fields got cancelled
already. As in the earlier case, we drop the surface terms involving
$\partial^+$. Thus, 
collecting together the results in eq.(\ref{Th+iqaq}) and eq.(\ref{Th+igl}),
we get, 
\begin{eqnarray}
\Theta^{+2} && = 2 i {\psi^+}^\dagger \partial^2 \psi^+ +
{ 1 \over 2} \partial^1({\psi^+}^\dagger \Sigma^3 \psi^+) \nonumber \\
&& ~~ + \partial^+ A^{1a} \partial^2 A^{1a} + \partial^+ A^{2a} \partial^2
A^{2a} + \partial^1 (A^{1a} \partial^+ A^{2a})\, ,
\end{eqnarray}
where we see that the interaction dependent terms get cancelled completely
among themselves.  
By a similar calculation,
\begin{eqnarray}
\Theta^{+1} && = 2 i {\psi^+}^\dagger \partial^1 \psi^+ -
{ 1 \over 2} \partial^2({\psi^+}^\dagger \Sigma^3 \psi^+) \nonumber \\
&& ~~ + \partial^+ A^{1a} \partial^1 A^{1a} + \partial^+ A^{2a} \partial^1
A^{2a} + \partial^2 (A^{2a} \partial^+ A^{1a}).
\end{eqnarray}
From the above two equations it is clear that $\Theta^{+1}$ and
$\Theta^{+2}$ agree with the free field theory form at the operator level.
This shows that
in light-front quantization, with $A^+=0$ gauge, ${\cal J}^3 = J^3$ (the
naive canonical form independent of interactions) at the
operator level, {\it provided} 
the fields vanish at the boundary. Using the form
of $\Theta^{+i}$ in eq.(\ref{j3in}), we obtain 
\begin{eqnarray}
J^3 = J^3_{f(o)}+ J^3_{f(i)}+ J^3_{g(o)}+J^3_{g(i)}\, ,
\end{eqnarray}
with 
\begin{eqnarray}
J^3_{f(o)} &&= \int dx^- d^2 x^\perp {\psi^+}^\dagger i ( x^1 \partial^2 -x^2
\partial^1) \psi^+ , \nonumber \\
J^3_{f(i)}&& = { 1 \over 2} \int dx^- d^2 x^\perp {\psi^+}^\dagger
\Sigma^3 \psi^+, \nonumber \\
J^3_{g(o)}&& = {1 \over 2} \int dx^- d^2 x^\perp \Big \{ 
x^1 [\partial^+A^1 \partial^2 A^1 + \partial^+A^2 \partial^2 A^2]
-x^2 [\partial^+A^1 \partial^1 A^1 + \partial^+A^2 \partial^1 A^2]
\Big \}, \nonumber \\
J^3_{g(i)} && = { 1 \over 2} \int dx^- d^2 x^\perp [ A^1 \partial^+ A^2 - 
A^2 \partial^+ A^1 ].
\label{j3decom}
\end{eqnarray}
The colour indices are implicit in these equations.

Using canonical commutation relations, we can explicitly find that,
\begin{eqnarray}
i \left [ J^{3}_{f(o)}, \psi^{+} (x) \right ] &&= (x^1 \partial^2 - x^2 \partial^1)
\psi^+(x), \label{commu1} \\
i \left  [ J^{3}_{f(i)},\psi^{+}(x) \right ] && = { 1 \over 2} \gamma^1
\gamma^2 \psi^{+}(x),  \label{commu2} \\
i \left [ J^{3}_{g(o)},A^{i}(x) \right ]  && =  (x^1 \partial^2 - x^2 \partial^1) 
A^{i}(x),  \label{commu3} \\
i \left [J^{3}_{g(i)}, A^{i} (x)\right ]  && = - \epsilon_{ij} A^{j} (x) .
\label{commu4}
\end{eqnarray}
Thus, these operators do qualify as angular momentum operators (generators of
rotations in the transverse plane) in the theory\cite{ks}.

To summarize,
the helicity operator constructed from manifestly gauge
invariant, symmetric, energy momentum tensor in QCD, in the gauge $A^+=0$,
and after the elimination of constraint variables, is equal to the naive
canonical form of the light-front helicity operator plus surface terms. In
the topologically trivial sector, we can legitimately require the dynamical
fields to vanish at the boundary. This eliminates the residual gauge degrees
of freedom and removes the surface terms. Thus we have a gauge fixed
Poincare generator which we consider in the following sections.

\section{ Orbital helicity distribution functions}
In this section, we use the gauge fixed helicity operator $J^3$ and its
physical separation as shown in the previous section and define various
structure functions containing the information regarding nucleonic helicity
structure. 
The polarized structure function $g_1(x,Q^2)$ is connected to 
the intrinsic light-front helicity  content of the quarks and anti-quarks as
is evident from the following expression.
We recall the expression for $g_1$ given earlier in Chapter 3,
\begin{eqnarray}
g_1(x,Q^2) = { 1 \over 8 \pi S^+} \int d \eta e^{ - i \eta x}
\sum_\alpha e_\alpha^2 
\langle PS \mid \Big [ {\overline \psi}_\alpha (\xi^- ) \gamma^+ \gamma^5
 \psi_\alpha(0) +
(\xi\leftrightarrow 0) \Big ] \mid PS \rangle 
\label{g1chd},
\end{eqnarray}
with $\eta={1\over 2}P^+\xi^-$. 
By following exactly the same procedure 
as that used in obtaining the sum rule for $F_2$, we get,
\begin{eqnarray}
\int_0^1 dx g_1(x,Q^2) = { 1 \over 2 S^+} \sum_\alpha e^2_\alpha  
\langle PS \mid  \psi^{+\dagger}_\alpha (0) 
\Sigma^3 \psi^+_\alpha(0) \mid PS \rangle 
\label{smg}
\end{eqnarray}
where $ \Sigma^3 = i \gamma^1 \gamma^2$.

One can define the intrinsic helicity distribution function $\Delta q(x,Q^2)$
to be the same as $g_1$ with the weight factors $e_\alpha =1$, 
\begin{eqnarray}
\Delta q(x,Q^2) = { 1 \over 8 \pi S^+} \int d \eta e^{ - i \eta x}
\langle PS \mid \Big [ {\overline \psi} (\xi^- ) \gamma^+ \gamma^5
 \psi(0) +
(\xi\leftrightarrow 0) \Big ] \mid PS \rangle 
\label{dqchd}.
\end{eqnarray}
It follows from eq.(\ref{smg})  that
\begin{eqnarray}
\int_0^1 dx \Delta q(x,Q^2) &&= { 1 \over 2 S^+}  
\langle PS \mid  \psi^{+\dagger}(0) 
\Sigma^3 \psi^+(0) \mid PS \rangle \nonumber\\
&&=
{\langle PS \mid J_{q(i)}^3  \mid PS \rangle \over 2 S^+(2\pi)^3\delta^3(0)}
\nonumber\\
&&=\pm{1\over {\cal N}}\langle PS \mid J_{q(i)}^3  \mid PS \rangle \, ,
\label{smdq}
\end{eqnarray}
where ${\cal N}=2(2\pi)^3P^+\delta^3(0)$. In the last step, we have taken
into account the fact that $S^+=\pm P^+$ for longitudinally polarized target
(only in which case the above formula makes sense).

In  analogy with  eq.(\ref{dqchd}),
we define the orbital helicity distribution for the fermion and gluons as
follows:
\begin{eqnarray}
\Delta q_L(x, Q^2)& =& { 1 \over 4 \pi P^+} \int d \eta e^{ -i  \eta x} 
\langle PS \mid \Big [ {\overline \psi}(\xi^-) \gamma^+ i (x^1 \partial^2
- x^2 \partial^1) \psi(0) + h.c.\Big ] \mid PS \rangle \label {fo}
\nonumber\\
\Delta g_L(x,Q^2) &=& {-1  \over 4 \pi P^+} \int d \eta 
		e^{ - i \eta x} \langle PS \mid \Big [ x^1 F^{+ \alpha}(\xi^-) 
		\partial^2 A_\alpha(0)  - x^2 F^{+ \alpha} (\xi^-) \partial^1 
		A_\alpha(0) \Big ] \mid PS \rangle. 
\nonumber\\
\label{go}
\end{eqnarray}
Here $ \mid P S \rangle $ denotes the
hadron state with momentum $P$ and helicity $S$.  
Similarly one can define the gluon intrinsic light-front helicity 
distribution  
\cite{jgd} as
\begin{eqnarray}
\Delta g(x,Q^2) = -{ i \over 4 \pi (P^+)^2 x} \int d \eta e^{ - i \eta x}
\langle PS \mid F^{+ \alpha} (\xi^-) {\tilde F}^+_{~~\alpha}(0) \mid PS \rangle.
\label {gi}
\end{eqnarray}
The dual tensor is given by 
\begin{eqnarray}
{\tilde F^{\mu \nu}} = { 1 \over 2} \epsilon^{\mu \nu \rho \sigma} F_{\rho
\sigma} ,~~~~
{\rm with} ~~~~ \epsilon^{+1-2} = 2.
\end{eqnarray}
Note that the above distribution functions are defined (as we always do) 
in the
light-front gauge $A^+=0$. 

The above distribution functions are defined in such a way 
that integration of the above distribution functions over $x$ is
directly related to the expectation values of the corresponding helicity
operators analogous to eq.(\ref{smdq}) as follows: 
\begin{eqnarray}
\int_0^1 dx  \Delta q_L(x,Q^2)=&&~ { 1 \over {\cal N}} \langle PS \mid
J^3_{q(o)} 
 \mid PS \rangle\nonumber\\
\int_0^1 dx  \Delta g(x,Q^2)=&&~ { 1 \over {\cal N}} \langle PS \mid  J^3_{g(i)} 
 \mid PS \rangle\nonumber\\
\int_0^1 dx  \Delta g_L(x,Q^2)=&&~ { 1 \over {\cal N}} \langle PS \mid
J^3_{g(o)} 
 \mid PS \rangle.
\end{eqnarray}
Eq.(\ref{go}) constitutes the new definition we proposed for the structure
functions containing the information regarding orbital helicity of quarks
and gluons. Now, in the next section we aim to demonstrate the utility of
these definitions by explicit calculations.

\section{Perturbative calculation of anomalous dimensions} 
Before showing the explicit calculations, we first concentrate a little more
on the various gauge fixed angular momentum operators as defined in
eq.(\ref{j3decom}).  We use the Fourier decomposition of the dynamical
quark and gluon fields  as given in Chapter 2, 
\begin{eqnarray}
\psi^+(x) = \int {dk^+ d^2 k^\perp \over 2 (2 \pi)^3 \sqrt{k^+}} \sum_s
\chi_s
\Big[ b(k,s) e^{-i k.x} + d^\dagger(k,-s) e^{ik.x} \Big ]
\end{eqnarray}
and
\begin{eqnarray}
A^\perp(x) = \int {dk^+ d^2 k^\perp \over 2 (2 \pi)^3 k^+} \sum_\lambda \Big
[
a(k,\lambda) \epsilon^\perp_\lambda e^{-i k.x} + a^\dagger(k,\lambda)
(\epsilon^\perp_\lambda)^* e^{ik.x} \Big ]\, ,
\end{eqnarray}
in terms of which the $J^3$s
become as follows:
\begin{eqnarray}
J^3_{f(o)} && = i \sum_s \int {dk^+ d^2 k^\perp \over 2 (2 \pi)^3 k^+} \Bigg
[
b^\dagger(k,s) \Big [ k^2 { \partial \over \partial k^1} - k^1 { \partial
\over \partial k^2} \Big ] b(k,s)
+ d^\dagger(k,s) \Big [ k^2 { \partial \over \partial k^1} - k^1 { \partial
\over \partial k^2} \Big ] d(k,s) \Bigg ], \nonumber \\
J^3_{f(i)} && = {1 \over 2} \sum_\lambda \lambda
\int {dk^+ d^2 k^\perp \over 2 (2 \pi)^3 k^+} \Bigg
[b^\dagger(k,\lambda) b(k,\lambda)+
 d^\dagger(k,\lambda) d(k,\lambda) \Bigg ], \nonumber \\
J^3_{g(o)} && = i \sum_{\lambda}  \int { dk^+ d^2 k^\perp \over 2 ( 2 \pi)^3
k^+} a^\dagger (k, \lambda) \Big [ k^2 {\partial \over \partial k^1} - k^1
{\partial \over \partial k^2} \Big ] a(k, \lambda), \nonumber \\
j^3_{g(i)} && =  \sum_{\lambda}  \lambda \int { dk^+ d^2 k^\perp \over 2 (
2 \
k^+} a^\dagger (k, \lambda)  a(k, \lambda).
\label{ksfj3}
\end{eqnarray}
In actual calculation we generally replace the momenta of individual quarks
and gluons (in the above operators) 
in terms of relative internal momenta and centre of mass momenta.
We first show how special features of light front transverse boost symmetry
simplifies the calculation and helps understanding some of the features of
the calculation very easily, which had been  
claimed to be very surprizing in the
literature\cite{jth}.
\subsection{Internal orbital helicity: Non-relativistic versus light-front
(relativistic) case}
Here we address the connection between 
non-relativistic situation and the light-front relativistic case.
We need to decompose the total orbital angular momentum of a composite
system as a sum of the orbital angular momentum associated with internal motion and
the orbital angular momentum associated with the center of mass motion. We
are interested only in the former and not in the latter.  
For illustrative purposes, consider a two
body system consisting of two particles with masses $m_1$ and $m_2$ and
momenta ${\bf k_1}$ and ${\bf k_2}$. Let ${\bf P}$ denote the total momentum.
In the non-relativistic case, let ${\bf q}$ denote the relative
momentum, i.e., $ {\bf q} = { m_2 {\bf k_1}  - m_1 {\bf k}_2 \over m_1 +m_2}$. 
It is well-known\cite{chengli} that the contribution of particle
one (two)
to the third component of internal orbital angular momentum is given by
\begin{eqnarray}
L^3_{1(2)} = i {m_{2(1)} \over m_1 + m_2} \Big [ q^2 {\partial \over \partial
q^1} - q^1 {\partial \over \partial q^2} \Big]. \label{nrl3}
\end{eqnarray} 

Next consider the light-front case. 
Let $k_1=(k_1^+,k_1^\perp)$ and $k_2=(k_2^+,k_2^\perp)$ denote the
single particle momenta and  $P=(P^+,P^\perp)$ denote
the total momentum of the two particle system, i.e.,
$k_1^{+,i}+k_2^{+,i}=P^{+,i}$. Light-front kinematics allows us to introduce
boost-invariant internal transverse momentum $q^\perp$ and longitudinal
momentum fraction $ x_i$ by
\begin{eqnarray}
k_1^\perp = q^\perp + x_1 P^\perp, ~~k_1^+ = x_1 P^+, ~~~~
k_2^\perp = -q^\perp + x_2 P^\perp, ~~k_2^+ = x_2 P^+.
\end{eqnarray}
Note that $x_1 + x_2 = 1$ and $ q^\perp = x_2 k_1^\perp - x_1 k_2^\perp$.
For the first particle, we have
\begin{eqnarray}
L^3_{1} && = i \Big [ k_1^2 { \partial \over \partial k_1^1} - k_1^1 { \partial
\over \partial k_1^2} \Big ] \nonumber \\
&& = ix_2 \Big [ q^2 { \partial \over \partial q^1} - q^1 { \partial \over
\partial q^2} \Big ] + 
i x_1 \Big [ P^2 { \partial \over \partial P^1} - P^1 { \partial \over
\partial P^2} \Big ]  \nonumber \\
&& ~~~~ +~i x_1 x_2 \Big [ P^2 { \partial \over \partial q^1} - P^1 { \partial \over
\partial q^2} \Big ] +
i \Big [ q^2 { \partial \over \partial P^1} - q^1 { \partial \over
\partial P^2} \Big ] .
\label{lor1}
\end{eqnarray} 
For the second particle, we have
\begin{eqnarray}
L^3_2 && = i \Big [ k_2^2 { \partial \over \partial k_2^1} - k_2^1 { \partial
\over \partial k_2^2} \Big ] \nonumber \\
&& = ix_1 \Big [ q^2 { \partial \over \partial q^1} - q^1 { \partial \over
\partial q^2} \Big ] + 
i x_2 \Big [ P^2 { \partial \over \partial P^1} - P^1 { \partial \over
\partial P^2} \Big ]  \nonumber \\
&& ~~~~-~i x_1 x_2 \Big [ P^2 { \partial \over \partial q^1} - P^1 { \partial \over
\partial q^2} \Big ] -
i \Big [ q^2 { \partial \over \partial P^1} - q^1 { \partial \over
\partial P^2} \Big ] .
\label{lor2}
\end{eqnarray} 
Total orbital helicity 
\begin{eqnarray}
L^3= L^3_1 + L^3_2 = 
i \Big [ q^2 { \partial \over \partial q^1} - q^1 {\partial \over
\partial q^2}\Big ] + i \Big [ P^2 { \partial \over \partial P^1 } - P^1 
{\partial \over \partial P^2} \Big ].
\end{eqnarray}
Thus we have decomposed the total orbital helicity of a two particle system
into internal orbital helicity and the orbital helicity associated with 
the ``center of mass motion".

Note that the internal orbital helicity carried by particle one is the total
internal helicity multiplied by the longitudinal momentum fraction carried
by particle two and vice versa. This factor can be understood 
by comparison with the
situation in non-relativistic dynamics and recalling the close analogy
between Gallilean relativity and light-front dynamics in the transverse
plane. 
In non-relativistic two-body problem, the center of mass coordinate is
defined by $ \stackrel{\rightarrow}{R} = {m_1 \stackrel{\rightarrow}{r_1} +
m_2 \stackrel{\rightarrow}{r_2} \over m_1 + m_2}$. The generator of Gallilean
boost is $ \stackrel{\rightarrow}{B} = - \sum_i m_i
\stackrel{\rightarrow}{r_i}$.
 Thus in non-relativistic dynamics, $ \stackrel{\rightarrow}{R} =
- {\stackrel{\rightarrow}{B} \over M}$ with $ M= m_1 + m_2$.
In light-front dynamics, the variable analogous to $ B^\perp$ is $E^\perp$,
the generator of transverse boost and the variable analogous to $M$ is   
$P^+$. Thus in light-front theory, the transverse center of mass coordinate 
$ R^\perp = {\sum_i k_i^+ r_i^\perp \over \sum_i k_i^+} = x_1 r_1^\perp +
x_2 r_2^\perp$. Thus we recognize that instead of ${m_2 \over m_1+ m2}$ 
(${m_1 \over m_1+m_2})$ in non-relativistic theory, $x_2$ ($x_1$) appears in
light-front theory. (See the the discussion at the end of this Chapter how
this helps understanding our result very easily.) 

By  comparing light-front (relativistic) and non-relativistic cases, 
we readily see that the
role played by particle masses in individual contributions to the third
component of internal
orbital angular momentum in non-relativistic dynamics is replaced by
longitudinal momentum fractions in relativistic (light-front) theory. 
This also shows that
the physical picture of the third component of internal orbital angular
momentum is drastically different in non-relativistic and relativistic cases.
We stress that it is only the latter, in which parton masses do not appear at
all, that  is of relevance to the nucleon
helicity problem. Lastly, we emphasize that it is 
the transverse boost invariance in light front dynamics that makes 
possible the separation of dynamics associated with the center of mass 
and the internal dynamics. 
{\it In equal-time relativistic theory, this separation
cannot be achieved at the kinematical level since boosts are dynamical}.
\subsection{Dressed Parton Calculations}
In this section, we evaluate the internal helicity distribution functions for
a dressed quark in perturbative QCD by replacing the hadron target by a
dressed quark target. We have provided the necessary details of the
calculation which may serve as the stepping stone for more realistic
calculation with meson target. From this simple calculation, we have 
illustrated how easily one can 
extract the relevant splitting
functions and evaluate the corresponding anomalous dimensions. 
Note that, since we are not interested in exhaustive calculation of various
anomalous dimensions and the purpose of this section being illustrative, 
we can safely drop the derivative of delta function which will 
come naturally in the following
calculations and work explicitly with forward matrix element.

The dressed quark state with fixed helicity can 
be expressed as
\begin{eqnarray}
	|k^+,k_\bot,\lambda \rangle &=& \Phi^{\lambda}(k) b^\dagger_\lambda
		(k)|0\rangle + \sum_{\lambda_1\lambda_2} \int {dk_1^+d^2
		k_{\bot 1}\over \sqrt{2(2\pi)^3 k_1^+}} 
                {dk_2^+d^2k_{\bot 2}\over \sqrt{2 (2\pi)^3k_2^+}} 
     \sqrt{2(2\pi)^3 k^+} \delta^3(k-k_1-k_2) \nonumber \\
	&&~~~~~~~~~~~~~~~~~~~~~~~~~~~~~~~\times \Phi^\lambda
		_{\lambda_1\lambda_2}(k;k_1,k_2)b^\dagger_{\lambda_1}
		(k_1) a^\dagger_{\lambda_2} (k_2) | 0 \rangle + 
		\cdots, \label{dsqs}
\end{eqnarray}
where the normalization of the state is determined by
\begin{equation}
	\langle {k'}^+,k'_\bot,\lambda' |k^+,k_\bot,\lambda \rangle
	= 2(2\pi)^3 k^+ \delta_{\lambda,\lambda'}\delta(k^+-{k'}^+)
	\delta^2(k_\bot-k'_\bot), 
\label{nork}
\end{equation}
We introduce the boost invariant amplitudes $\psi_1^\lambda$ 
and $ \psi^\lambda_{\sigma_1,
\lambda_2}(x,\kappa^\perp)$ respectively by  $\Phi^\lambda(k)=\psi_1^\lambda$ and 
 $\Phi^\lambda_{\lambda_1\lambda_2}(k;k_1,k_2)  
= { 1 \over \sqrt{P^+}}  \psi^\lambda_{\sigma_1
\lambda_2}(x,\kappa^\perp)$. From the light-front QCD Hamiltonian, to lowest
order in perturbation theory, we have (see Appendix C for details),
\begin{eqnarray}
	\psi^\lambda_{\sigma_1\lambda_2}(x,\kappa_\bot) &=& - 
	{g \over \sqrt{2 (2 \pi)^3}}
		T^a {1\over \sqrt{1-x}} 
		{x(1-x)\over \kappa_\bot^2 + m_q^2(1-x)^2}
		\chi^\dagger_{\sigma_1} \Bigg\{2{\kappa_\bot^i \over
		1-x} \nonumber \\
	&& ~~~~~~~~~~~~~ + {1\over x}(\sigma_\bot\cdot \kappa_\bot)
		\sigma^i -im_q\sigma^i{1-x\over x}\Bigg\}
		\chi_\lambda \varepsilon^{i*}(\lambda_2)~\psi_1^\lambda .
 \label{psip}
\end{eqnarray}
Here $x$ is the longitudinal momentum fraction carried by the quark.
We shall ignore the $m_q$ dependence in the above wave function which can
lead to higher twist effects in orbital helicity.
In the following we take the helicity of the dressed
quark to be + $ {1 \over 2}$. Note, due to transverse boost invariance,
without loss of generality, we take the transverse momentum of the initial
quark to be zero. Kinematical nature of the transverse boost invariance also
suggests that the wave-functions 
$\psi^\lambda_{\sigma_1\lambda_2}(x,\kappa_\bot)$ are independent of the
target momenta $k^i$ (transverse component of centre of mass momenta), as is
also evident from eq.(\ref{psip}). Therefore, in the calculation of
structure functions involving orbital motions, only the first term in
eq.(\ref{lor1}) or, eq.(\ref{lor2}) contributes. 

First, we evaluate the intrinsic helicity distribution functions $\Delta
q(x,Q^2)$. As in the calculation of unpolarized structure functions, we use
the Fourier decomposition of the dynamical fields  and use the
dressed quark state eq(\ref{dsqs}) to work out $\Delta q(x,Q^2)$  given 
in eq.(\ref{dqchd}). Explicit
calculation using the standard commutation relation for  
creation  and
annihilation operators of dynamical quarks and gluons fields, gives
\begin{eqnarray}
        \Delta q(x,Q^2) =  \Big \{ {1\over 2}\mid\psi^{\uparrow}_1
	\mid^2\delta (1-x)
         + \sum_{\sigma_1, \lambda_2}\sigma_1 
	\int d^2 \kappa^\perp
                \mid \psi^{\uparrow}_{\sigma_1,\lambda_2}(x,\kappa^\perp)
        \mid^2 \Big \}\, , 
\end{eqnarray}
where we have used $S^+=k^+$ for $\lambda={1\over 2}$ in our case. 
Here and in the following, 
we have replaced the superscript $\lambda$ by $\uparrow$ in the wave-functions
$\psi$'s, since in our calculation it is always $+{1\over 2}$.

In the massless case, using eq.(\ref{psip}) with $m=0$, we arrive at
\begin{eqnarray}
         \sum_{\sigma_1,\lambda_2} \sigma_1\int d^2 \kappa^\perp \mid
                \psi^{\uparrow}_{\sigma_1,\lambda_2}(x,\kappa^\perp)\mid^2 
        = {1\over 2}{g^2 \over (2 \pi)^3}  C_f { 1 + x^2 \over 1-x}
        \int d^2 \kappa^\perp { 1 \over (\kappa^\perp)^2} 
	\mid \psi_1\mid^2\, ,
\label{qdensu}
\end{eqnarray}
where $C_f = {N^2 -1 \over 2N}$.
It is instructive to compare this equation with the corresponding one in 
unpolarized case. 
The transverse momentum integral in Eq. (\ref{qdensu}) is divergent at
both limits of integration.  We regulate the lower limit by $\mu$ and the
upper limit by $Q$. Thus we have
\begin{eqnarray}
        \Delta q(x,Q^2) ={1\over 2} \mid\psi^{\uparrow}
		_1\mid^2  \Big [ \delta (1-x) \,
                + \, {\alpha_s \over 2 \pi}C_f{1+x^2 \over 1-x} \ln{Q^2
                \over \mu^2} \Big].
\label{nqconp}
\end{eqnarray}
Note that $\mu$ has to be large enough so that perturbative calculation is
not invalidated.
Now, $\mid\psi_1\mid^2$ is evaluated using the normalization condition
eq.(\ref{nork}) and proceeding exactly the same way as in the case of
unpolarized structure function $F_2^q$, we get to order $\alpha_s$,
\begin{eqnarray}
        \Delta q(x,Q^2)  = {1\over 2}\Big [\delta(1-x) + {\alpha_s \over 2 \pi}
                C_f \ln{Q^2 \over \mu^2}\Big[ {1 +x^2 \over (1-x)_{+}} +
                {3 \over 2} \delta(1-x) \Big] \Big ]\, .
\label{Dqq}
\end{eqnarray}
Notice that
\begin{equation}
\int^1_0 \Delta q(x,Q^2) = {1\over 2}.
\label{sdqfi}
\end{equation}

Next, we evaluate the gluon intrinsic helicity
distribution function $\Delta g(x,Q^2)$ as 
given in eq.(\ref{gi}) in the dressed quark state.
Using the the Fourier decomposition of the dynamical fields and the dressed
quark target given in eq.(\ref{dsqs}) it easy to see that the 
non vanishing contribution in $\Delta g(x,Q^2)$
comes from the quark-gluon state. A straightforward evaluation gives, 
\begin{eqnarray}
\Delta g (1-x,Q^2) && = \sum_{\sigma_1, \lambda_2} \lambda_2  ~
\int d^2 \kappa^\perp~
{\psi^{\uparrow}_{\sigma_1 \lambda_2}}^*(x, \kappa^\perp) 
{\psi^{\uparrow}_{\sigma_1 \lambda_2}}(x, \kappa^\perp)
\nonumber \\
&& = {\alpha_s \over 2 \pi} C_f \ln{Q^2 \over \mu^2} ~x^2 (1-x)^2
{ 1 \over 1-x} \Big [ { 1 \over x^2 (1-x)^2} - { 1 \over (1-x)^2} \Big ]
\mid \psi_1\mid^2.
\end{eqnarray}
The first (second) term inside the square bracket arises from the 
state with gluon helicity +1 (-1).
To order $\alpha_s$ calculation, we take  $\mid \psi_1\mid^2$ to be unity. So, 
we have the gluon intrinsic helicity contribution in the dressed quark
state, to order $\alpha_s$,
\begin{eqnarray}
\Delta g(1-x,Q^2) = {\alpha_s \over 2 \pi} C_f \ln{Q^2 \over \mu^2}  ~(1+x).
\label{gih}
\end{eqnarray}
Note that the gluon distribution function has the argument $(1-x)$ since we
have assigned $x$ to the quark in the dressed quark state.

Next, we proceed in a same way to evaluate the quark orbital helicity
distribution function given in  eq.(\ref{fo}) in the dressed quark state. 
Again the non vanishing contribution comes from the quark-gluon state. We get,
\begin{eqnarray}
\Delta q_L(x,Q^2) && = \sum_{\sigma_1, \lambda_2}   ~\int d^2 \kappa^\perp
~(1-x)~
{\psi^{\uparrow}_{\sigma_1 \lambda_2}}^*(x, \kappa^\perp) 
(- i {\partial \over \partial \phi})
{\psi^{\uparrow}_{\sigma_1 \lambda_2}}(x, \kappa^\perp)  \nonumber \\
&& = -{\alpha_s \over 2 \pi} C_f \ln{Q^2 \over \mu^2}  
(1-x) x^2 (1-x)^2
{ 1 \over 1-x} \Big [ { 1 \over x^2 (1-x)^2} - { 1 \over (1-x)^2} \Big ]
\mid \psi_1\mid^2.
\end{eqnarray}
The first (second) term inside the square bracket arises from the 
state with gluon helicity +1 (-1). Note that for convenience in calculation,
we have changed the variable $\kappa^\perp$: $(\kappa^1,
\kappa^2)\rightarrow (\mid\kappa^\perp\mid, \phi)$. 
Thus we have the quark orbital helicity contribution in the dressed quark
state to order $\alpha_s$ (and therefore, taking $\mid \psi_1\mid^2=1$),
\begin{eqnarray}
\Delta q_L(x,Q^2) = - {\alpha_s \over 2 \pi} C_f \ln{Q^2 \over \mu^2}   ~(1-x)
(1+x).
\label{qoh}
\end{eqnarray}
Similarly, we get the gluon orbital helicity distribution defined in 
eq.(\ref{go}) in the dressed
quark state as  
\begin{eqnarray}
\Delta g_L(1-x,Q^2) && = \sum_{\sigma_1, \lambda_2}   ~\int d^2 \kappa^\perp
~~x~
{\psi^{\uparrow}_{\sigma_1 \lambda_2}}^*(x, \kappa^\perp) 
(- i {\partial \over \partial \phi})
{\psi^{\uparrow}_{\sigma_1 \lambda_2}}(x, \kappa^\perp)  \nonumber \\
&&= -{\alpha_s \over 2 \pi} C_f \ln{Q^2 \over \mu^2}  ~ x
(1+x)\, ,
\label{goh}
\end{eqnarray}
where we have already taken $\mid \psi_1\mid^2= 1$ here for the order
$\alpha_s$ calculation.

We note that the helicity is conserved at the quark gluon vertex. 
For the initial quark of zero transverse momentum, total helicity of the
initial state is the intrinsic helicity of the initial quark, namely, $+ { 1
\over 2}$ in our case.
Since we have neglected quark mass effects, the final quark also has the
same 
intrinsic helicity $+ { 1\over 2}$ as is also evident from eq.(\ref{sdqfi}). 
Thus total helicity conservation implies
that the contributions from gluon intrinsic helicity and quark and gluon
internal orbital helicities have to cancel. This is readily verified using eqs.
(\ref{gih}), (\ref{qoh}), and (\ref{goh}).

From eqs. (\ref{gih}), (\ref{qoh}) 
and (\ref{goh}) we extract the relevant splitting
functions. The splitting functions are 
\begin{eqnarray}
P_{SS(gq)}(1-x) ~ &&= C_f~(1+x) ,\nonumber \\
P_{LS(qq)} (x) ~ &&= -~C_f~(1-x^2), \nonumber \\
P_{LS(gq)}(1-x) ~ &&= -C_f ~ x~(1+x).
\label{splorb}
\end{eqnarray}
We define the anomalous dimension $ A^n = \int_0^1 dx x^{n-1} P(x)$.
Thus, we can easily work out  the set of
corresponding anomalous dimensions using
eq.(\ref{splorb}) and these are given by 
\begin{eqnarray} 
A^n_{SS(gq)} = C_f ~ { n+2 \over n(n+1)}, ~~
A^n_{LS(qq)} = -~ C_f ~ { 2 \over n(n+2)}, ~~ A^n_{LS(gq)} = -~ C_f ~{ n+4
\over n (n+1) (n+2)}.
\end{eqnarray}
It is to be noted that these anomalous dimensions agree with those given 
in the recent work of 
H\"{a}gler and Sch\"{a}fer\cite{sch}.
\section { Verification of helicity  sum rule}
Helicity sum rule for the fermion target is given by
\begin{eqnarray}
{ 1 \over {\cal N}} \langle PS \mid \Big [ J^3_{q(i)} + J^3_{q(o)} +
J^3_{g(i)} + J^3_{g(o)} \Big ] \mid PS \rangle = \pm { 1 \over 2}.
\end{eqnarray}
For boson target RHS of the above equation should be replaced by the
corresponding helicity.

Here we verify the correctness of our definitions of distribution functions
in the context of
helicity sum rule for a dressed quark as well as
a dressed gluon target perturbatively. This is necessary as a consistency
check of our results as well. We use the
operator $J^3$s given in eq.(\ref{ksfj3}) and calculate their matrix elements
in the dressed quark and gluon states. For simplicity, we take the external
transverse momenta of the target to be zero so that there is no net angular
momentum associated with the center of mass of the target.

Here we present the final results omitting all the  
details of calculations (see Appendix C). 
For a dressed quark target having helicity $+{1\over2}$ we get,
\begin{eqnarray}
{1\over {\cal N}}\langle P , \uparrow \mid J^3_{f(i)} \mid P, \uparrow
\rangle_q = && \int dx  \Big[ {1\over 2}\delta(1-x) + {\alpha\over 2\pi}C_f
\ln{Q^2\over \mu^2}\big[ {1+x^2 \over (1-x)_+}
+{3\over2}\delta(1-x)\big]\Big]\nonumber\\
= &&~~~{1\over2}\nonumber\\
{1\over {\cal N}}\langle P , \uparrow \mid J^3_{f(o)} \mid P, \uparrow
\rangle_q = && - {\alpha\over 2\pi}C_f
\ln{Q^2\over \mu^2}\int dx   ~(1-x)~(1+x)\nonumber\\
{1\over {\cal N}}\langle P , \uparrow \mid J^3_{g(i)} \mid P, \uparrow
\rangle_q = &&~~~  {\alpha\over 2\pi}C_f
\ln{Q^2\over \mu^2}\int dx  ~ (1+x)\nonumber\\
{1\over {\cal N}}\langle P , \uparrow \mid J^3_{g(o)} \mid P, \uparrow
\rangle_q = && - {\alpha\over 2\pi}C_f
\ln{Q^2\over \mu^2}\int dx ~ x ~(1+x).
\end{eqnarray}
Adding all the contributions, we get,
\begin{equation}
{1\over {\cal N}}\langle P , \uparrow \mid J^3_{f(i)}
+J^3_{f(o)}+J^3_{g(i)}+J^3_{g(o)}
\mid P, \uparrow
\rangle_q =~{1\over2}\, .\label{srlqu}
\end{equation}
For a dressed gluon having helicity $+1$, the corresponding expressions 
are worked out to be the
following.
\begin{eqnarray}
{1\over {\cal N}}\langle P , \uparrow \mid J^3_{f(i)} \mid P, \uparrow
\rangle_g = &&~~~0\nonumber\\
{1\over {\cal N}}\langle P , \uparrow \mid J^3_{f(o)} \mid P, \uparrow
\rangle_g = &&  {\alpha\over 2\pi}N_fT_f
\ln{Q^2\over \mu^2}\int dx ~[x^2+ (1-x)^2]\nonumber\\
{1\over {\cal N}}\langle P , \uparrow \mid J^3_{g(i)} \mid P, \uparrow
\rangle_g = &&~~~ \psi_1^*\psi_1 \nonumber\\
=&&~~~1-  {\alpha\over 2\pi}N_fT_f
\ln{Q^2\over \mu^2}\int dx~  [x^2+ (1-x)^2]\nonumber\\
{1\over {\cal N}}\langle P , \uparrow \mid J^3_{g(o)} \mid P, \uparrow
\rangle_g = &&~~~0 
\end{eqnarray}
Adding all the contributions, we get,
\begin{equation}
{1\over {\cal N}}\langle P , \uparrow \mid J^3_{f(i)}
+J^3_{f(o)}+J^3_{g(i)}+J^3_{g(o)}
\mid P, \uparrow
\rangle_g =~1\, .\label{srlgl}
\end{equation}
Thus, eq.(\ref{srlqu}) and eq.(\ref{srlgl}) clearly show that the sum
rules are satisfied.  
Note that in evaluating the above expression, we have used the Fourier
decomposition of the dynamical fields and the 
Fock-expansion for the target states. For the dressed quark we have used
eq.(\ref{dsqs}), while for gluon we have used similar expansion but ignored
two-gluon Fock sector for simplicity.   

Before concluding, let us summarize what we have presented  in this Chapter.
We have presented a detailed analysis of the
light-front helicity operator (generator of rotations in the transverse
plane) in QCD.  
We have explicitly shown that,
the operator constructed from manifestly gauge
invariant, symmetric energy momentum tensor in QCD, in the gauge $A^+=0$,
and after the elimination of constraint variables, is equal to the naive
canonical form of the light-front helicity operator plus surface terms. In
the topologically trivial sector, we can legitimately require the dynamical
fields to vanish at the boundary. This eliminates the residual gauge degrees
of freedom and removes the surface terms. 

Next, we have defined non-perturbative quark and gluon orbital helicity 
distribution functions as Fourier
transform of forward hadron matrix elements of appropriate bilocal operators
with bilocality only in the light-front longitudinal direction.
We have calculated these distribution functions by replacing the hadron
target by a dressed parton providing all the necessary details. 
From these simple calculations we have 
illustrated the utility of the newly defined distribution functions in the 
calculation of
splitting functions and hence the anomalous dimensions in perturbation theory. 
We have also verified the helicity sum rule explicitly to the first non-trivial
order in perturbation theory. 
 
We also have compared and contrasted the expressions 
for internal
orbital helicity in non-relativistic and light-front (relativistic) cases. 
Our calculation shows that the role played by particle masses in the internal
orbital angular momentum in the non-relativistic case is replaced by the
longitudinal momentum fraction in the relativistic case.
Although four terms appear in the expression of $L_3$ for
individual particles in two body system, only the term proportional to the
total internal $L_3$ contributes due to transverse boost invariance of the
multi-parton wave-function in light-front dynamics. We also note the occurrence 
of the longitudinal
momentum fraction $x_2$($x_1$) multiplied by the total internal $L_3$ in the 
expressions
of $L_3$ for particle one(two). This explains why one needs to take first moment with
respect to $x$ as well as $(1-x)$ for the respective distributions in
obtaining the helicity sum rule\cite{jth}.

It should be emphasized that 
our explicit demonstration here, that the gauge-fixed light-front helicity 
operator 
is exactly equal to the naive
canonical form, is facilitated by the
fact that in light-front theory only transverse gauge fields are dynamical
degrees of freedom. The conjugate momenta (color electric fields) are
constrained variables in the theory. Thus we 
were able to show explicitly that the
resulting gauge fixed operator is free of interactions.  
The question naturally arises as
to whether this result is valid in other gauges also. Several years ago, in
the context of magnetic monopole solutions, 
it has been shown\cite{cgw} that in
Yang-Mills-Higgs system, quantized in the axial gauge $A_3=0$ using the
Dirac procedure, the angular momentum operator constructed from manifestly
gauge invariant symmetric energy momentum tensor differs from the canonical
one only by surface terms. 
In the study of QCD in $A_3 =0$ gauge, it has been shown\cite{bg} that
in the presence of surface terms, Poincare algebra holds only in the
physical subspace.  
The situation in $A^0=0$ gauge or
in covariant gauges where unphysical degrees of freedom are present is to be
investigated. Another interesting problem to be studied 
is the helicity conservation in
the topologically non-trivial sector of QCD and its implications, if any,
for deep inelastic scattering.   

\newpage
\section*{\large\bf {Summary and Conclusion}}
In this thesis, we have tried to put 
forth some of the aspects of light-front
field theory through their successful application in a physical problem. The
physical problem that we have chosen is DIS, which is strongly 
believed to have a
convenient description in light-front language but seldom pursued
consistently. We have developed a light-front QCD Hamiltonian
description of the DIS structure functions consistently working in the
light-front co-ordinates and light-front gauge and using light-front
QCD Hamiltonian as dictating the 
underlying dynamics. In retrospect, the importance of our work can be
summarized from two different viewpoints as follows. 

As far as DIS is concerned, our approach is convenient in the sense that it
closely follows our physical intuitions as is the case in parton model where
DIS has a simple 
partonic interpretation. 
Our approach differs from the usual covariant  one in an
essential way that it addresses directly the structure functions, which are
experimental objects, instead of its moments which come naturally in the
usual way. More importantly, our approach gives a complete description of the
structure functions in the sense that it has the potential of incorporating
the non-perturbative contents of the structure functions. 
We have shown introducing a new factorization scheme that the
non-perturbative contents of the DIS structure functions can be obtained by 
solving the
light-front bound state equations, which seems viable due to the ongoing
research activity for solving QCD bound state problem in light-front using the
similarity renormalization scheme. On the other hand, 
the perturbative contents can be extracted by calculating 
the dressed parton structure functions as we showed by explicitly
working them out to the leading order in QCD coupling. 
In contrast, the usual approaches are so designed that one 
only parametrizes the nonperturbative information and the whole
concentration is put on the $Q^2$-evolution of the moments of the structure
functions using perturbation theory. 

Simplicity of our approach becomes evident when we try to
describe structure functions in the context of the nucleonic helicity by
defining new structure functions in our approach. The ambiguity of gauge
invariance and interaction dependence in defining various parts of the 
helicity operator for
quarks and gluons in the usual way are absent in our gauge fixed theory
and the well known helicity operator in light-front seems to provide the
consistent physical information as we have shown explicitly. 

On the other hand, although the study in light-front field theory has been
around for quite some decades, there is still some doubt whether it produces
the {\it same} 
result as the more familiar approach, since the formulation and the
methods used in two cases are so different. 
Perturbative calculations of the dressed parton structure functions presented 
here show that,
to the leading order in coupling, our approach yields equivalent results as in
the usual way. This constitute a proof of equivalence of light-front field
theory and the more familiar equal time field theory, even though the  
differences in formulation and the methods used in the two cases make
the equivalence not so apparent. Our investigation in the context of
coupling constant renormalization in light-front QCD hamiltonian methods
shows the importance of Gallilean boost symmetry in understanding the
correctness of any higher order calculation using time($x^+$)-ordered
old-fashioned perturbation theory. 

Our study, of course, leaves open several issues that needs further
investigation. Firstly, to add more substance to our proof of
equivalence between light-front field theory and the equal time one, the
dressed parton structure function calculation can be extended to the next
higher order. This may turn out not only to 
serve as a check between various methods
but also to 
give us the opportunity to investigate the intricacies involved in
the two-loop calculation using light-front perturbation theory.\footnote {
We prefer not to put any epilogue simply because it is a dissertation of an
on going research work. And, in general, the quest for knowledge goes on and
on. }

\newpage
\appendix
\section{Notation, Conventions, and Useful Relations}
\newcommand{\be}{\begin{equation}}
\newcommand{\ee}{\end{equation}}
\noindent Light-Front variables are defined in terms of 
${x}^{\mu} = (x^{0}, x^{3}, x^{1}, x^{2})$ as   
\begin{equation} 
x^{\pm}= x^{0} \pm x^{3} \; , \; \; \; x^{\perp}= (x^{1},x^{2})  . 
\end{equation}
Let us denote the four-vector $x^\mu$ by
\be x^{\mu} = (x^{+},x^{-},x^{\perp}) . \ee
Scalar product $x^\mu x_\mu$ is defined as 
\begin{equation}
 x.y = {1 \over 2} x^{+}y^{-}+{1 \over 2}x^{-}y^{+}-
x^{\perp}.y^{\perp}  . 
\end{equation}
The metric tensors are as follows
\be g^{\mu \nu} = \left(\begin{array}{lrrr}0 & 2 & 0 & 0 \\
                          2 &\,\, 0 & 0 & 0 \\
			  0 & 0 & -1 & 0 \\
			  0 & 0 & 0 & -1 \end{array}\right) \, ,
\,\,\,\quad\quad 
g_{\mu \nu} = \left(\begin{array}{lrrr}0 & {1 \over 2} & 0 & 0 \\
                          {1 \over 2} & \,\,0 & 0 & 0 \\
			  0 & 0 & -1 & 0 \\
			  0 & 0 & 0 & -1 \end{array}\right) \, , \ee
so that
  \be
 x_{-}= {1 \over 2} x^{+}  , \; \; x_{+} = {1 \over 2} x^{-} . \ee 			  
\noindent Similarly for the partial derivatives:
\be \partial^{+}= 2 \partial_{-}= 2 {\partial \over \partial x^{-}}  .\ee
\be \partial^{-}= 2 \partial_{+}= 2 {\partial \over \partial x^{+}}  . \ee
\noindent Four-dimensional volume element:
\be d^{4}x= dx^{0} d^{2}x^{\perp} dx^{3}  = {1 \over 2} dx^{+} dx^{-} 
d^{2}x^{\perp}  .\ee
\noindent Lorentz invariant volume element in momentum space:
\be [d^{3}k] = {dk^{+} \; d^2 k^{\perp} \over 2 (2 \pi)^{3} k^{+}} . \ee
\noindent The step function 
\begin{eqnarray}
  \theta (x) && = 0, \; \; \; x < 0 \nonumber \\ 
           &&   = 1, \; \; \; x > 0 .  
\end{eqnarray}
The antisymmetric step function 
\be \epsilon(x) = \theta(x) - \theta(-x)  . \ee
\be {\partial \epsilon \over \partial x} \; \; =
2 \; \delta(x)   \ee
where $\delta(x)$ is the Dirac delta function.
\be \mid x \mid \; \; = \; \; x \; \epsilon(x)  . \ee
\noindent We define the integral operators
\be {1 \over \partial^{+}} f(x^{-})= {1 \over 4}\, \int \, dy^{-}
\epsilon(x^{-}-y^{-}) \, f(y^{-})  , \ee
\be  ({1 \over \partial^{+}})^{2} f(x^{-}) =  { 1 \over 8} \, \int \,
dy^{-} \mid x^{-} - y^{-} \mid \, f(y^{-})  . \ee
\noindent Unless otherwise specified, we
choose the Bjorken and Drell convention for the gamma matrices:
\be \gamma^{0} = \beta = \left(\begin{array}{cr} I &  0 \\
                                    0 & -I \end{array}\right) ,\,\,\,
\quad\quad{\vec \gamma} = \left(\begin{array}{cr}0 &\, {\vec \sigma} \\
                                 -{\vec \sigma} & 0 \end{array}\right) \, , \ee
where ${\vec \sigma}$s are Pauli matrices
\be \sigma_{1}=\left(\begin{array}{cr}0  & \, 1 \\
                        1  &  0 \end{array}\right)  ,\,\,
\quad\quad \sigma_{2}= \left(\begin{array}{cr}0 & -i \\
                        i & 0 \end{array}\right)  ,\,\, 
\quad\quad \sigma_{3} =\left(\begin{array} {cr}1 & 0 \\
                          0 & -1 \end{array}\right) . \ee										 
\be \gamma^{5}\; = \; i \gamma^{0} \gamma^{1} \gamma^{2} \gamma^{3} =
\pmatrix{0 & I \cr
         I & 0 \cr} \ee
\be {\vec \alpha}  \; = \; \gamma^{0} {\vec \gamma} . \ee
\be \gamma^{\pm} \; \; = \gamma^{0} \pm \gamma^{3}  . \ee
Projection operators are defined as 
\be  \Lambda^{\pm}  \; =  \; {1 \over 4} \gamma^{\mp} \gamma^{\pm}
 = {1 \over 2} \gamma^{0} \gamma^{\pm} \; = \; {1 \over 2}
(I \pm \alpha^{3})  . \ee
such that
\be \Lambda^{+} \; \; +  \;\; \Lambda^{-} \;\;  = \; \; I\, ,\quad 
(\Lambda^{\pm})^2 \; \; = \; \; \Lambda^{\pm}  \, ,\quad 
\Lambda^+\Lambda^-\;\;=\;\;0\, ,\quad 
(\Lambda^{\pm})^{\dagger} \; \; = \; \; \Lambda^{\pm} \, 
   . \ee
and they satisfy the following relations
\be \gamma^{\perp} \; \Lambda^{\pm} \; \; 
= \; \; \Lambda^{\pm} \gamma^{\perp}  \, ,\quad
 \gamma^{0} \; \Lambda^{\pm} \; \; = \; \; \Lambda^{\mp} \gamma^{0}
 \, ,\quad   	 
 \gamma^{5} \; \Lambda^{\pm} \; \; 
= \; \; \Lambda^{\pm} \gamma^{5} \, . \ee
\section{  
An expansion of $T^{\mu\nu}$ in inverse power of light-front energy of
	 the virtual photon}
In this appendix we show how the virtual photon hadron forward Compton 
scattering can be obtained as an expansion in the inverse power of light
front energy of the virtual photon $q^-=q^0-q^3$. This is the light front
version of what is known as Bjorken-Johnson-Low limit. Explicitly, the forward 
virtual photon hadron Compton scattering amplitude is given by,
\begin{eqnarray}	
	T^{\mu \nu} &=& i \int d^4\xi e^{iq\cdot \xi} \langle PS | 
		T(J^\mu(\xi) J^\nu(0)) |PS \rangle \nonumber\\
              &=&  i \int d^4\xi \big( {1 \over iq^-} \partial^- e^{iq\cdot \xi} \big) \langle PS | 
		\theta(\xi^+)\big[ J^\mu(\xi), J^\nu(0))\big] |PS \rangle 
\end{eqnarray}
Here, in the second line we have changed the time ordered product into a
commutator and introduced a derivative operation
$\partial^-\equiv 2 {\partial \over \partial
\xi^+}$ on the
exponential which does not alter the expression at all. Now, doing a partial
integration we get,
\begin{eqnarray}
T^{\mu \nu} &=& - {1\over q^-}\int d^4\xi e^{iq\cdot \xi} \partial^-\langle PS|
\theta(\xi^+)\big[J^\mu(\xi), J^\nu(0)\big] |PS \rangle \nonumber\\
&=& - {1\over q^-}\int d^4\xi e^{iq\cdot \xi} \langle PS|2
\delta(\xi^+)\big[J^\mu(\xi), J^\nu(0)\big] |PS \rangle\nonumber\\
& & \quad\quad - {1\over q^-}\int d^4\xi e^{iq\cdot \xi} \langle PS|
\theta(\xi^+)\big[\partial^- J^\mu(\xi), J^\nu(0)\big] |PS \rangle
\label{inter}
\end{eqnarray}
First term is integrated using the delta function and the fact that $d^4\xi
\equiv {1\over 2}d\xi^+d\xi^-d^2\xi^\perp$ and with the second term we
follow exactly the same procedure applied earlier, to obtain the following
result.
\begin{eqnarray}
T^{\mu \nu} &=& - {1\over q^-}\int d\xi^-d^2\xi^\perp 
e^{iq\cdot \xi} \langle PS|
\big[J^\mu(\xi), J^\nu(0)\big]_{\xi^+=0} |PS \rangle \nonumber\\
& & \quad - {1\over q^-}\int d^4\xi e^{iq\cdot \xi} \langle PS|2
\delta(\xi^+)\big[i\partial J^\mu(\xi), J^\nu(0)\big] |PS \rangle\nonumber\\
& & \quad\quad - {1\over q^-}\int d^4\xi e^{iq\cdot \xi} \langle PS|
\theta(\xi^+)\big[ \partial^-(i\partial^- J^\mu(\xi)), J^\nu(0)\big] 
|PS \rangle
\end{eqnarray}
Notice that the first line of the RHS contains the equal-$x^+$ commutator
and constitutes the first term in the BJL expansion, whereas second and
third line ressemble with that of eq.(\ref{inter}).
Thus, iterating the above procedure we finally get an expansion of
$T^{\mu\nu}$ in the inverse power of $q^-$ as follows.
\begin{equation} \label{lfccap}
	T^{\mu \nu} = - \sum_{n=0}^\infty \Big({1\over q^-}
		\Big)^{n+1} \int d\xi^- d^2 \xi_\bot e^{iq\cdot \xi}
	  \langle PS | [(i\partial_\xi^-)^nJ^\mu(\xi), J^\nu(0)]_{
		\xi^+=0}| PS\rangle \, ,
\end{equation}
The above expansion shows that the time-ordered matrix element can 
be expanded in terms of an infinite series of equal light-front time 
commutators.
\section{ Details of one loop calculations}
In this appendix, we discuss  how the vertex functions (VF) and
energy denominators (ED) in the light-front 
Hamiltonian perturbation theory  are used in one
loop calculations providing necessary details. 
In particular, we shall work out the VF and ED 
relevant in the context of $F_2$ calculation and then show various
summation procedure which may be necessary in other cases as well. 

Recall that the dressed quark state $\mid P \sigma\rangle_q$ truncated at
the two particle level, is given by,
\begin{eqnarray}
        \mid P \sigma \rangle_q = && \sqrt{{\cal N}_q} \Bigg \{
                b^\dagger(P,\sigma) \mid 0 \rangle \nonumber \\
        && ~ + \sum_{\sigma_1,\lambda_2} \int {dk_1^+ d^2 k_1^\perp \over
                \sqrt{2 (2 \pi)^3 k_1^+}} \int {dk_2^+ d^2 k_2^\perp
                \over \sqrt{2 (2 \pi)^3 k_2^+}} \psi_2(P,\sigma
                \mid k_1, \sigma_1; k_2 , \lambda_2)   \nonumber \\
        && ~~~~~~~~~~~~~~~~~~~ \times \sqrt{2 (2 \pi)^3 P^+}
                \delta^3(P-k_1-k_2) b^\dagger(k_1,\sigma_1)
                a^\dagger(k_2,\lambda_2) \mid 0 \rangle \Bigg \},
\label{apdsq}
\end{eqnarray}
which satisfies the light-front version of the Schroedinger equation,
\begin{equation}
P^-_{QCD} |P \sigma \rangle = {P_\bot^2 + M^2 \over P^+}
                | P \sigma \rangle  \, ,
\label{apsch}
\end{equation}
where $P^-_{QCD}=P^-_0+V$ with  $P^-_0$ and $V$ being the free and interaction
parts of the LFQCD Hamiltonian respectively. Introduce the Jacobi momenta
$(x_i,\kappa^\perp_i)$ 
\begin{eqnarray}
k_i^+ = x_i P^+, \,  k_i^\perp = \kappa_i^\perp + x_i P^\perp
\end{eqnarray}
so that
\begin{eqnarray}
\sum_i x_i = 1, \sum_i \kappa_i^\perp =0.
\end{eqnarray}
Also, introduce the boost invariant
amplitude $ \Phi_2$ as
\begin{eqnarray}
\sqrt{P^+} \psi_2(k_i^+, k_i^\perp) = &&  \Phi_2(x_i,
\kappa_i^\perp).
\end{eqnarray}
Now substituting the dressed quark state eq.(\ref{apdsq}) into
eq.(\ref{apsch}) and taking a projection on a bare one quark- one gluon
state $\mid 2\rangle \equiv 
b^\dagger(p_1,\sigma_1)a^\dagger(p_2,\lambda_2)\mid 0\rangle$ we get, 
\begin{equation}
\langle 2\mid \big [{M^2 + (P^\perp)^2\over P^+} -P^-_0\big ] \mid
P\sigma\rangle = \langle 2\mid V_{qqg}\mid P \sigma\rangle
\label{approj}
\end{equation}
Notice that the operator on the LHS being free of interaction $\mid
2\rangle$ projects out $\psi_2$ and to the order $g$ RHS gets contribution
only from $V_{qqg}$. Straight-forward calculation starting from
 eq.(\ref{approj}) using
the Fourier decomposition of the dynamical fields and using the standard
commutation relation for dynamical quarks and gluon fields, we get,
\begin{eqnarray}
     &&  \Big[ {M^2+ (P^\perp)^2 \over P^+} - {m^2 + (p_1^\perp)^2 
      \over p_1^+} - {(p_2^\perp)^2 \over p_2^+} \Big] \psi_2(P, 
	\sigma \mid p_1,\sigma_1; p_2,\lambda_2) =  \nonumber \\
        &&~~~~~~~~~~ { g \over \sqrt{2 (2 \pi)^3}} T^a {1 \over
                \sqrt{p_2^+}}\chi^\dagger_{\sigma_1} \Big[ 2 {p_2^\perp \over
                p_2^+} - {\sigma^\perp.p_1^\perp - im \over p_1^+}
                \sigma^\perp - \sigma^\perp {\sigma^\perp.P^\perp + im
                \over P^+}  \Big] \chi_\sigma .
		{(\epsilon^\perp_{\lambda_2})}^*.
\end{eqnarray}
where $M$ and $m$ are the masses of the dressed quark and bare quark
respectively. 
The factor multiplying the $\psi_2$ is the known as the ED and the RHS
constitute the relevant VF here. 
We rewrite the above equation in terms of Jacobi momenta
($ p_i^+ = x_i P^+, \kappa^\perp_i =p^\perp_i + x_i P^\perp$) and the
wave-functions $\Phi_i$ which are functions of Jacobi momenta. Using the
notation $x=x_1, \kappa_1 = \kappa$ and using the facts $x_1+x_2=1,
\kappa_1+\kappa_2=0$, we have
\begin{eqnarray}
        && \Phi^\sigma_{2\,\sigma_1, \lambda_2}
		(x,\kappa^\perp; 1-x, - \kappa^\perp)
                = { 1 \over \Big[ M^2 - {m^2 +(\kappa^\perp)^2 \over x } -
                {(\kappa^\perp)^2 \over 1-x} \Big] } \nonumber \\
        &&~~~~~~~~~~\times   { g \over \sqrt{2 (2 \pi)^3}} T^a {1\over \sqrt
		{1-x}}
                \chi^\dagger_{\sigma_1} \Big[ - 2 {\kappa^\perp \over 1-x} -
                {\sigma^\perp.\kappa^\perp  -i m \over x} \sigma^\perp -
                \sigma^\perp i m \Big] \chi_{\sigma} .{(\epsilon^\perp_{
                \lambda_2})}^*\, . 
\end{eqnarray}

Now, we have mostly used $\Phi_2$ in the massless cases $M=m=0$, 
where it simplifies to
\begin{eqnarray}
         \Phi^\sigma_{\sigma_1, \lambda_2}
		(x,\kappa^\perp)
                = -  { g \over \sqrt{2 (2 \pi)^3}} T^a {1\over \sqrt
		{1-x}}{ x(1-x) \over (\kappa^\perp)^2 }
                \chi^\dagger_{\sigma_1} \Big[  2 {\kappa^\perp \over 1-x} +
                {\sigma^\perp.\kappa^\perp  \over x} \sigma^\perp 
                \Big] \chi_{\sigma} .{(\epsilon^\perp_{
                \lambda_2})}^*. 
\label{apvf}
\end{eqnarray}
The above example shows how easily one can obtain various two-particle
wave-functions in terms of VF and ED. 

Now, to calculate various structure functions with a dressed quark target, we 
encountered this wave-function $\Phi_2$ where we needed to perform different 
summation over helicities. For example,
\begin{eqnarray}
{\rm for}~~& F^{|q\rangle}_{2q}(x,Q^2) 
&\quad \rightarrow {1\over 2} \sum_\sigma
\sum_{\sigma_1,\lambda_2} \mid \Phi^\sigma_{\sigma_1, \lambda_2}\mid^2\, ,
\label{aps1}\\
{\rm for}~~& \Delta q(x,Q^2) &\quad \rightarrow 
\sum_{\sigma_1,\lambda_2}\sigma_1 \mid \Phi^\sigma_{\sigma_1, \lambda_2}\mid^2\, ,
\label{aps2}\\
{\rm for}~~& \Delta g(1-x,Q^2) &\quad \rightarrow 
\sum_{\sigma_1,\lambda_2}\lambda_2 \mid \Phi^\sigma_{\sigma_1, \lambda_2}\mid^2\, .
\label{aps3}
\end{eqnarray}
Here we provide the necessary details how these are actually worked out 
after calculating $\Phi_2$ explicitly for various helicity configurations. 
Here, we denote $\sigma =+{1\over 2}$ as $(\uparrow)$ and so on which are
self-indicating. 
To perform the summation over $\sigma_1$ in the above
expressions, we notice that 
in $\Phi_2$, $2\times2$ matrix in helicity space sandwiched between
$\chi^{\dagger}_{\sigma_1}$ and $\chi_{\sigma}$ is a diagonal matrix. Thus,
the matrix element will be zero unless $\sigma_1=\sigma$ irrespective of the
gluon polarization $\epsilon_{\lambda_2}$. 
That implies, taking $\sigma=+{1\over 2}$
\begin{eqnarray}
\Phi^\uparrow_{\downarrow\uparrow}(x,Q^2)&& =0\nonumber\\
\Phi^\uparrow_{\downarrow\downarrow}(x,Q^2)&& =0\, .
\label{apf1}
\end{eqnarray}
For convenience, we change the variable as $\{\kappa_1\rightarrow |\kappa| 
{\cos
\phi}{\rm ,}~ \kappa_2\rightarrow |\kappa| {\sin
\phi}\}$, so that $\kappa_\perp^2=|\kappa|^2$ and $
\kappa\cdot\epsilon^*_\uparrow = \kappa_1-i\kappa_2= |\kappa|~e^{-i\phi}$,
and so on.
For $\sigma_1=\sigma=+{1\over 2}$ and $\lambda_2=1$, 
$\Phi_2$ can be written as 
\begin{eqnarray}
\Phi^\uparrow_{\uparrow\uparrow}&&=(\cdot\cdot\cdot) \chi^\dagger_
\uparrow\Big [ 
{2(\kappa\cdot\epsilon^*_\uparrow)\over 1-x} +
{(\kappa\cdot\sigma)(\sigma\cdot\epsilon^*_\uparrow)\over x} \Big]
\chi_{\uparrow}\nonumber\\
&&=(\cdot\cdot\cdot){1\over\sqrt 2}~ \chi^\dagger_{\uparrow}\Big [ 
{2|\kappa|~e^{-i\phi}\over 1-x} +
{|\kappa|~e^{-i\phi} + \sigma^3|\kappa|~e^{-i\phi}\over x} \Big]
\chi_\uparrow\nonumber\\
&&=(\cdot\cdot\cdot){1\over\sqrt 2}~  
{|\kappa|~e^{-i\phi}\over x(1-x)} \Big [
2x + (1-x)(1+1) \Big]
\nonumber\\
&&= (\cdot\cdot\cdot){\sqrt 2} ~ 
{|\kappa|~e^{-i\phi}\over x(1-x)} 
\label{apf2}
\end{eqnarray}
Similarly, for $\lambda_2=-1$, we have,
\begin{eqnarray}
\Phi^\uparrow_{\uparrow\downarrow}&&=(\cdot\cdot\cdot) 
\chi^\dagger_\uparrow\Big [ 
{2(\kappa\cdot\epsilon^*_\downarrow)\over 1-x} +
{(\kappa\cdot\sigma)(\sigma\cdot\epsilon^*_\downarrow)\over x} \Big]
\chi_\uparrow\nonumber\\
&&=(\cdot\cdot\cdot){1\over\sqrt 2}~ \chi^\dagger_{\uparrow}\Big [ 
{2|\kappa|~e^{+i\phi}\over 1-x} +
{|\kappa|~e^{+i\phi} + \sigma^3|\kappa|~e^{+i\phi}\over x} \Big]
\chi_\uparrow\nonumber\\
&&=(\cdot\cdot\cdot){1\over\sqrt 2}~  
{|\kappa|~e^{+i\phi}\over x(1-x)} \Big [
2x + (1-x)(1-1) \Big]
\nonumber\\
&&= (\cdot\cdot\cdot){\sqrt 2}~   
{x|\kappa|~e^{-i\phi}\over x(1-x)} 
\label{apf3}
\end{eqnarray}
Here we have used $\chi$'s and polarization vector $\epsilon$'s as are given
earlier and the fact that $\chi^\dagger_\uparrow \sigma^3\chi_\uparrow = 1$ 
and $\chi^\dagger_\downarrow \sigma^3\chi_\downarrow = - 1$, while 
$\chi^\dagger_{\alpha} \chi_{\beta} = \delta_{\alpha\beta}$. The omitted
common factor in the above expressions is $(
\cdot\cdot\cdot)= -  { g \over \sqrt{2 (2 \pi)^3}} T^a {1\over \sqrt
{1-x}}{ x(1-x) \over (\kappa^\perp)^2 }$.

Now, from eq.(\ref{apf1}), eq.(\ref{apf2}) and eq.(\ref{apf3}), it is
straight forward to perform the required summations as follows.
For $F^{|q\rangle}_{2q}(x,Q^2)$,
\begin{eqnarray}
 {1\over 2} \sum_\sigma
\sum_{\sigma_1,\lambda_2} \mid \Phi^\sigma_{\sigma_1, \lambda_2}\mid^2\, ,
&& =  {1\over 2} \sum_\sigma (
\cdot\cdot\cdot)^2 \Big [{2|\kappa|^2\over x^2(1-x)^2}
+{2|\kappa|^2 x^2\over x^2(1-x)^2}\Big]\nonumber\\
~~&& = {g^2\over 2(2\pi)^3} C_F {1+x^2\over 1-x} \Big(
{2\over \kappa_\perp^2} \Big)\, .
\label{aps12}\\
\end{eqnarray}
Note that here in the unpolarized case averaging over initial spin
$\sigma=\pm {1\over 2}$ 
does not matter, since both are same. Next, in the polarized case of 
$\Delta q(x,Q^2)$, 
\begin{eqnarray}
\sum_{\sigma_1,\lambda_2}\sigma_1 
\mid \Phi^\uparrow_{\sigma_1, \lambda_2}\mid^2\, &&=
+{1\over 2}\Big\{ \mid \Phi^\uparrow_{\uparrow \uparrow}\mid^2
+ \mid \Phi^\uparrow_{\uparrow \downarrow}\mid^2\Big\}\nonumber\\
&&={1\over 2}~{g^2\over 2(2\pi)^3} C_F {1+x^2\over 1-x} \Big({2\over 
\kappa_\perp^2}\Big) .
\label{aps22}\\
\end{eqnarray}
Since the contribution from $\sigma_1=\downarrow$ is zero, the result in
the last equation is simply helicity $+{1\over 2}$ times the expression in
unpolarized case( eq.(\ref{aps12})). It trivially follows that if we
considered the dressed quark to be in helicity $-{1\over 2}$ state, i.e.,
$\sigma=\downarrow$ the result would have been negative of what we have in
eq.(\ref{aps22}).
On the other hand, for $\Delta g(1-x,Q^2)$, 
\begin{eqnarray}
\sum_{\sigma_1,\lambda_2}\lambda_2 \mid \Phi^\uparrow_{\sigma_1, 
\lambda_2}\mid^2 &&=
\Big\{ (+1) \mid \Phi^\uparrow_{\uparrow \uparrow}\mid^2
+~(-1)\mid \Phi^\uparrow_{\uparrow \downarrow}\mid^2\Big\}\nonumber\\
&&=(
\cdot\cdot\cdot)^2\Big \{(+1){2|\kappa|^2\over x^2(1-x)^2}
+(-1){2|\kappa|^2 x^2\over x^2(1-x)^2}\Big\}\nonumber\\
&& = {g^2\over 2(2\pi)^3} C_F (1+x)\Big( {2\over \kappa_\perp^2}\Big)\, . 
\label{aps32}
\end{eqnarray}
\section{Manifest boost symmetry of energy denominators and vertices}
In this appendix we verify the Gallilean boost invariance of vertices and
energy differences that occur in light-front time-ordered loop diagrams. 
First consider the canonical vertex given in eq.(5.4). 
Let $P^+$ and $P^\perp$  denote total longitudinal and transverse momentum in
the problem. We introduce the momentum fractions $x_i$ and the relative
transverse momenta $\kappa_i^\perp$ by 
\begin{eqnarray}
 p_2^+ = x P^+~,~ p_2^\perp =
\kappa_1^\perp + x P^\perp, ~~ q^+ = (1-x)P^+~,~ q^\perp = - \kappa_1^\perp +
(1-x) P^\perp.
\end{eqnarray}
The longitudinal momentum fractions $x_i$ and the relative
transverse momenta $\kappa_i^\perp$ obey the constraints $\sum x_i =1$ and
$\sum \kappa_i^\perp = 0$. The canonical vertex takes the form
\begin{eqnarray}
{\cal V}_1 ~= ~g ~T^a ~ \sqrt{x}~ \chi^\dagger_{s_{2}}~
 \Big [  2 {\kappa_1^\perp \over 1-x}
+ {\sigma^\perp
.\kappa_1^\perp \over x} \sigma^\perp  +im \Big( 1  - {1 \over
x} \Big) \Big ] ~\chi_{s_{1}} ~ .~
(\epsilon^\perp_\lambda)^*. 
\end{eqnarray}
In terms of the internal momenta, the boost invariance of the quark-gluon
vertex is clearly manifest.

Next consider loop diagrams. 
As an example we consider the diagram shown in Fig. 2(a).
Parameterize the single particle momenta in terms of the internal momenta 
as follows. 
\begin{eqnarray}
k_3^+ = y P^+~,~ k_3^\perp = \kappa_2^\perp + y P^\perp, ~~
k_1^+=(1-y)P^+~,~ k_1^\perp = - \kappa_2^\perp + (1-y) P^\perp.
\end{eqnarray}
Then 
\begin{eqnarray}
k^+ = k_1^+ - q^+ = (x-y) P^+, ~~ k^\perp = k_1^\perp - q^\perp =
\kappa_1^\perp - \kappa_2^\perp  + (x-y) P^\perp.
\end{eqnarray}
The energy difference appearing in the two energy denominators are, then,
\begin{eqnarray} 
p_1^- - k_1^- - k_3^- = &&- {(\kappa_2^\perp)^2 \over P^+} \Big ( {1 \over y} +
{1 \over 1-y } \Big ), \nonumber \\
~~ p_1^- - k_3^- - k^- - q^- = && - { 1 \over P^+} \Big
[ {(\kappa_2^\perp)^2 \over y} + {(\kappa_1^\perp)^2 \over 1-x} +
{(\kappa_1^\perp - \kappa_2^\perp)^2 \over x-y} \Big ].
\end{eqnarray}
The vertex factors are 
\begin{eqnarray}
 -2  {k_3^\perp \over k_3^+} + \sigma^\perp {\sigma^\perp .k^\perp \over
k^+} + {\sigma^\perp . p_2^\perp \over p_2^+} \sigma^\perp
= &&{1 \over P^+} \Big [- 2 {\kappa_2^\perp  \over y} +
\sigma^\perp  {\sigma^\perp . (\kappa_1^\perp-\kappa_2^\perp) \over x-y} + 
{\sigma^\perp .
\kappa_1^\perp \over x} \sigma^\perp \Big ], 
\nonumber \\
 -2  {q^\perp \over q^+} + \sigma^\perp {\sigma^\perp .k_1^\perp \over
k_1^+} + {\sigma^\perp . k^\perp \over k^+} \sigma^\perp 
= &&{1 \over P^+} \Big [ 2 {\kappa_1^\perp \over 1-x} - \sigma^\perp 
{\sigma^\perp . \kappa_2^\perp \over 1-y} + {\sigma^\perp .(\kappa_1^\perp- 
\kappa_2^\perp)
\over x-y} \sigma^\perp \Big ], \nonumber \\
-2  {k_3^\perp \over k_3^+} + \sigma^\perp {\sigma^\perp .p_1^\perp 
\over p_1^+} + {\sigma^\perp . k_1^\perp \over k_1^+} \sigma^\perp
= &&{ 1 \over P^+} \Big [ - 2 { \kappa_2^\perp \over y} -
{\sigma^\perp . \kappa_2^\perp \over 1-y} \sigma^\perp \Big ].
\end{eqnarray}
Thus the vertices and energy denominators appearing in Fig. 2(a) are
manifestly invariant under the Gallilean boosts in the transverse plane and
this is a general property of any $x^+$ ordered diagram in light-front
perturbation theory.


\newpage
{\flushleft {\huge\bf List of Publications}}
\vskip 1cm 

\noindent Following 
is the list of publications connected with our collaborative work.
Main results of this thesis are published in a slightly different form in
1-4 of the list.

\begin{enumerate}
\item {NONPERTURBATIVE DESCRIPTION OF DEEP INELASTIC STRUCTURE FUNCTIONS IN
LIGHT FRONT QCD.\\
By A. Harindranath, Rajen Kundu and Wei-Min Zhang;\\
Published in Phys. Rev. {\bf D59}:094012, 1999;\\
e-Print Archive: hep-ph/9806220.}

\item { DEEP INELASTIC STRUCTURE FUNCTIONS IN LIGHT FRONT QCD: RADIATIVE
CORRECTIONS.\\
By A. Harindranath, Rajen Kundu and Wei-Min Zhang;\\
Published in Phys. Rev. {\bf D59}:094013, 1999;\\
e-Print Archive: hep-ph/9806221.}

\item { UTILITY OF GALILEAN SYMMETRY IN LIGHT FRONT PERTURBATION THEORY: A
NONTRIVIAL EXAMPLE IN QCD.\\
By A. Harindranath and  Rajen Kundu;\\ 
Published in Int. Jour. Mod. Phys. {\bf A13}:4591-4604, 1998;\\
e-Print Archive: hep-ph/9802309.}

\item { ORBITAL ANGULAR MOMENTUM IN DEEP INELASTIC SCATTERING.\\
By A. Harindranath and  Rajen Kundu;\\
Published in Phys. Rev. {\bf D59}:116013, 1999;\\
e-Print Archive: hep-ph/9802406.}

\item { SUM RULE FOR THE TWIST FOUR LONGITUDINAL STRUCTURE FUNCTION.\\
By A. Harindranath, Rajen Kundu, Asmita Mukherjee and  
James P. Vary;\\
Published in Phys. Lett. {\bf B417}:361-368, 1998;\\
e-Print Archive: hep-ph/9711298.}

\item {TWIST FOUR LONGITUDINAL STRUCTURE FUNCTION IN LIGHT FRONT QCD.
By A. Harindranath, Rajen Kundu, Asmita Mukherjee and  James P.
Vary;\\
Published in Phys. Rev. {\bf D58}:114022, 1998;\\
e-Print Archive: hep-ph/9808231.}  

\end{enumerate}

\vskip .5cm
------------------------------------------------------------------------------
\vskip 2cm
{\centerline {\large\bf Acknowledgement}}
\vskip 1cm
It is very hard, if not impossible, to single out a few of  
all those who, by their comments, advice and active support, helped me
out of all the difficult situations --- be it academic or nonacademic --- 
that I
encountered during my research career. I would like to take this opportunity
to express my deep gratitude to all of them. 

Specially, I would like to thank my supervisor Prof. A. Harindranath for
providing all the inspiration and guidance to carry out the research work
pertaining to this thesis. I am grateful to him for giving me the necessary
exposure to the frontiers of Light-Front Field Theory and for prompt help
whenever I needed it most. 

I would also like to thank my collaborator Prof. W. M. Zhang for sharing his
views and illuminating discussions that I had with him. I am thankful to
Prof. J. P. Vary and A. Mukherjee for their active participation to make the
research program a success. 

A special debt of gratitude is due to them from whom I got the inspiration,
in the first place, to study physics during my college days. The 
support that I always got 
from my parents and other family members as well as 
my friends during hard times of my research career 
cannot be expressed with my limited
vocabulary.  Lastly, I would like to thank all the members of Theory Group of
SINP for their  cooperation during my research work.
\vskip 2.5cm
\noindent Calcutta, \hspace{8.6cm} Rajen Kundu\\
\noindent 1 June,1999 \hspace{8cm} Theory Group, SINP

\end{document}